 \documentclass[final,5p,times,twocolumn,authoryear]{elsarticle}

\usepackage{amssymb}
\usepackage{lipsum}

\usepackage{subfigure}
\usepackage{subcaption}
\usepackage{verbatim}
\usepackage{multirow}
\usepackage{amsmath}
\usepackage{color}
\usepackage{ulem}

\journal{XXX}

\begin{document}

\begin{frontmatter}



\title{Advection-based multiframe iterative correction for pressure estimation from velocity fields}


\author[a]{Junwei Chen}\corref{mycorrespondingauthor}
\cortext[mycorrespondingauthor]{Corresponding author}
\ead{junwei.chen@uc3m.es}
\author[a]{Marco Raiola}
\ead{mraiola@ing.uc3m.es}
\author[a]{Stefano Discetti}
\ead{sdiscett@ing.uc3m.es}
\address[a]{Department of Aerospace Engineering, Universidad Carlos III de Madrid, Avda. Universidad 30, 28911, Leganes, Spain}


\begin{abstract}
A novel method to improve the accuracy of pressure field estimation from time-resolved Particle Image Velocimetry data is proposed. This method generates several new time-series of velocity field by propagating in time the original one using an advection-based model, which assumes that small-scale turbulence is advected by large-scale motions. Then smoothing is performed at the corresponding positions across all the generated time-series. The process is repeated through an iterative scheme. The proposed technique smears out spatial noise by exploiting time information. Simultaneously, temporal jitter is repaired using spatial information, enhancing the accuracy of pressure computation via the Navier-Stokes equations. We provide a proof of concept of the method with synthetic datasets based on a channel flow and the wake of a 2D wing. Different noise models are tested, including Gaussian white noise and errors with some degree of spatial coherence. Additionally, the filter is evaluated on an experimental test case of the wake of an airfoil, where pressure field ground truth is not available. The result shows the proposed method performs better than conventional filters in velocity and pressure field estimation, especially when spatially coherent errors are present. The method is of direct application in advection-dominated flows, although its extension with more advanced models is straightforward.
\end{abstract}




\end{frontmatter}




\section{Introduction}
\label{introduction}

Particle Image Velocimetry (PIV) has established in the last decade as a non-intrusive technique for instantaneous pressure field estimation \citep{van2013piv}. The availability of spatially and temporally resolved velocity field measurements allows extracting the pressure from the integration of the Navier-Stokes equation. This process is well known to be strongly sensitive to noise errors in the measurement of the velocity fields since the finite difference amplifies it \citep{azijli2016posteriori,pan2016error,mcclure2017instantaneous,liu2020error,zhang2022uncertainty}. High-quality PIV fields are fundamental to obtain accurate pressure field measurements. This is not always possible, due to the presence of well-known uncertainty sources in time-resolved PIV measurements \citep{sciacchitano2019uncertainty}.

Generally speaking, noise suppression in velocity field measurements is often tackled through data filtering. Typical solutions include convolution in the space-temporal domain, based for instance on Gaussian kernels or the Savitzky-Golay (SG) filter, filters in the Fourier domain \citep{foucaut2002some} or in the domain defined by the Proper Orthogonal Decomposition (POD) eigenfunctions \citep{Raiola2015POD,brindise2017proper,epps2019singular}. These solutions are often combined with data assimilation, in which the governing equations can be used to regularize the outcome. The complete Navier-Stokes equations can be directly plugged in the optimization process of the velocity field correction, allowing a direct estimation of the pressure, as in the adjoint approach proposed by \cite{lemke2016adjoint} for compressible flows and in the super-resolution technique developed by \cite{he2024four}. Partly due to the considerable computation cost of the adjoint approach, a simpler alternative, based on solenoidal filtering to impose only divergence-free conditions to incompressible flows, has been also explored \citep{silva2013minimization,schiavazzi2014matching,wang2016Divergence}. Other popular regularization approaches are the vortex-in-cell (VIC) method \citep{christiansen1973numerical,schneiders2014Time,scarano2022dense}, based on vorticity transport for an inviscid incompressible fluid
, and the B-Spline or radial basis function regression from particle tracks \citep{gesemann2015particle,sperotto2022meshless}.

The main drawback of the methods based on leveraging the governing equations to compute the pressure is that they require 3D data to be properly implemented. It is well known, however, that under conditions of advection-dominated flow, simpler constraints can be imposed to extract pressure (see e.g. \citealp{de2012pressure, van2019pressure}). Our main hypothesis is that the same principle could also be leveraged to regularize the data.

In this work an Advection-based Multiframe Iterative Correction (AMIC) is proposed to regularize time-resolved PIV fields in advection-dominated flows. Using an advection model for the fluctuating part, the velocity field can be propagated to previous and successive frames. This generates several parallel time-series of velocity fields, enabling filtering across the parallel fields at corresponding space-time position. An iterative correction is applied for better numerical stability. The correction results in an improvement of both the velocity and pressure estimation.

The proposed AMIC method, as well as the methods to which it is compared, will be introduced in Sec. \ref{sec:method}. Two synthetic validation datasets will be described in Sec. \ref{sec:data}. After that, an elbow of errors in iteration and the result will be discussed in Sec. \ref{sec:loop} and Sec. \ref{sec:result}. Then, performance evaluation that can also be used in the absence of a ground truth will be introduced in Sec. \ref{sec:index} for the two synthetic datasets. The discussed methods and tools are further tested on a real experimental dataset of an airfoil wake in Sec. \ref{sec:experiment}. Finally, the conclusions are drawn.

\section{Method}
\label{sec:method}
\subsection{Advection-based Multiframe Iterative Correction}

\begin{figure}
	\centering 
	\includegraphics[width=0.5\textwidth]{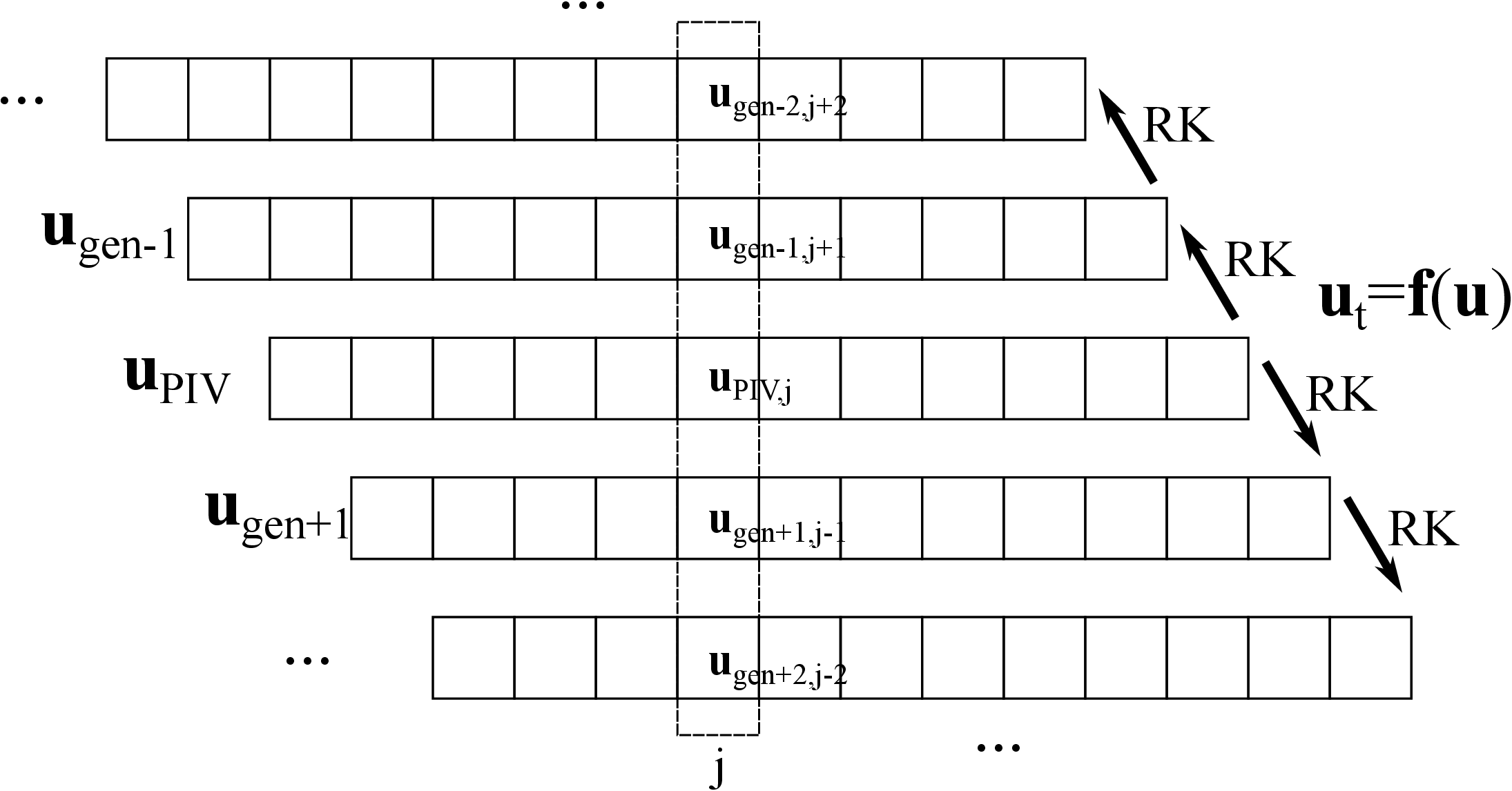}	
	\caption{The sketch of generating parallel time-series of the velocity field from the original one by propagating forward and backward in time using the governing equation (RK stands for Runge-Kutta) and the update across different time-series.} 
	\label{fig:sketch}
\end{figure}

In the proposed method, each frame of the velocity field can be propagated from the neighbouring frames, either in forward or backward direction, as shown in Fig. \ref{fig:sketch}. This propagation creates several parallel time-series of velocity field $\mathbf{u}_{gen}$ from the original PIV time-series $\mathbf{u}_{PIV}$, thus adding one more dimension to the dataset. Then the smoothing takes place in this added dimension, i.e. using for each grid point the velocity data computed from forward and backward projection of adjacent (in time) snapshots. We prove that this approach reduces detail loss compared to other methods applying the filter in the spatial or temporal domain.

Different physics-based models can be used for the velocity field propagation, e.g. full Navier-Stokes equation (requiring solving pressure together or alternately) or vorticity transport equations (e.g. VIC). In this specific work an advection model based on Taylor's frozen turbulence hypothesis is introduced. The hypothesis assumes small-scale flow motions are advected by the large-scale motions. This can be described as the material derivative of small-scale motions being negligible,
\begin{equation}
    \frac{D\mathbf{u}'}{D t} = \frac{\partial \mathbf{u}'}{\partial t} +(\mathbf{u}_c\cdot\nabla)\mathbf{u}' \approx 0,
\end{equation}
where the $\mathbf{u}_c$ is the local convective velocity, and the $\mathbf{u}'$ is the fluctuating velocity from Reynolds decomposition. The convective velocity depends mostly on large-scale motions. A method to extract it from instantaneous fields is reported in the appendix. Then, assuming that the timescale of large-scale motions is much larger than that of small-scale motions, the approximate temporal derivative of the velocity field can be
\begin{equation}
    \frac{\partial\mathbf{u}}{\partial t}=-(\mathbf{u}_c\cdot\nabla)\mathbf{u}'.
    \label{eqn:TH}
\end{equation}

The temporal derivative of the velocity field for the propagation can thus be obtained from a single snapshot, this is the governing equation of the advection-based model. This hypothesis considers the fluctuation as advecting passively with a local convective velocity, as previously discussed in \cite{de2012pressure}, while rotation and strain are not taken into consideration.

The advection model of \cite{de2012pressure} is used to directly estimate pressure. The main novelty of AMIC is to push further this principle by enforcing it for data regularization. The velocity fields are then propagated bidirectionally using the 4$^{th}$-order Runge-Kutta method, generating a couple of sequences of velocity fields with the same time separation of the PIV acquisition. In the following, the sequences $\mathbf{u}_{gen-1}$, $\mathbf{u}_{gen+1}$, $\mathbf{u}_{gen-2}$, $\mathbf{u}_{gen+2}$, ... will be used to refer to the time-series obtained by propagating the original time-series $\mathbf{u}_{PIV}$ of one step backwards and forward and 2 steps backwards and forward, respectively, as shown in Fig. \ref{fig:sketch}. Then the $j^{th}$ frame in the time-series $\mathbf{u}_{PIV,j}$ is updated by the weighted average from the original and propagated time-series,
\begin{equation}
    \mathbf{u}_{PIV,j} \leftarrow (1-2\sum_i^N\lambda_i)\mathbf{u}_{PIV,j} + \sum_i^N (\lambda_i \mathbf{u}_{gen-i,j+i} + \lambda_i\mathbf{u}_{gen+i,j-i}),
\end{equation}
where the weight coefficients $\lambda_i$ are chosen so to comply with $0<\lambda_i<1$ and $(1-2\sum_i^N\lambda_i) > 0$, $N$ is the number of elements in $\lambda_i$. This process is implemented through under-relaxation Modified Richardson iteration \citep{richardson1911ix} in the additional dimension. The value of $\lambda_i$ is tuned with a trial and error process to balance the need of a fast update rate and the distortion caused by the propagation through a long time interval. The time-series propagation step and the correction step described above are then iterated to improve smoothing. A criterion to select the number of iterations is provided in section \ref{sec:loop}.

\subsection{Baseline filtering methods for comparison}

The AMIC method is compared with two widely used filtering techniques: the Savitzky-Golay filtering, and optimal POD truncation. 

The Savitzky-Golay (SG) filter \citep{savitzky1964smoothing} is a widely used tool for smoothing data based on local fitting with low-degree local polynomials. The filter operates by a convolution on the data with a kernel derived from the linear least squares method when the data points are equally spaced. In this paper a $5\times5\times5\times5$ 4D (3D space + time), $2^{nd}$ order polynomial filter for time-resolved 3D field and a $5\times5\times5$ 3D (2D space + time), $2^{nd}$ order polynomial filter for time-resolved 2D field are applied to the velocity field as a reference for the proposed AMIC method.

The POD truncation works on the assumption that, for a large enough dataset, the noise is nearly uniformly distributed in the POD spectrum, while the signal decays quickly as the mode number increases. Similarly to \cite{Raiola2015POD}, a cut-off condition is set based on the squared ratio of consecutive singular values $(\sigma_{i+1}/\sigma_i)^2>99\%$. To remove the effects of its fluctuations, in this work the cutoff is set when the threshold is passed by at least 3 consecutive modes. One of the limits of this method is that it requires abundant statistically independent snapshots to converge. This condition is often not fulfilled by time-resolved volumetric PIV experiments.


\section{Dataset for validation}
\label{sec:data}

\subsection{Channel flow}

The performance of the proposed method is tested on a synthetic data set from JHTDB (Johns Hopkins Turbulence Databases) \citep{li2008JHTDB}. The simulation is carried out in a channel of $8\pi h\times2h\times3\pi h$ with $h$ being the half-channel height and at a friction-based Reynolds number $Re_\tau\approx1000$. The subdomain for validation is extracted with a size $h\times h \times 0.15h$ extending from the wall to the centre of the channel, and the velocity and pressure field are interpolated to a $88\times88\times12$ grid. The temporal separation of the data used for validation is the same as the database storage, i.e. $0.0065$ non-dimensional time units. The convective time of 1 is based on the bulk velocity $U_b$ and the half-channel height $h$. A time series of $100$ frames of the velocity and pressure fields is used for comparison, although POD was trained on a larger dataset ($2400$ frames) to improve convergence.

\begin{figure}[ht]
\flushleft
\begin{minipage}{0.96\linewidth}
    \,\,\,\,\,\,\,\,\,\,
    \includegraphics[width=0.4\linewidth,trim=4mm 132mm 4mm 4mm,clip]{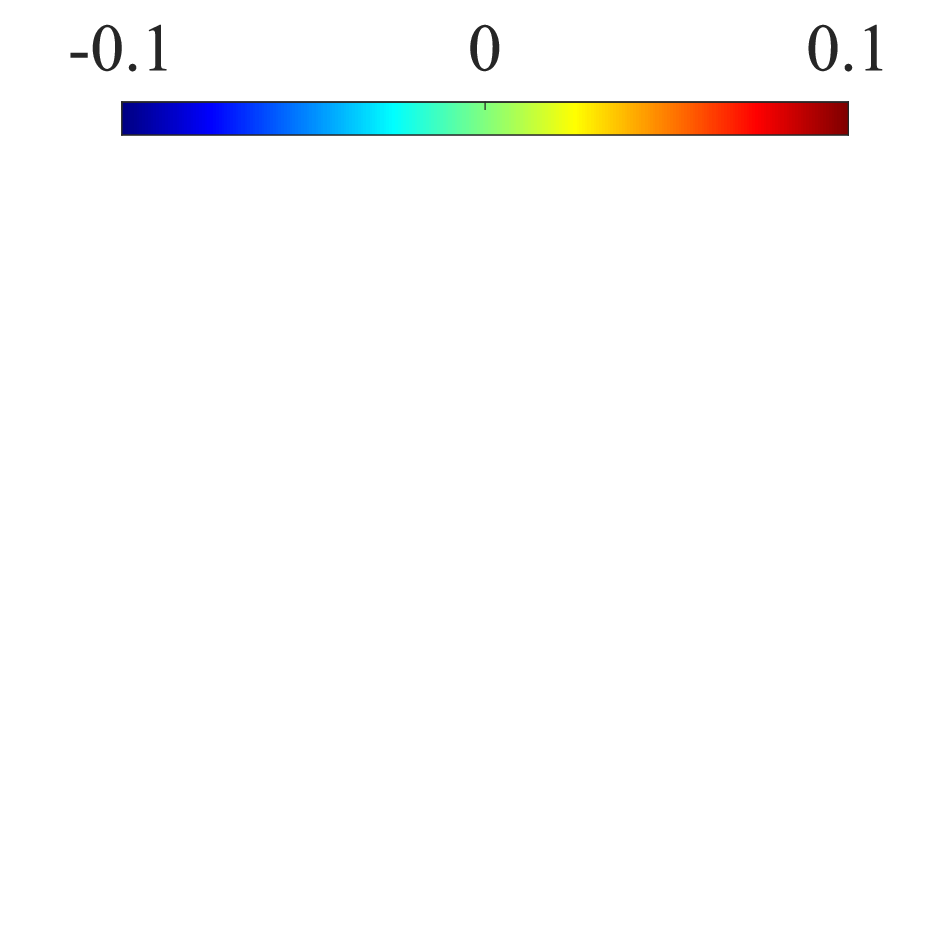}
    \,\,\,\,\,\,
    \includegraphics[width=0.4\linewidth,trim=4mm 132mm 4mm 4mm,clip]{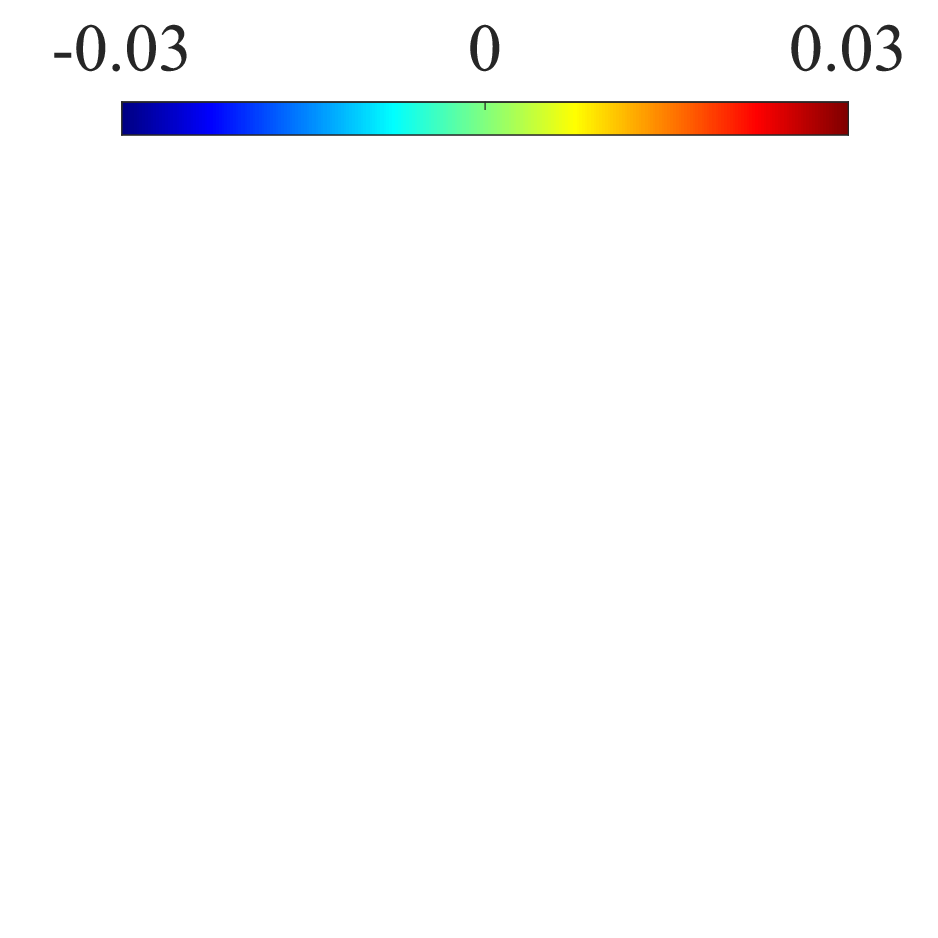}
\end{minipage}
\flushleft
\begin{minipage}{0.03\linewidth}
    \rotatebox{90}{\hspace{5mm}clean field}
\end{minipage}
\begin{minipage}{0.96\linewidth}
    \begin{subfigure}
        \centering
        \includegraphics[width=0.45\linewidth,trim=0mm 24mm 8mm 16mm,clip]{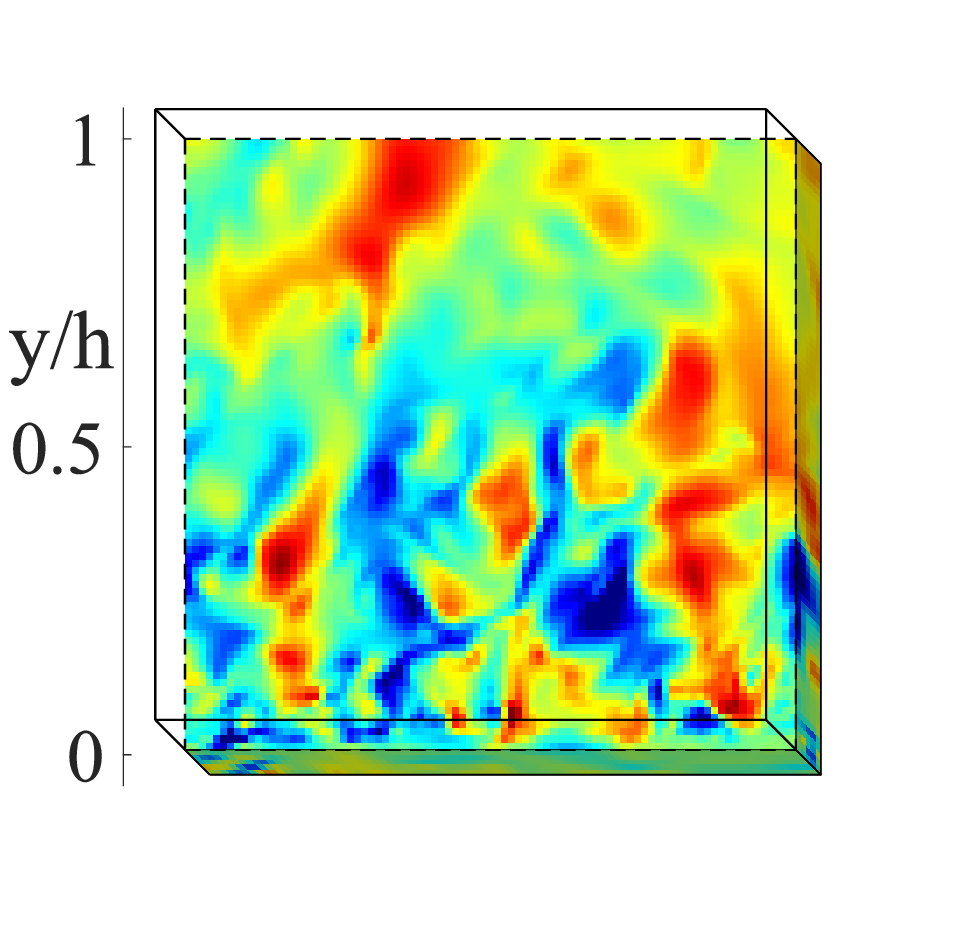}
    \end{subfigure}
\end{minipage}
\flushleft
\begin{minipage}{0.03\linewidth}
    \rotatebox{90}{\hspace{5mm}GWN}
\end{minipage}
\begin{minipage}{0.96\linewidth}
    \begin{subfigure}
        \centering
        \includegraphics[width=0.45\linewidth,trim=0mm 24mm 8mm 16mm,clip]{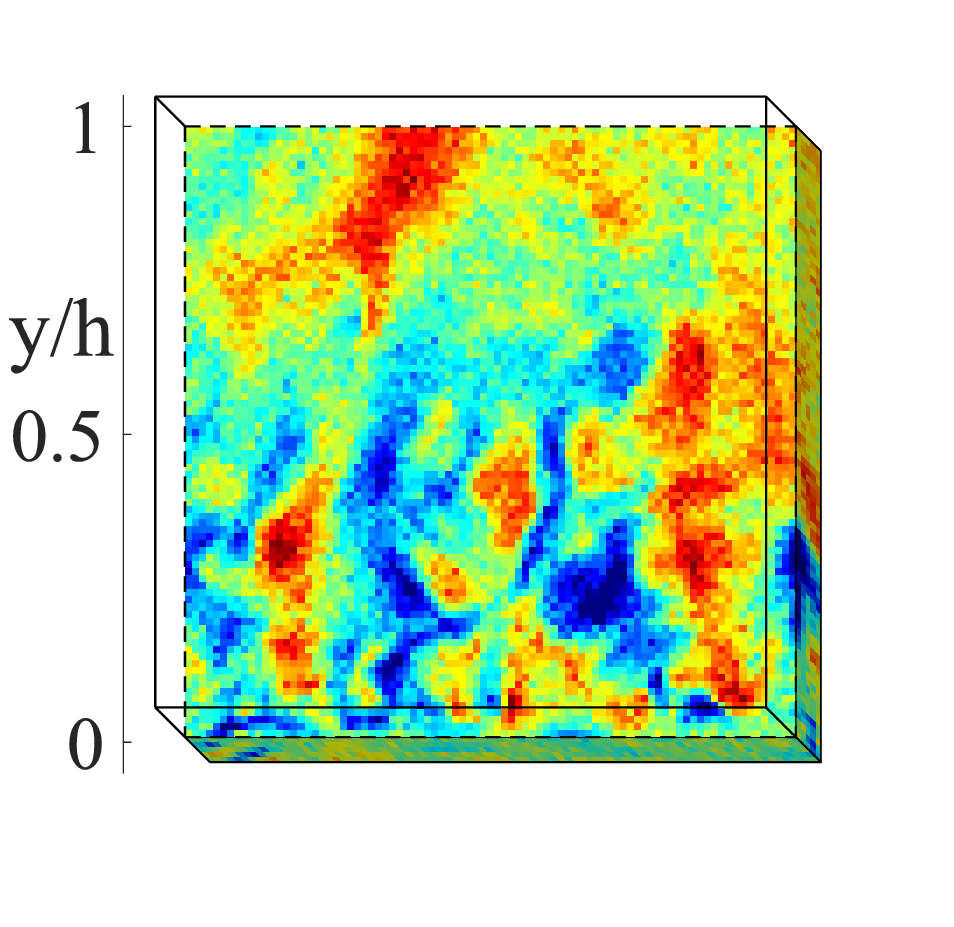}
    \end{subfigure}
    \begin{subfigure}
        \centering
        \includegraphics[width=0.45\linewidth,trim=0mm 24mm 8mm 16mm,clip]{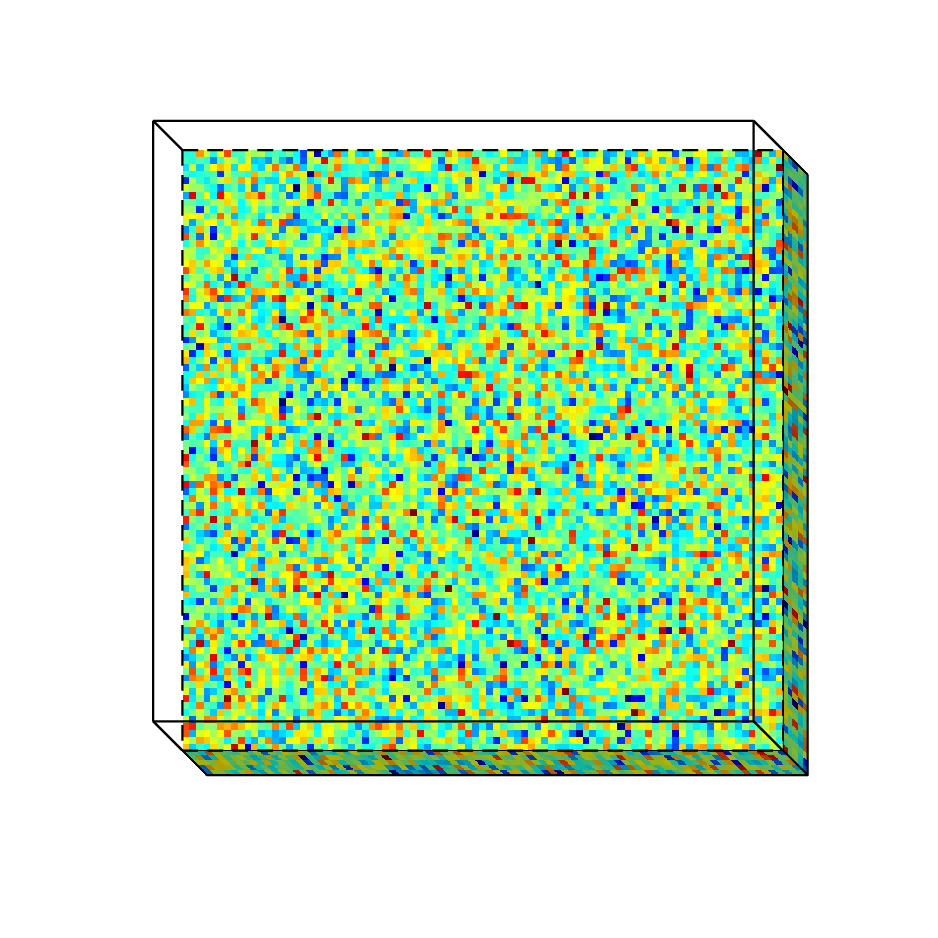}
    \end{subfigure}
\end{minipage}
\flushleft
\begin{minipage}{0.03\linewidth}
    \rotatebox{90}{\hspace{5mm}CGWN}
\end{minipage}
\begin{minipage}{0.96\linewidth}
    \begin{subfigure}
        \centering
        \includegraphics[width=0.45\linewidth,trim=0mm 0mm 8mm 20mm,clip]{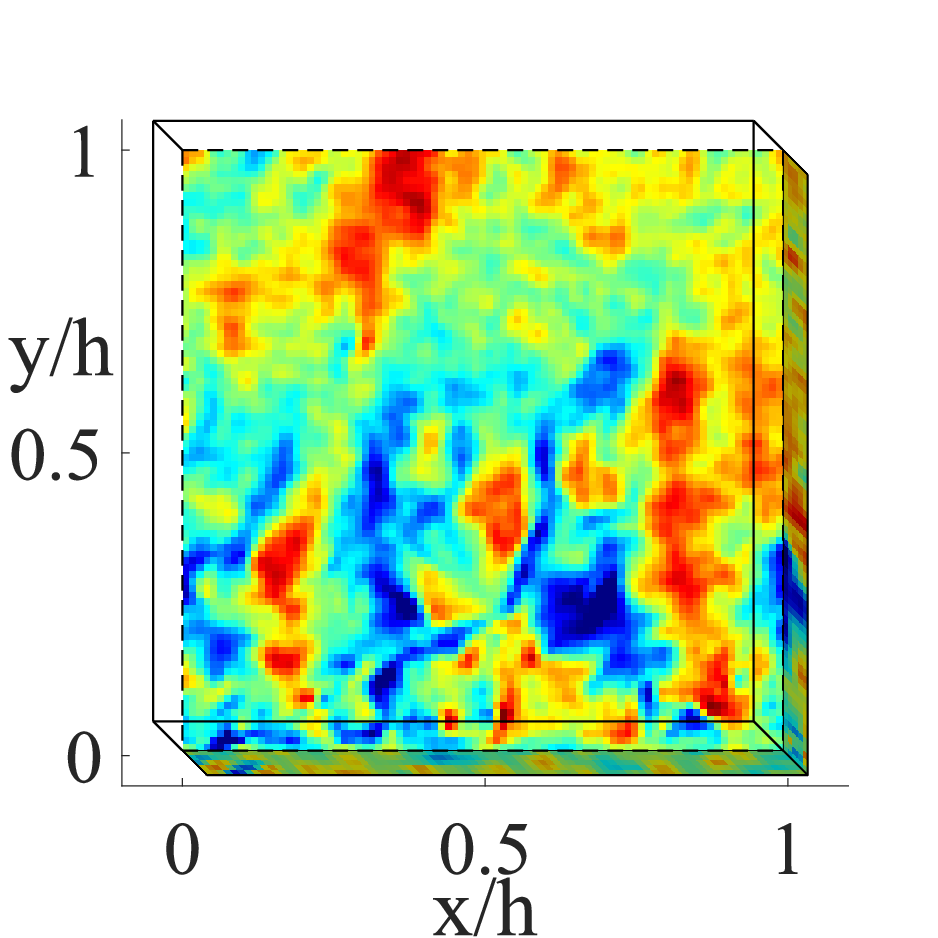}
    \end{subfigure}
    \begin{subfigure}
        \centering
        \includegraphics[width=0.45\linewidth,trim=0mm 0mm 8mm 20mm,clip]{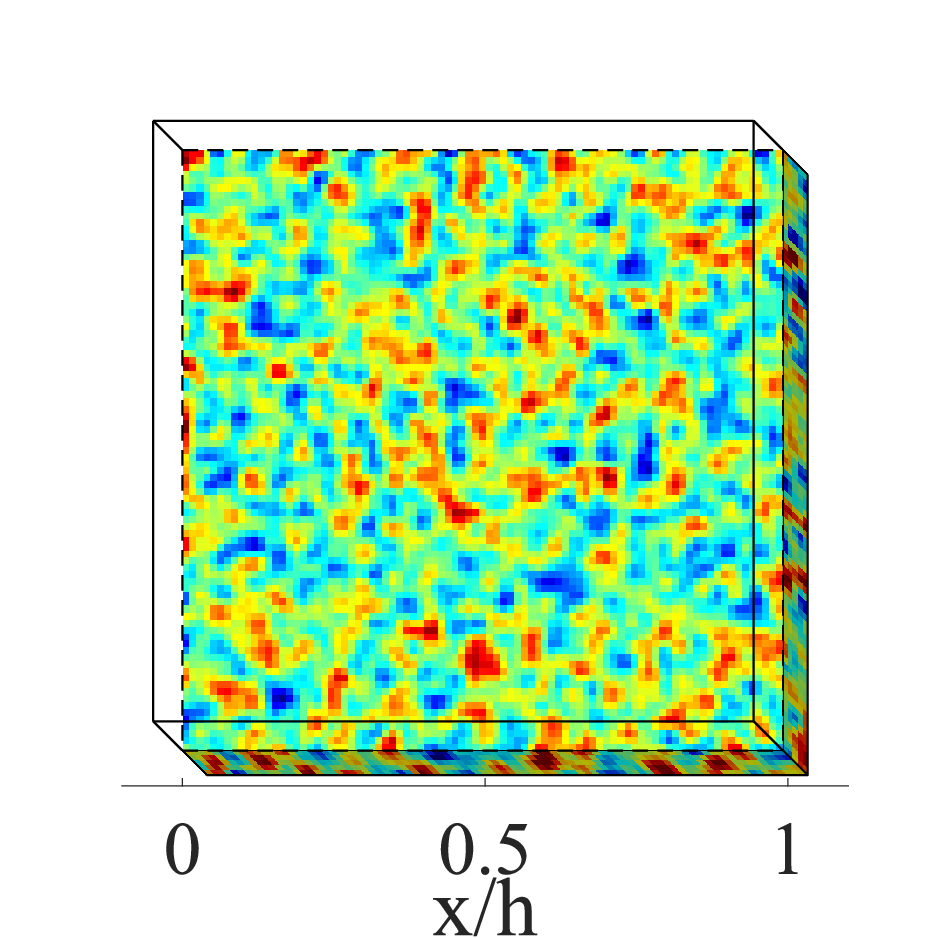}
    \end{subfigure}
\end{minipage}
\caption{Wall-normal velocity field (left column) and superposed generated noise (right column). From top to bottom, original, original+GWN, and original+CGWN fields are shown.}
\label{fig:noise}
\end{figure}

Two types of noise are added to the volumetric velocity field of the channel flow to simulate a PIV experiment. The first one is Gaussian white noise (GWN) with a standard deviation of $25\%$ of the root mean square (RMS) value of the fluctuating velocity. Since overlapping interrogation windows are widely used in PIV to obtain a finer vector spacing, noise might exhibit some degree of spatial correlation related to local image and flow features. To simulate this effect, the convolution of Gaussian white noise (CGWN) is generated by spatially smoothing the GWN with a $\sigma=1$ Gaussian kernel in the 3D space. In order to have the same standard deviation as the GWN, the CGWN is multiplied by an empirical scale factor of $6$\footnote{This is equivalent to a problem of the expectation of the standard deviation of the convolution over a distribution $s_i$ using a kernel $X$. Thus the reciprocal of the scaling factor can be computed from the root-sum-square of a $5\times5\times5$ Gaussian kernel with zero mean and unit standard deviation, which is close to $1/6$. An alternative method based on Cholesky decomposition of the covariance matrix can be found in \cite{azijli2015solenoidal, mcclure2017instantaneous}}. The two types of noise as well as the wall-normal component of velocity before and after adding noise are shown in Fig. \ref{fig:noise}. The figure shows a 2D slice of the mid $xy$-plane in the 3D field (with $x, y$ being respectively the streamwise and wall-normal direction). Slices of the $zx$-plane and $yz$-plane on the edges of the volume are partially shown (with $z$ being the spanwise direction). In the figures, the wall is located at $y = 0$.

\subsection{Synthetic airfoil wake}

\begin{figure}[htb]
\flushleft
\begin{minipage}{0.96\linewidth}
    \hspace{6mm}
    \includegraphics[width=0.92\linewidth,trim=4mm 100mm 4mm 0mm,clip]{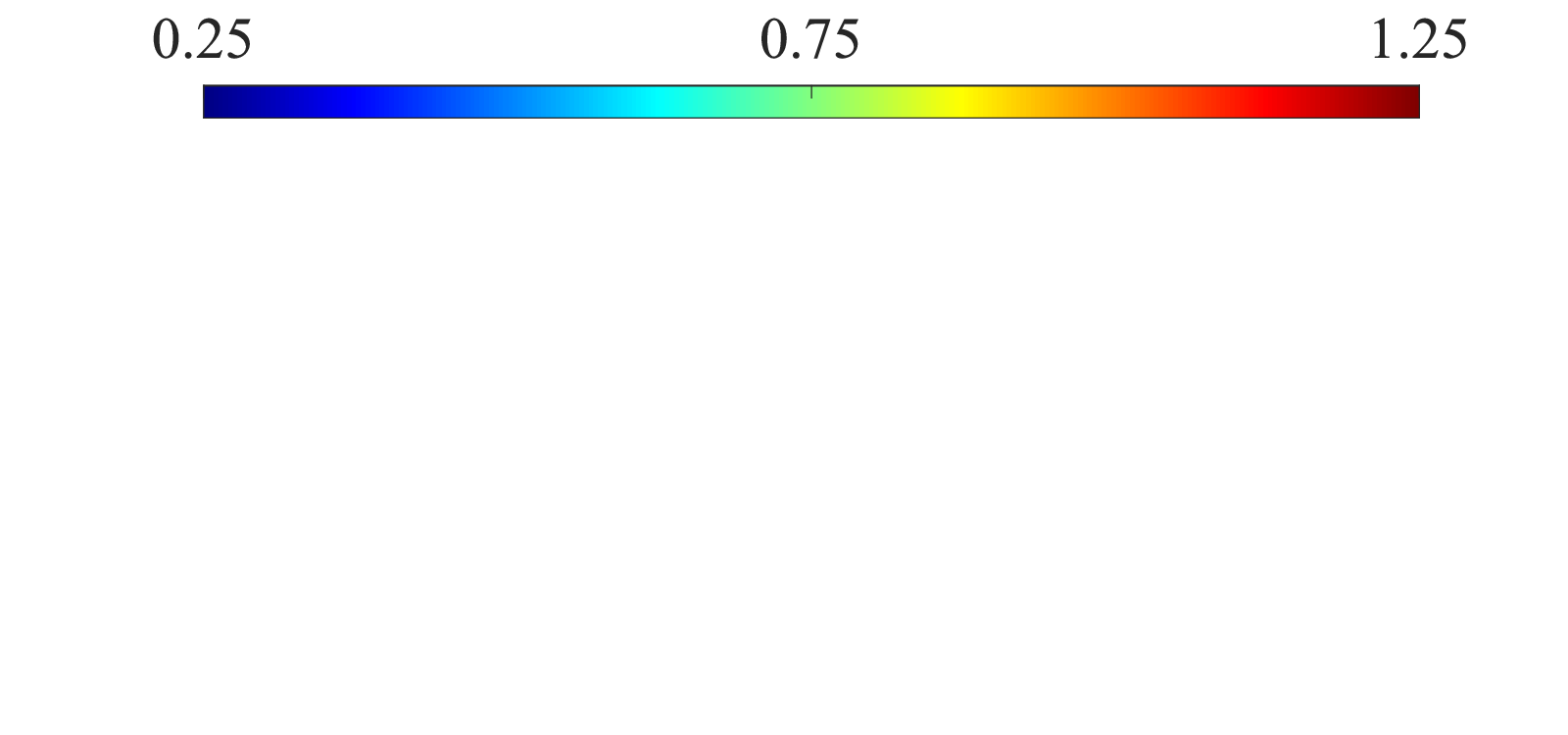}
\end{minipage}
\flushleft
\begin{minipage}{0.03\linewidth}
    \rotatebox{90}{\hspace{0mm}clean field}
\end{minipage}
\begin{minipage}{0.96\linewidth}
    \begin{subfigure}
        \centering
        \includegraphics[width=0.96\linewidth,trim=0mm 24mm 6mm 16mm,clip]{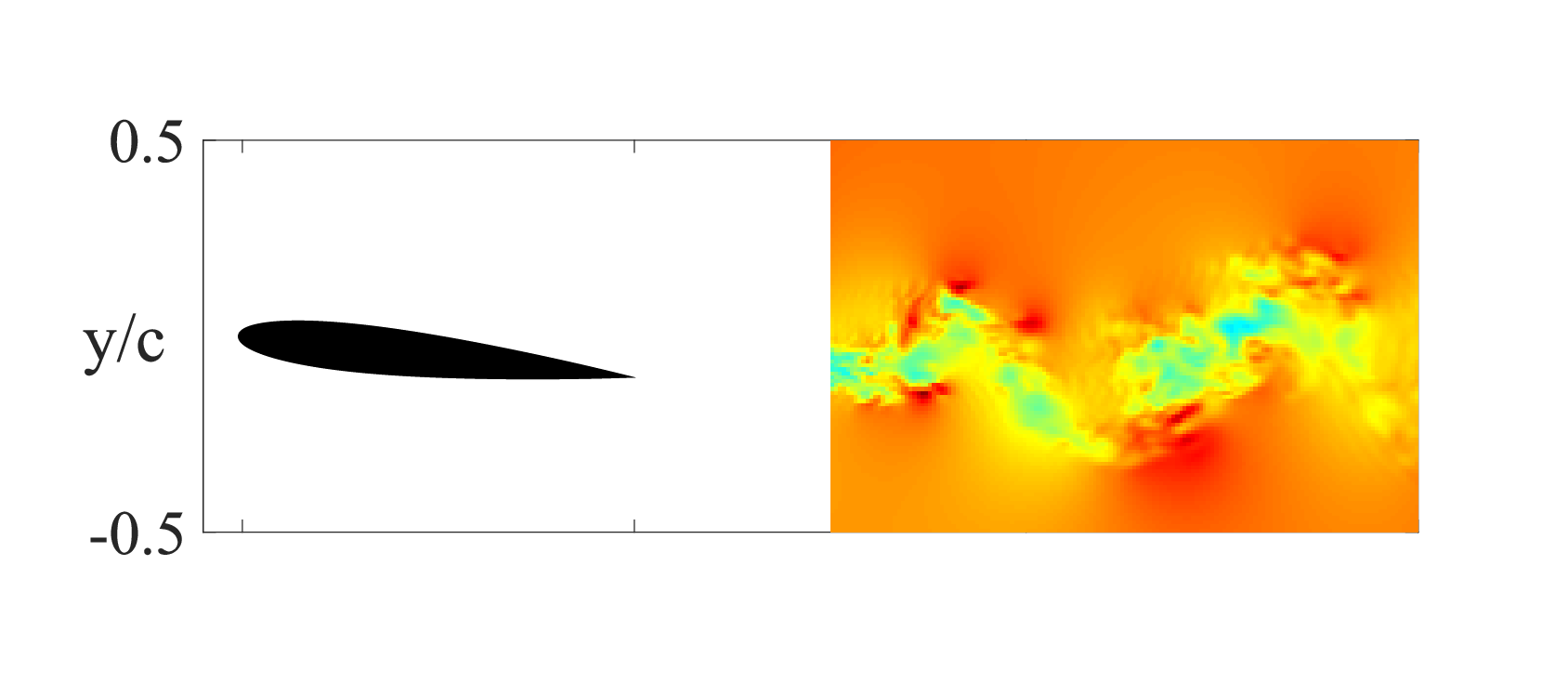}
    \end{subfigure}
\end{minipage}
\flushleft
\begin{minipage}{0.03\linewidth}
    \rotatebox{90}{\hspace{0mm}GWN}
\end{minipage}
\begin{minipage}{0.96\linewidth}
    \begin{subfigure}
        \centering
        \includegraphics[width=0.96\linewidth,trim=0mm 24mm 6mm 16mm,clip]{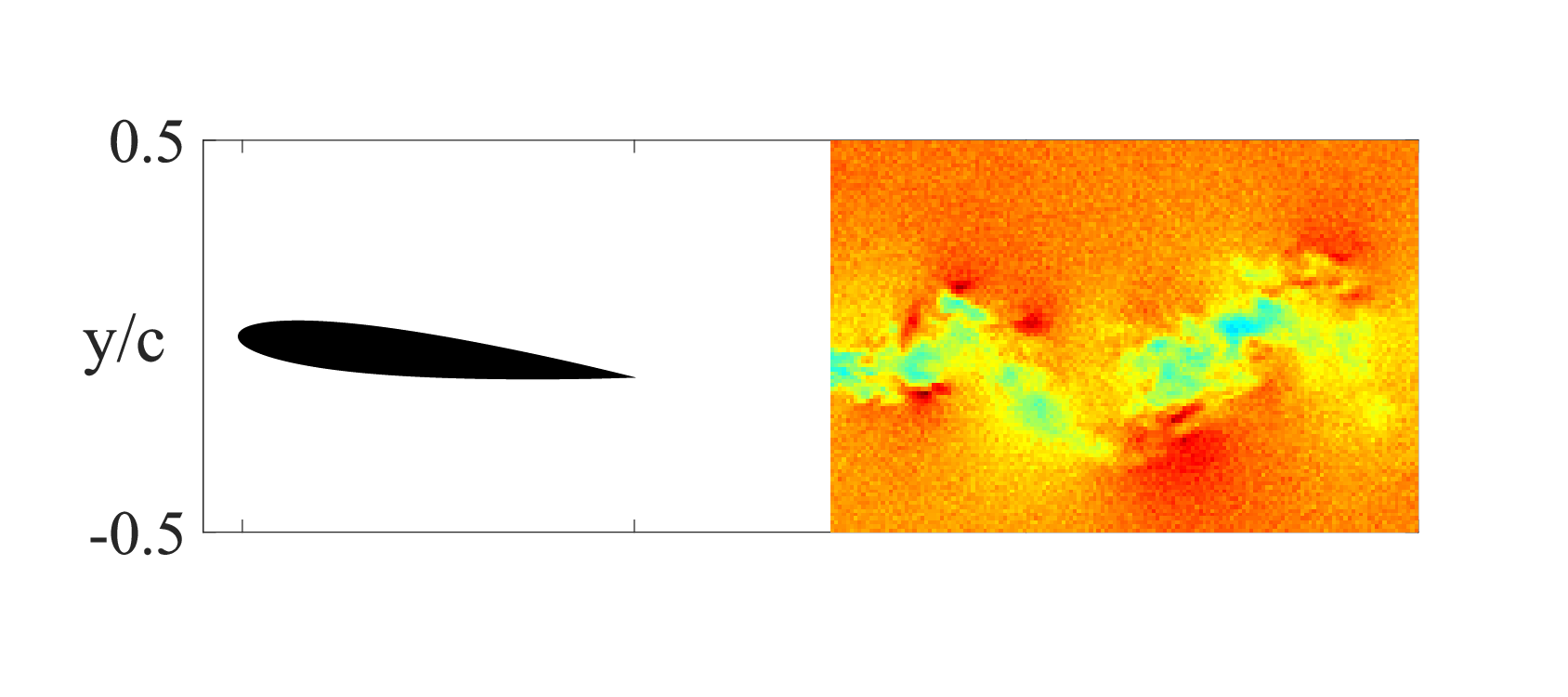}
    \end{subfigure}
\end{minipage}
\flushleft
\begin{minipage}{0.03\linewidth}
    \rotatebox{90}{\hspace{7mm}CGWN}
\end{minipage}
\begin{minipage}{0.96\linewidth}
    \begin{subfigure}
        \centering
        \includegraphics[width=0.96\linewidth,trim=0mm 0mm 6mm 20mm,clip]{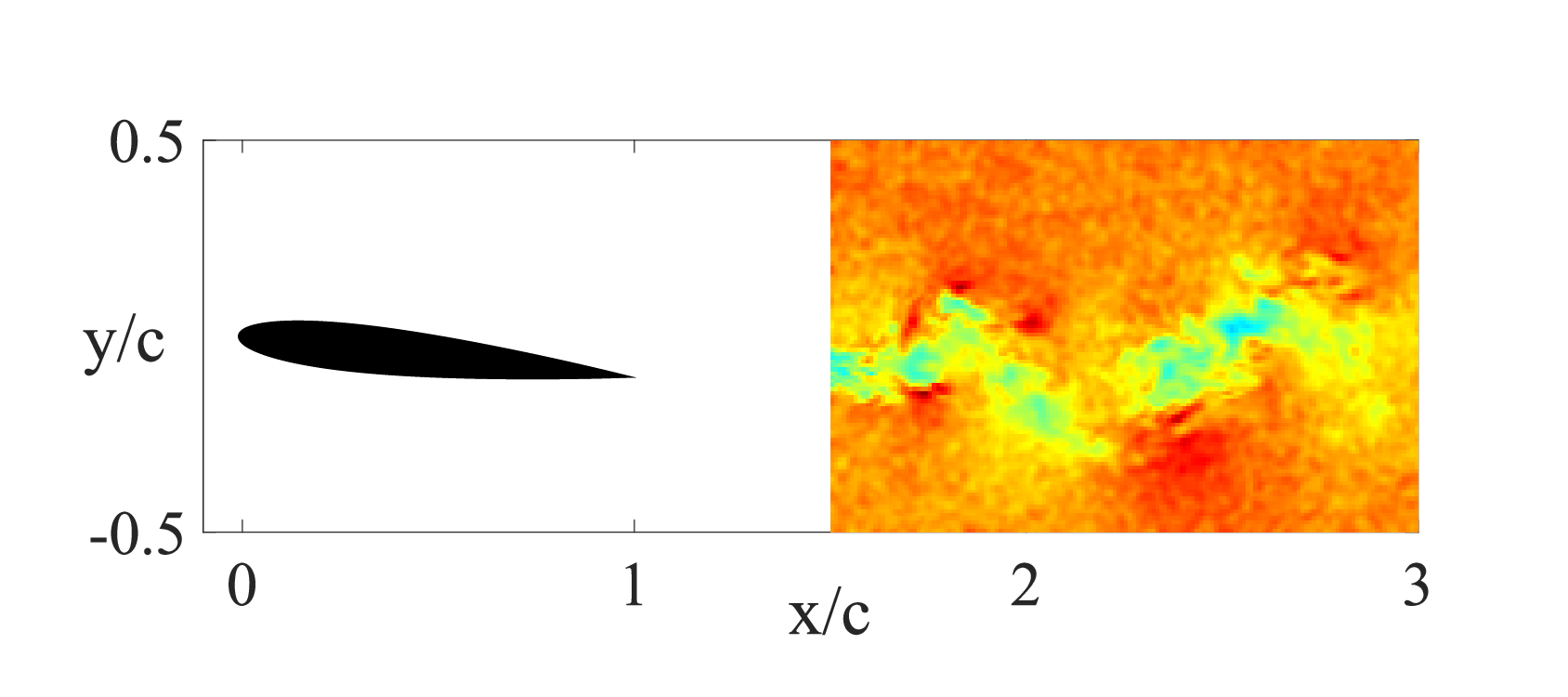}
    \end{subfigure}
\end{minipage}
\caption{Normalized streamwise velocity field in the wake of an airfoil from LES \citep{towne2023database}. From top to bottom, original, original+GWN, and original+CGWN field are shown.}
\label{fig:AF_noise}
\end{figure}

A second synthetic dataset is extracted from a Large Eddy Simulation of a turbulent wake downstream of a NACA 0012 airfoil subject to a uniform flow with freestream velocity $U_\infty$ and an angle of attack of $6^\circ$ \citep{towne2023database}. The Reynolds number is $Re = 23000$, and the Mach number is $Ma = 0.3$. The simulation is spanwise-periodic and the 3-component velocity field on the middle profile in spanwise is stored.

The 2-component velocity field on the middle plane is interpolated to an $101\times151$ grid in $x/c\in[1.5,3]$ and $y/c\in[-0.5,0.5]$ in Cartesian coordinate for validation, with $c$ being the chord length and the leading edge locating at $(0,0)$, as displayed in Fig. \ref{fig:AF_noise}. $320$-frame time-series are prepared for validation with the temporal increment at $0.104$, the same as in the original database. 
For the assessment against POD filtering, a larger database of $3200$ frames is used to improve convergence of the POD modes.

\section{Performances and stopping criterion of AMIC}
\label{sec:loop}

\begin{figure*}[h]
\begin{subfigure}
    \centering
    \hspace{18mm}
    \includegraphics[width=0.4\linewidth,trim=4mm 20mm 10mm 0mm,clip]{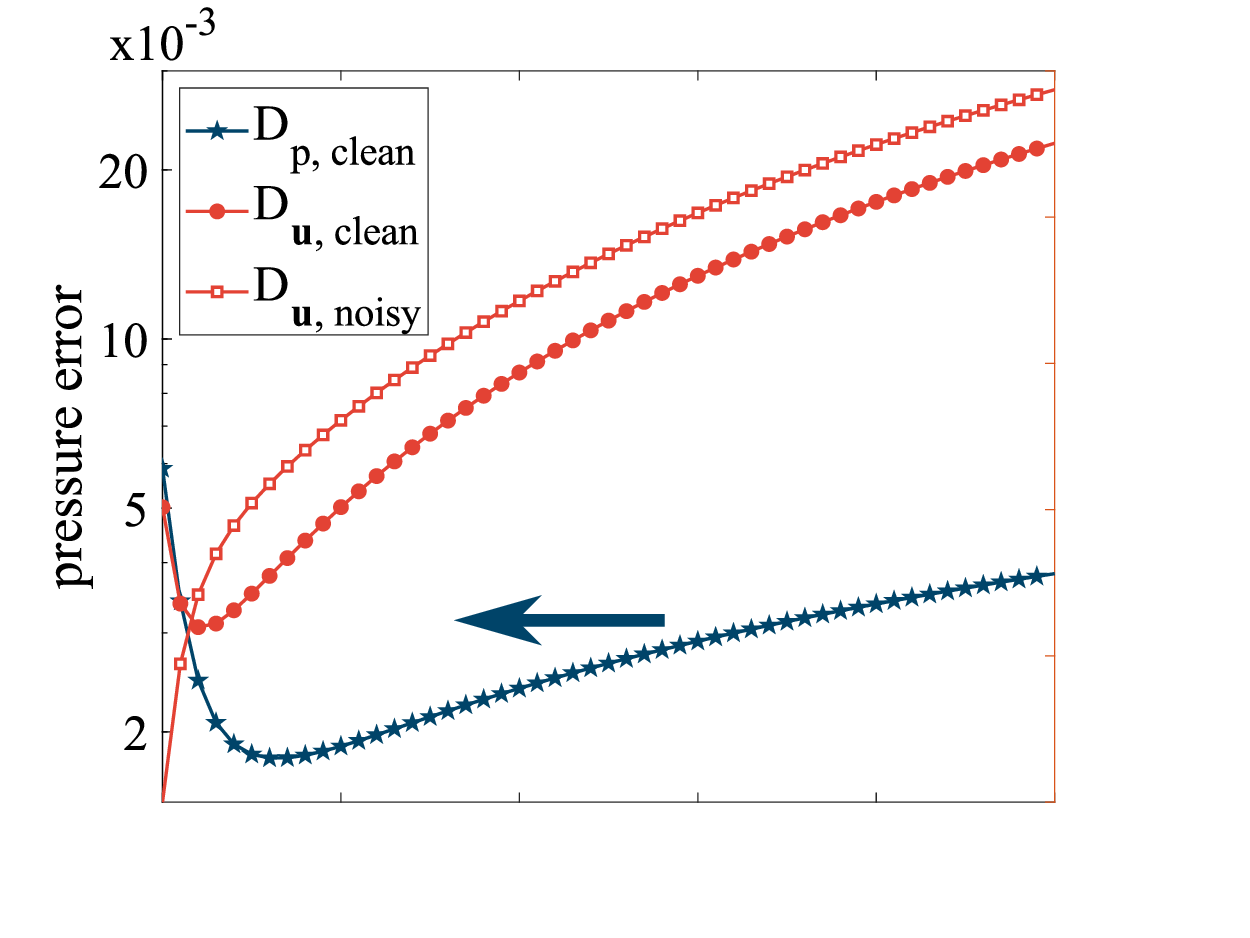}
\end{subfigure}
\begin{subfigure}
    \centering
    \includegraphics[width=0.4\linewidth,trim=14mm 20mm 0mm 0mm,clip]{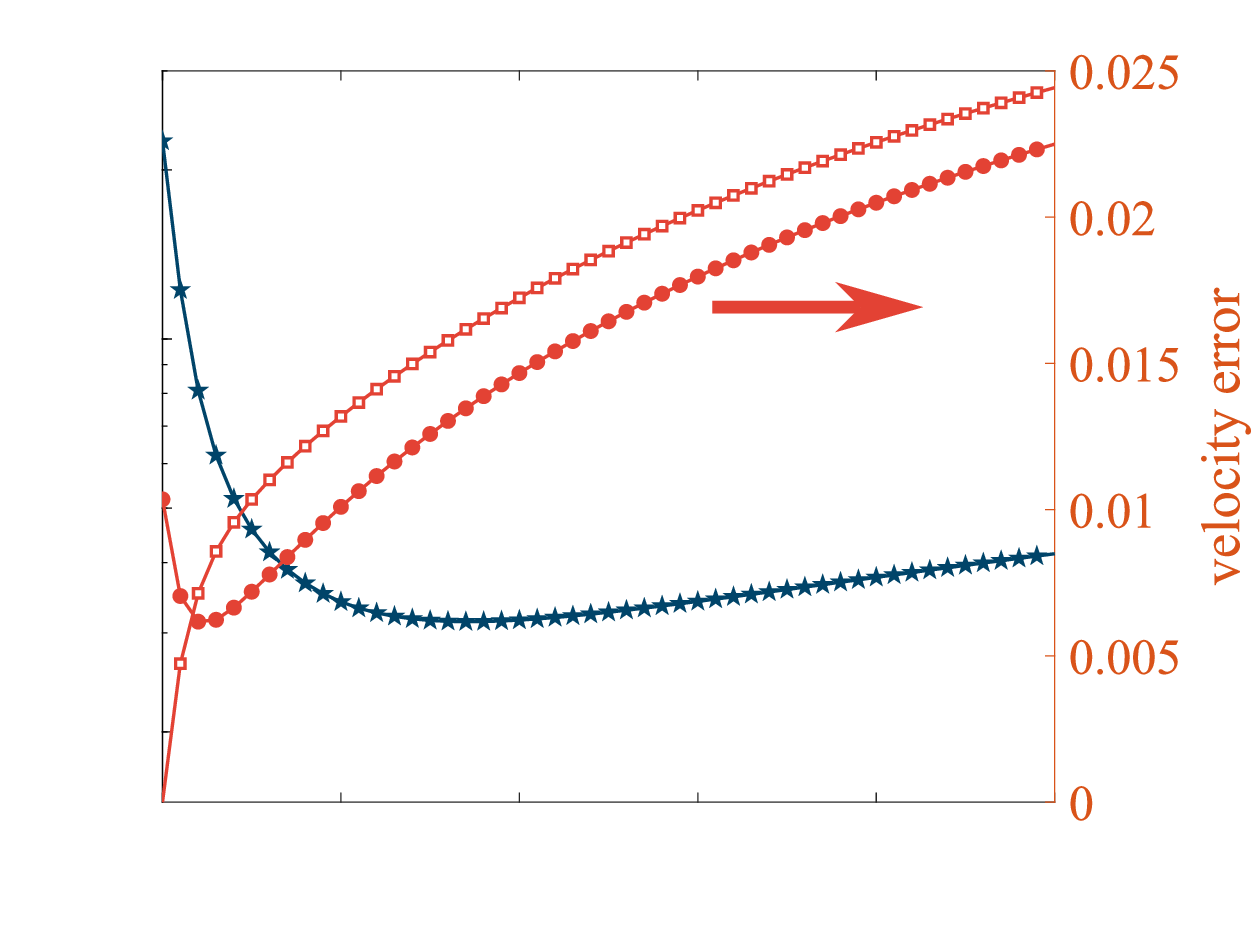}
\end{subfigure}
\flushleft
\begin{subfigure}
    \centering
    \hspace{18mm}
    \includegraphics[width=0.4\linewidth,trim=4mm 0mm 10mm 0mm,clip]{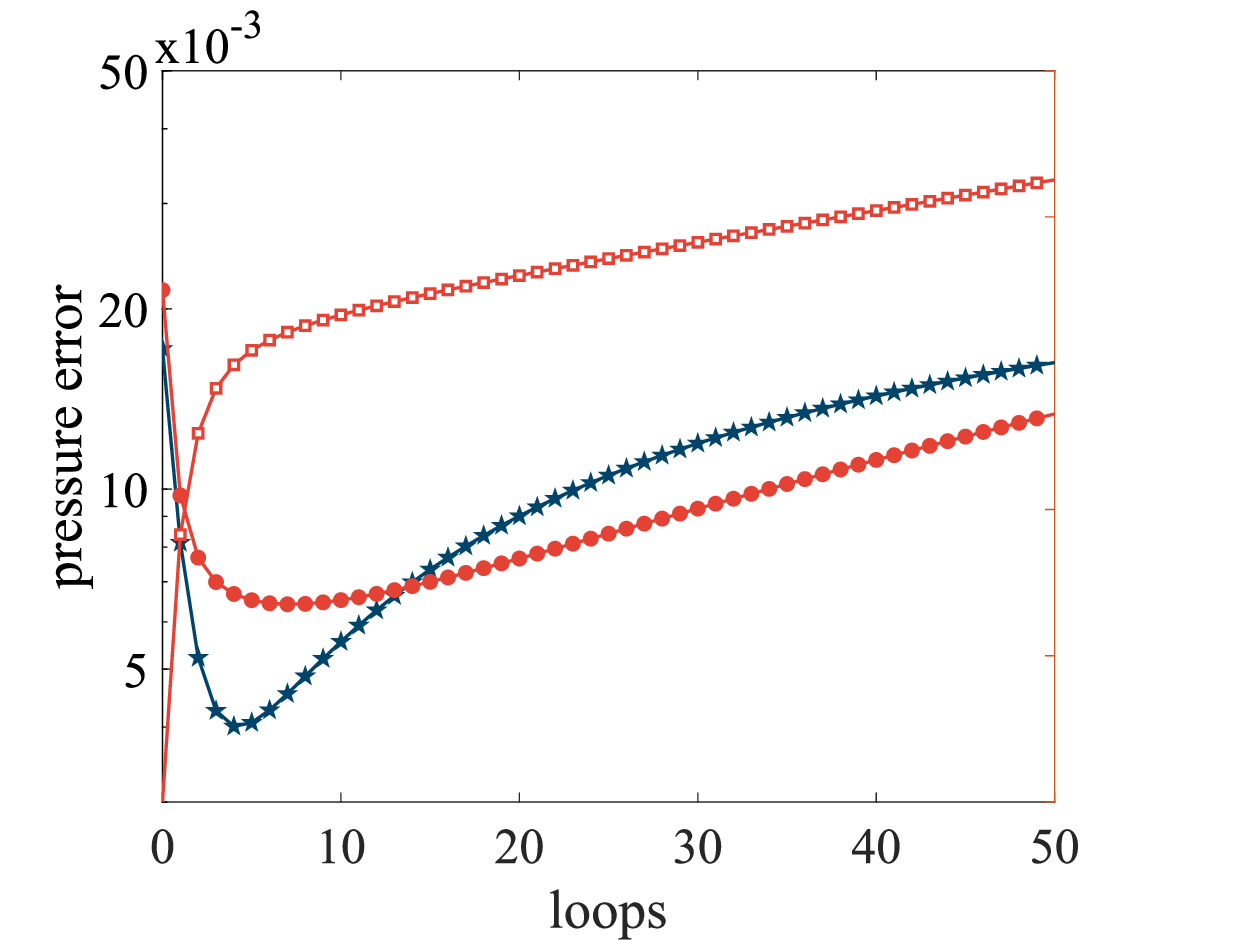}
\end{subfigure}
\begin{subfigure}
    \centering
    \includegraphics[width=0.4\linewidth,trim=14mm 0mm 0mm 0mm,clip]{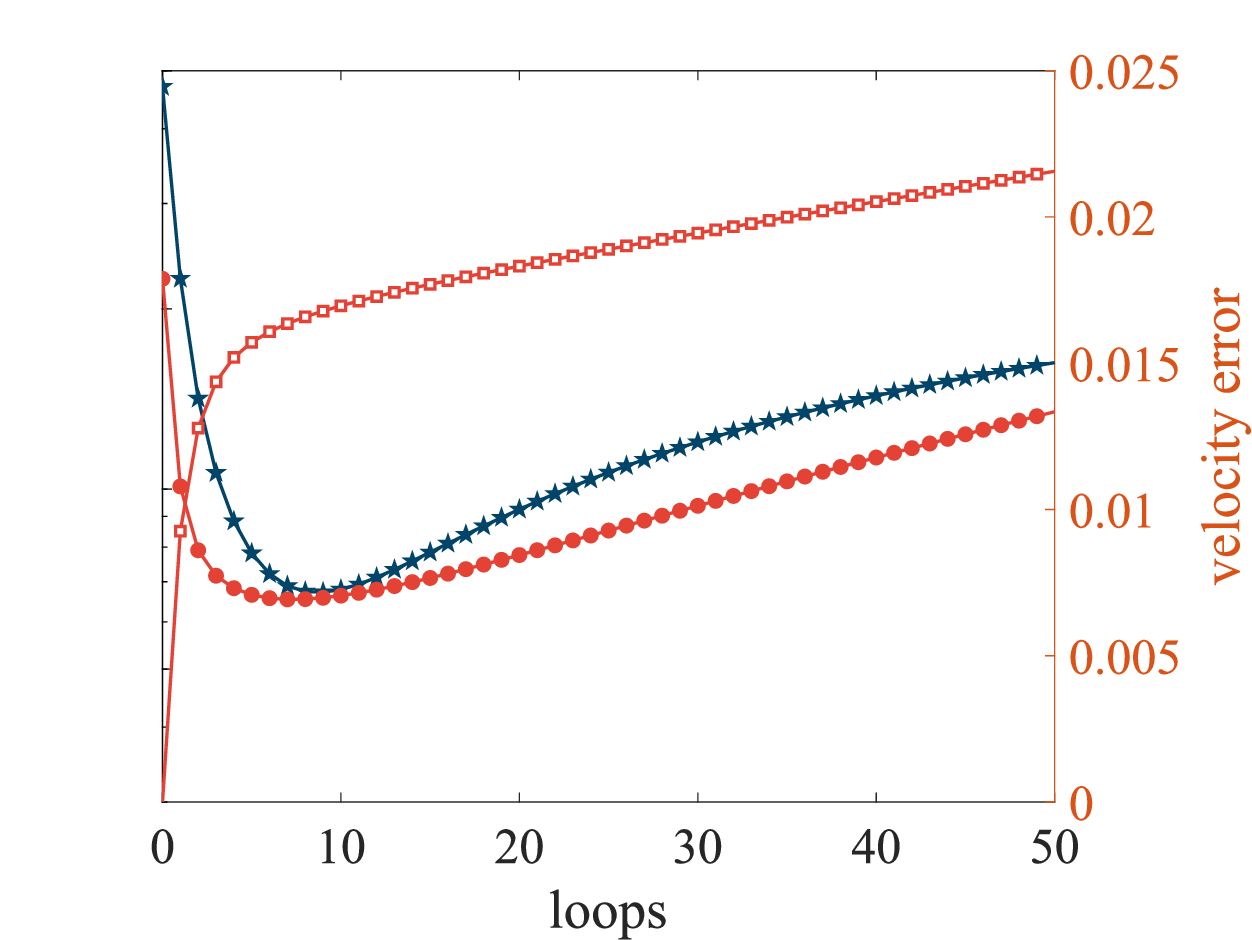}
\end{subfigure}
\caption{The error of pressure field in the iteration of AMIC using the clean field as baseline (blue curves with stars), and the error of velocity field using the clean field (vermillion curves with circles) and the noisy field (vermillion curve with squares) as baseline, with upper row for the 3D synthetic channel case and lower row for the 2D airfoil wake case and the left column for GWN and right column for CGWN.}
\label{fig:loops}
\end{figure*}

The number of iterations in AMIC is chosen to balance the trade-off between insufficient correction and over-smoothing of the velocity field. Fig. \ref{fig:loops} shows how the errors change over $50$ iterations for both dataset when the velocity field is superposed with both GWN and CGWN. The $\lambda$ for AMIC is set at $(0.15,0.15)$ for the channel flow and $(0.15,0.1)$ for the wake of an airfoil. The pressure computation is based on integrating its gradients from Navier-Stokes equation, using the 2D or 3D version of the route-free iterative method in \citep{chen2022Pressure} with the computing acceleration from NVIDIA RTX 4090. Using the same integration process used for the database also on the original fields ensures that differences cannot be ascribed to truncation errors due to the choice of the mesh. 

The error in this paper is defined by the Root-Mean-Square error over the domain and frames, then normalized by the characteristic velocity $U$ or dynamic pressure $\frac{1}{2}\rho U^2$,
\begin{equation}
    \epsilon = \left\{\begin{aligned}
    &\frac{D_{\mathbf{u}, ref}}{U}=\frac{\|\mathbf{u}-\mathbf{u}_{ref}\|_2}{U}
    &\qquad\text{, for velocity fields}\\
    &\frac{D_{p, ref}}{\frac{1}{2}\rho U^2}=\frac{\|p-p_{ref}\|_2}{\frac{1}{2}\rho U^2}
    &\qquad\text{, for pressure fields}
    \end{aligned}\right.
    \label{eq: error}
\end{equation}
with
\begin{equation}
    \begin{aligned}
    &D_{\mathbf{u}, ref}=\left(\frac{1}{n_tn_cn_p}\sum\limits_{i=1}^{n_t}\sum\limits_{i=1}^{n_cn_p}(u_{ij}-u_{ij,ref})^2\right)^{1/2}\\
    &D_{p, ref}=\left(\frac{1}{n_tn_p}\sum\limits_{i=1}^{n_t}\sum\limits_{i=1}^{n_p}(p_{ij}-p_{ij,ref})^2\right)^{1/2}
    \end{aligned}
\end{equation}
where the $n_t$, $n_p$ and $n_c$ stand for the number of frames, nodes in the domain and number of components. Subscripts indicate individual entries. In the previous equation, $\mathbf{u}$ and $p$ stand for the velocity and pressure field whose error is computed, while $\mathbf{u}_{ref}$ and $p_{ref}$ stand for the reference velocity and pressure field, respectively. Therefore the error $D_{p, clean} = \|p-p_{clean}\|_2$ is defined as the error of the pressure field at each iteration against the pressure $p_{clean}$ computed from the clean dataset  before adding noise.
 Similarly, the error of velocity field is defined as $D_{\mathbf{u}, clean} = \|\mathbf{u}-\mathbf{u}_{clean}\|_2$, with the meaning of subscripts like $D_{p, clean}$, and the distance of the velocity field to the noisy one is denoted as $D_{\mathbf{u}, noisy} = \|\mathbf{u}-\mathbf{u}_{noisy}\|_2$, where $\mathbf{u}_{noisy}$ is the velocity field after adding noise. It must be remarked that the latter is the only metric that would be directly available in an experiment.

For the channel flow, the errors in the velocity fields drop to their lowest point after $2$ iterations, while the pressure field errors hit their minimum after $6$ iterations for GWN and $18$ iterations for CGWN. Meanwhile, for the airfoil wake, the velocity field errors reach their lowest after $7$ iterations for both GWN and CGWN, whereas the pressure field errors bottom out after $4$ iterations for GWN and $9$ iterations for CGWN. Additionally, $D_{\mathbf{u}, noisy}$ shows a monotonous increase, with a steeper slope in the initial iterations then stabilizes at a milder slope. This hints at how AMIC works, the correction smooths out the noise while introducing distortion to the velocity field or smearing out the details. The noise reduction fades out as the velocity field becomes more aligned with Eq. \ref{eqn:TH}, while the distortion is introduced at a nearly constant rate. Consequently, $D_{\mathbf{u}, clean}$ first reaches a minimum and then increases when the distortion rate surpasses the noise reduction rate. The observed delay in reaching the minimum for $D_{p, clean}$ compared to $D_{\mathbf{u}, clean}$ can be ascribed to AMIC's ongoing process of aligning the temporal and spatial information of the velocity fields. This alignment continues to improve the accuracy of the pressure computation even after most of the noise in the velocity fields has been removed.

In real experiments, the true velocity and pressure fields are unavailable due to PIV uncertainty, making $D_{\mathbf{u}, clean}$ and $D_{p, clean}$ unknown for determining the optimal number of iterations. However, $D_{\mathbf{u}, noisy}$ is still available, thus we introduce the velocity correction index,
\begin{equation}
    I_{\mathbf{u}, correction}(i) = \frac{D_{\mathbf{u}, noisy}(i)-D_{\mathbf{u}, noisy}(i-1)}{D_{\mathbf{u}, noisy}(i-1)-D_{\mathbf{u}, noisy}(i-2)}
    \label{eq:noiter}
\end{equation}
where $i$ is the current iteration loop. The index represents the ratio of how much the velocity field is corrected in two successive loops. Similar to the criterion applied in \cite{Raiola2015POD}, the iteration is set to stop when the $I_{\mathbf{u}, correction}(i)$ starts to exceed $0.9$. In the channel flow test, the iterations stop at the $7th$ loop for GWN and at the $8th$ loop for CGWN. For the airfoil wake simulations, the iterations ends at the $10th$ loop for GWN and at the $11th$ loop for CGWN. Such criterion provides near-optimal pressure estimation and slightly deteriorated velocity estimation. This stopping criterion will be applied throughout the remainder of this paper.

It should be stressed out that the update rate of the velocity field also depends on the choice of $\lambda$. Generally, a larger $\lambda$ accelerates the correction process, but an over-relaxation setting ($(1-2\sum_i^N\lambda_i) < 0$) may lead to instability. In another aspect, a $\lambda$ with a larger kernel incorporates more adjacent frames into the filtering process, making the method more robust to noise. On the downside, this would also result in the smoothing of high-frequency motions. The performance of four additional $\lambda$ settings is listed in Tab. \ref{tab: lambda} for the channel case under GWN. Furthermore, the choice of $\lambda$ may influence the optimal threshold for the stopping criterion, though it seems to have a minor effect on the final error (see Tab. \ref{tab: lambda}).
\begin{table*}[htbp]
    \centering
    \caption{The performance with various setting of $\lambda$ for the channel case under GWN, with $N_{stop}$  being the loop to stop the correction when the correction index in Eq. \ref{eq:noiter} exceeds $0.9$.}
    \label{tab: lambda}
    \begin{tabular}{c|c|c|c|c}
        \hline
        $\lambda$ & $N_{stop}$ & $D_{\mathbf{u},clean}$ at $N_{stop}$ & $D_{p,clean}$ at $N_{stop}$ & $min(D_{p,clean})$ \\
        \hline
        (0.3, 0.3)      & 5  & 0.00912 & 0.00371 & 0.00362 \\
        (0.075, 0.075)  & 12 & 0.00729 & 0.00371 & 0.00368 \\
        (0.3)           & 9  & 0.00668 & 0.00384 & 0.00378 \\
        (0.1, 0.1, 0.1) & 10 & 0.0115  & 0.00403 & 0.00371 \\
        \hline
    \end{tabular}
\end{table*}

\begin{figure}[h]
\begin{minipage}{0.49\linewidth}\,\,\,\,\,\centering GWN\end{minipage}
\begin{minipage}{0.49\linewidth}\centering CGWN\,\,\,\,\,\,\,\,\end{minipage}
\flushleft
\begin{minipage}{0.96\linewidth}
    \,\,\,\,\,\,\,\,\,\,
    \includegraphics[width=0.4\linewidth,trim=4mm 132mm 4mm 4mm,clip]{fig/bar1.eps}
    \,\,\,\,\,\,
    \includegraphics[width=0.4\linewidth,trim=4mm 132mm 4mm 4mm,clip]{fig/bar1.eps}
\end{minipage}
\flushleft
\begin{minipage}{0.03\linewidth}
    \rotatebox{90}{\hspace{5mm}SG}
\end{minipage}
\begin{minipage}{0.96\linewidth}
    \begin{subfigure}
        \centering
        \includegraphics[width=0.45\linewidth,trim=0mm 24mm 8mm 20mm,clip]{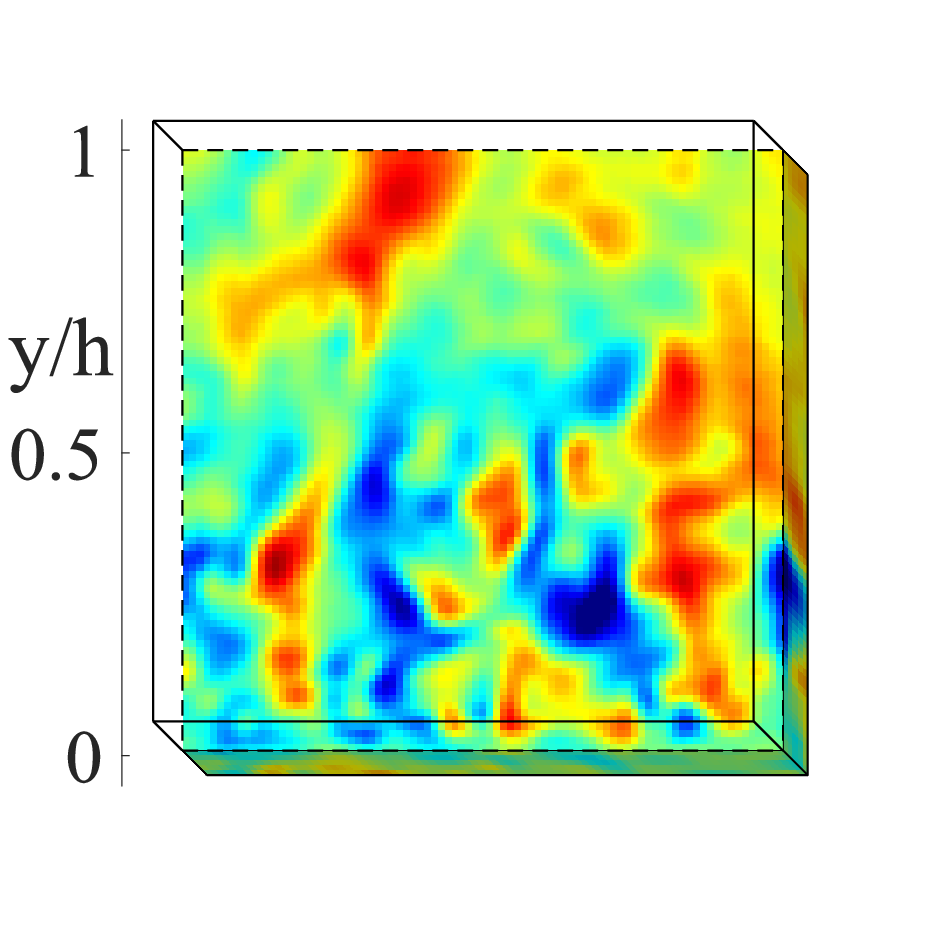}
    \end{subfigure}
    \begin{subfigure}
        \centering
        \includegraphics[width=0.45\linewidth,trim=0mm 24mm 8mm 20mm,clip]{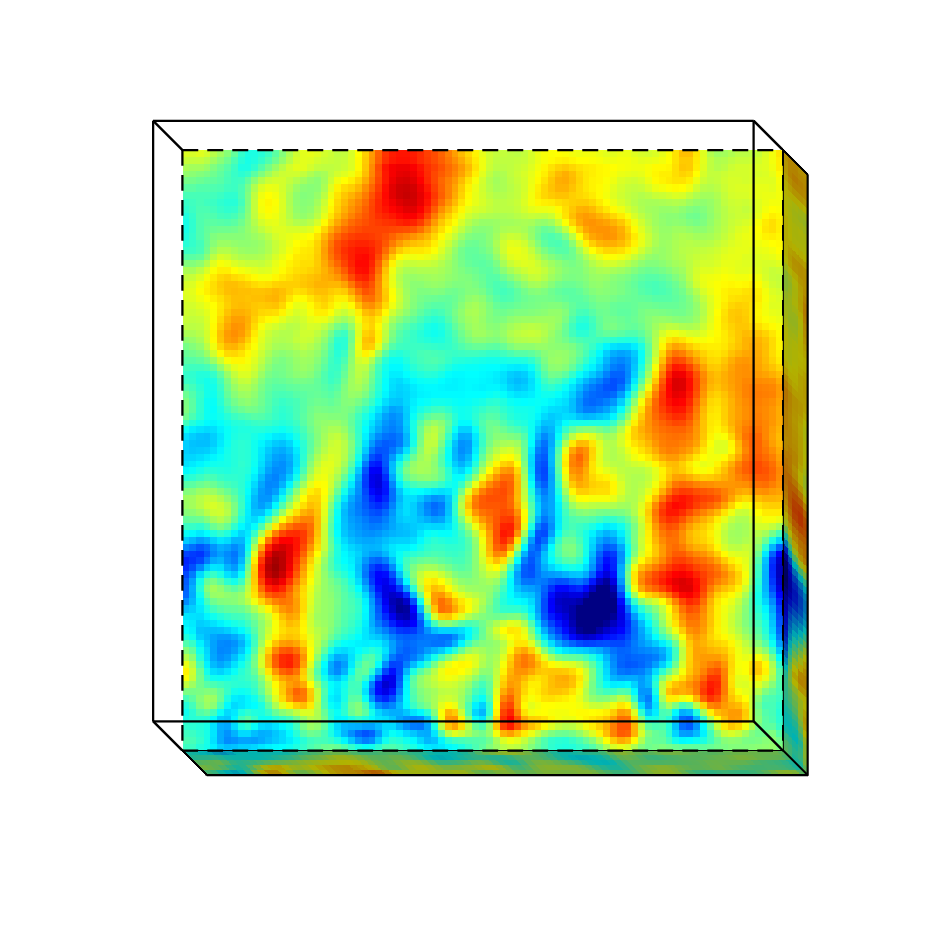}
    \end{subfigure}
\end{minipage}
\flushleft
\begin{minipage}{0.03\linewidth}
    \rotatebox{90}{\hspace{5mm}POD}
\end{minipage}
\begin{minipage}{0.96\linewidth}
    \begin{subfigure}
        \centering
        \includegraphics[width=0.45\linewidth,trim=0mm 24mm 8mm 20mm,clip]{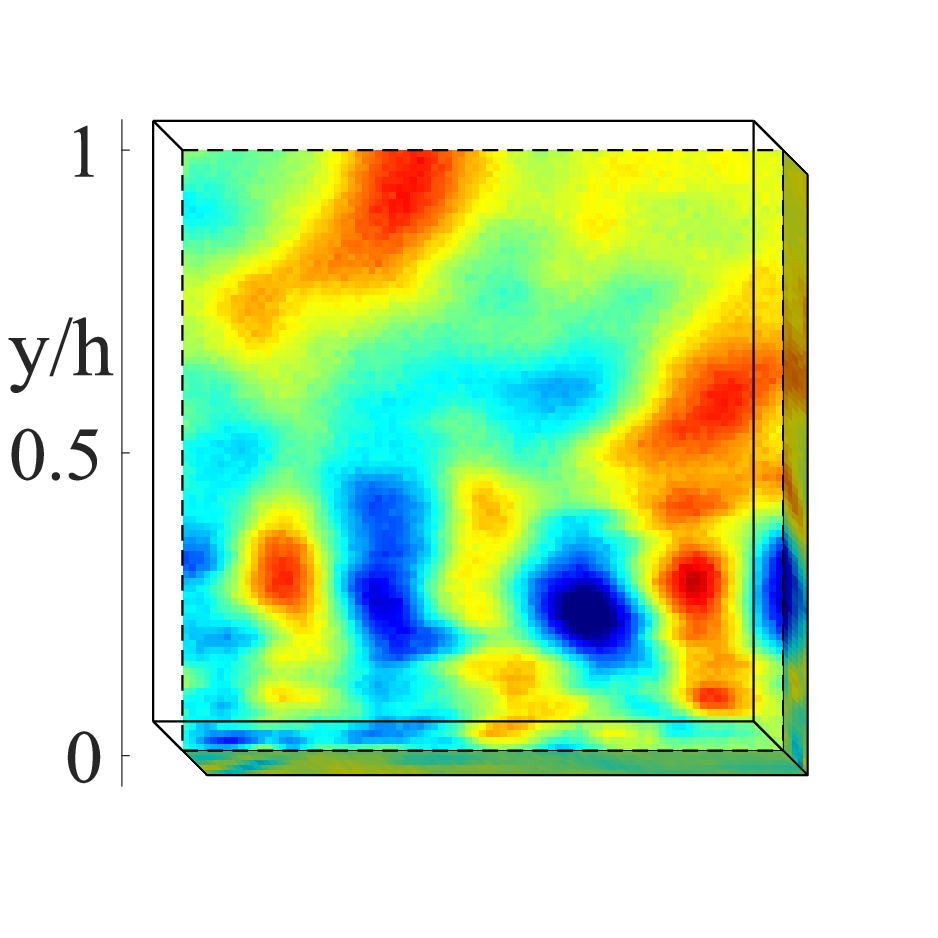}
    \end{subfigure}
    \begin{subfigure}
        \centering
        \includegraphics[width=0.45\linewidth,trim=0mm 24mm 8mm 20mm,clip]{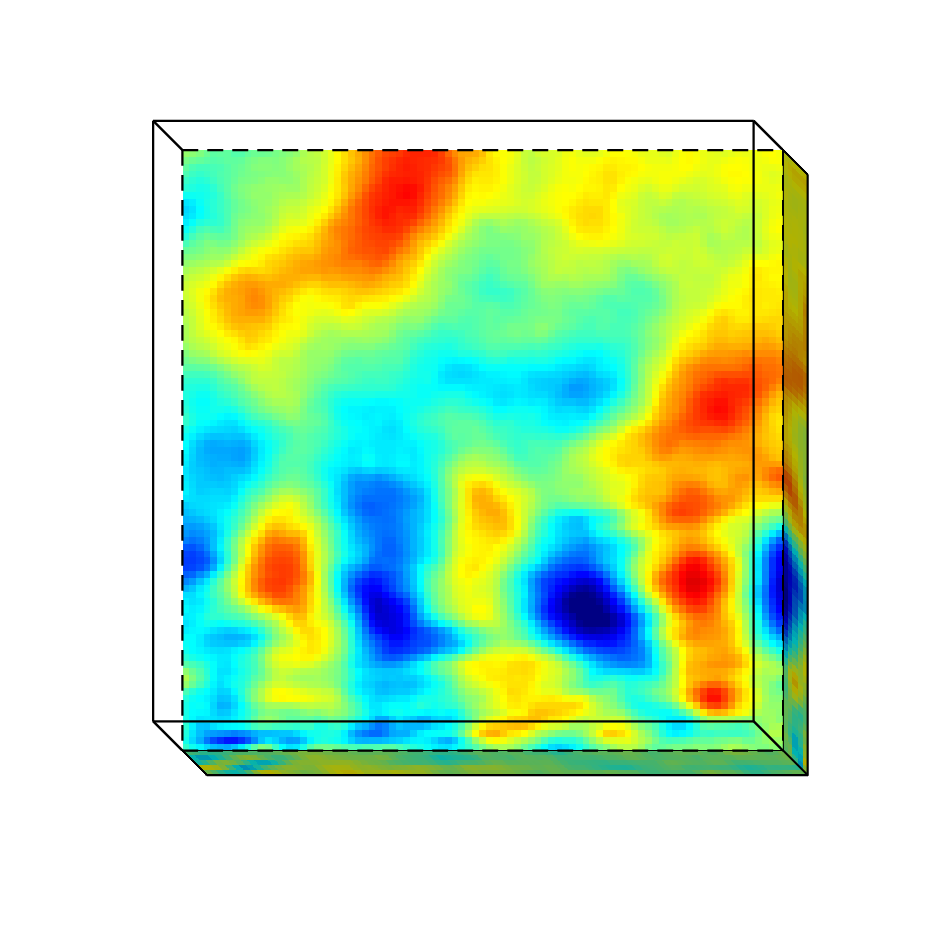}
    \end{subfigure}
\end{minipage}
\flushleft
\begin{minipage}{0.03\linewidth}
    \rotatebox{90}{\hspace{5mm}AMIC}
\end{minipage}
\begin{minipage}{0.96\linewidth}
    \begin{subfigure}
        \centering
        \includegraphics[width=0.45\linewidth,trim=0mm 0mm 8mm 20mm,clip]{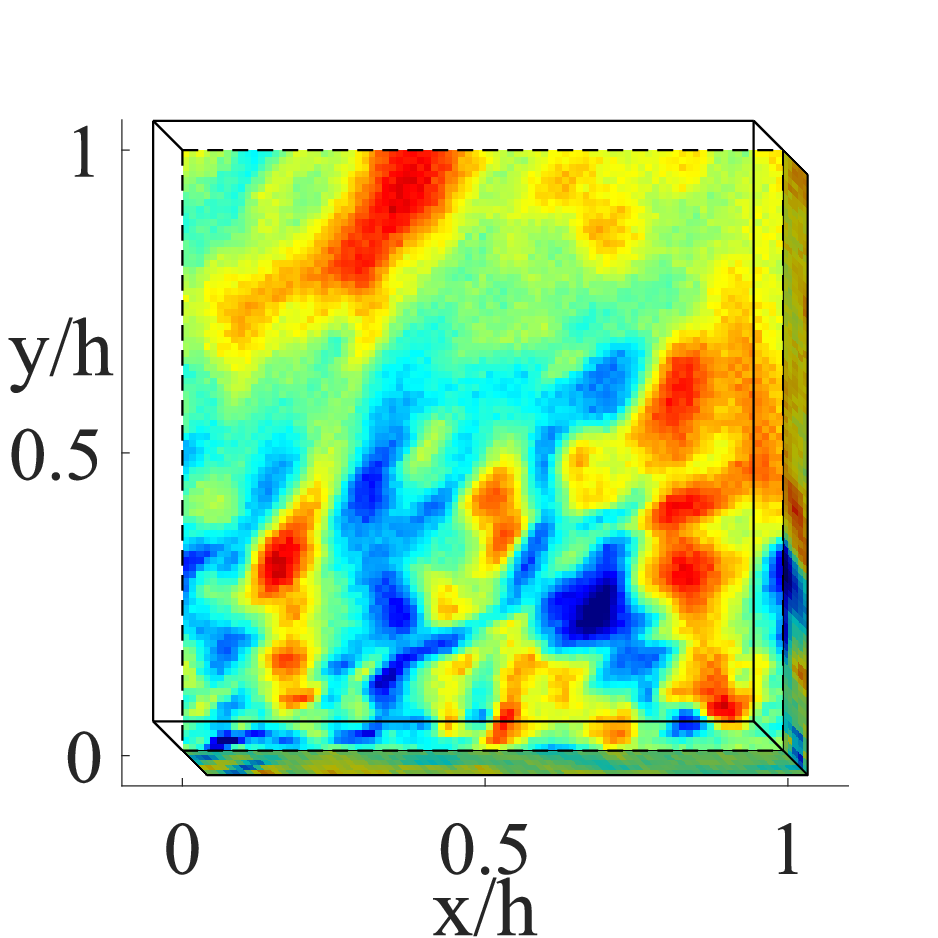}
    \end{subfigure}
    \begin{subfigure}
        \centering
        \includegraphics[width=0.45\linewidth,trim=0mm 0mm 8mm 20mm,clip]{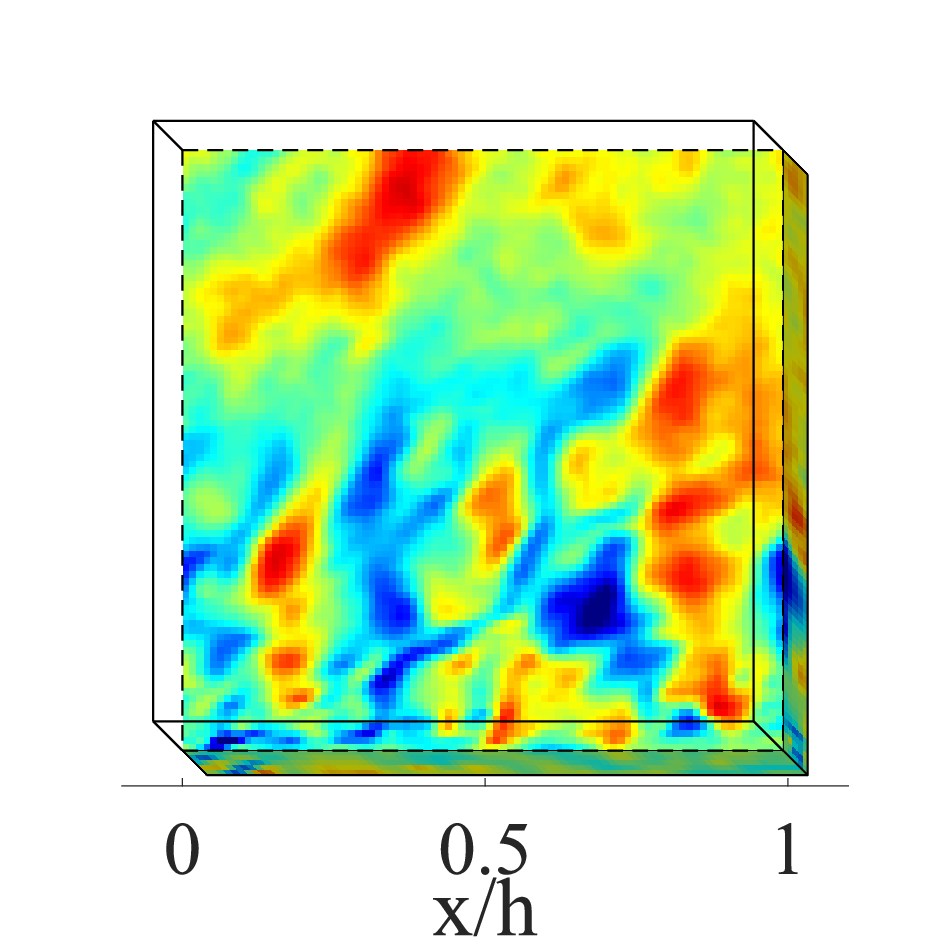}
    \end{subfigure}
\end{minipage}
\caption{Wall-normal component of the velocity fields after filtering/correcting with (top to bottom) SG, POD and AMIC. The figure shows the processing result of two types of noise: GWN (the left column) and CGWN (the right column). The exact field for reference is included in Figure \ref{fig:noise}.}
\label{fig:v}
\end{figure}

\section{Results on synthetic cases}
\label{sec:result}

Fig. \ref{fig:v} presents the wall-normal velocity component of the instantaneous channel flow field shown in Fig. \ref{fig:noise} after processing it with a SG filter, POD truncation and AMIC. The parameters for the SG filter and the POD truncation are mentioned in Sect. \ref{sec:method}, while those of AMIC are discussed in Sect. \ref{sec:loop}. The processing time for AMIC is less than $1s$ per frame on a 16-core processor. The SG filter shows good performance in velocity reconstruction under GWN since it recovered most of the flow structures, while the field reconstructed from CGWN still shows a weak noise residual. The POD truncation, instead, performs remarkably worse: due to the high spectral richness of the flow and the limited size of the dataset, the cutoff cancels out also fine-scale details of the true field. A less aggressive filtering would recover some of those scales but include more noise contamination in the reconstructed fields. AMIC performs quite well for both noise types, though with slightly larger small-scale residual noise if compared to the SG filter under GWN.

\begin{figure}[h]
	\centering 
        \includegraphics[width=0.35\textwidth,trim=4mm 4mm 16mm 20mm,clip]{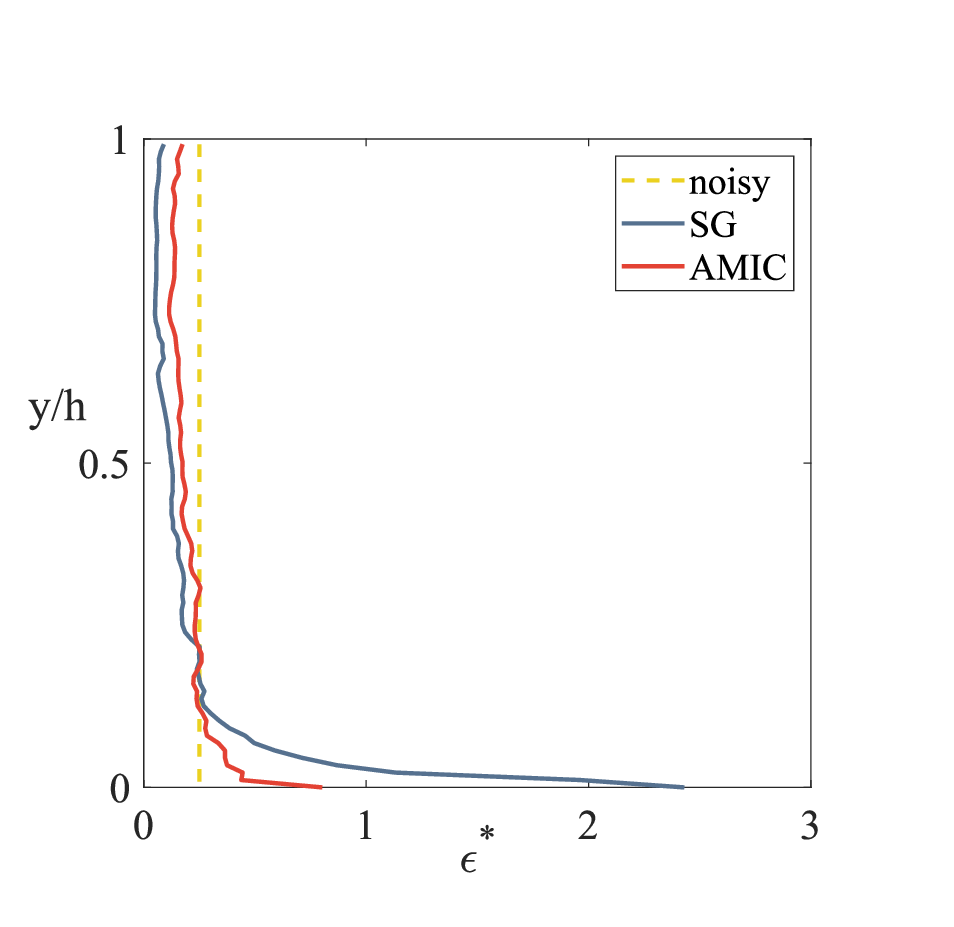}
	\caption{The profile of RMS error along the wall-normal direction in the estimation of streamwise velocity using SG filter and AMIC, on the noise type of GWN. The error $\epsilon^*$ is normalized by the standard deviation of the corresponding clean fields.}
	\label{fig:profile}
\end{figure}

\begin{figure}[h]
    \hspace{1mm}
    \begin{subfigure}
        \centering
        \includegraphics[width=0.48\linewidth,trim=0mm 0mm 8mm 6mm,clip]{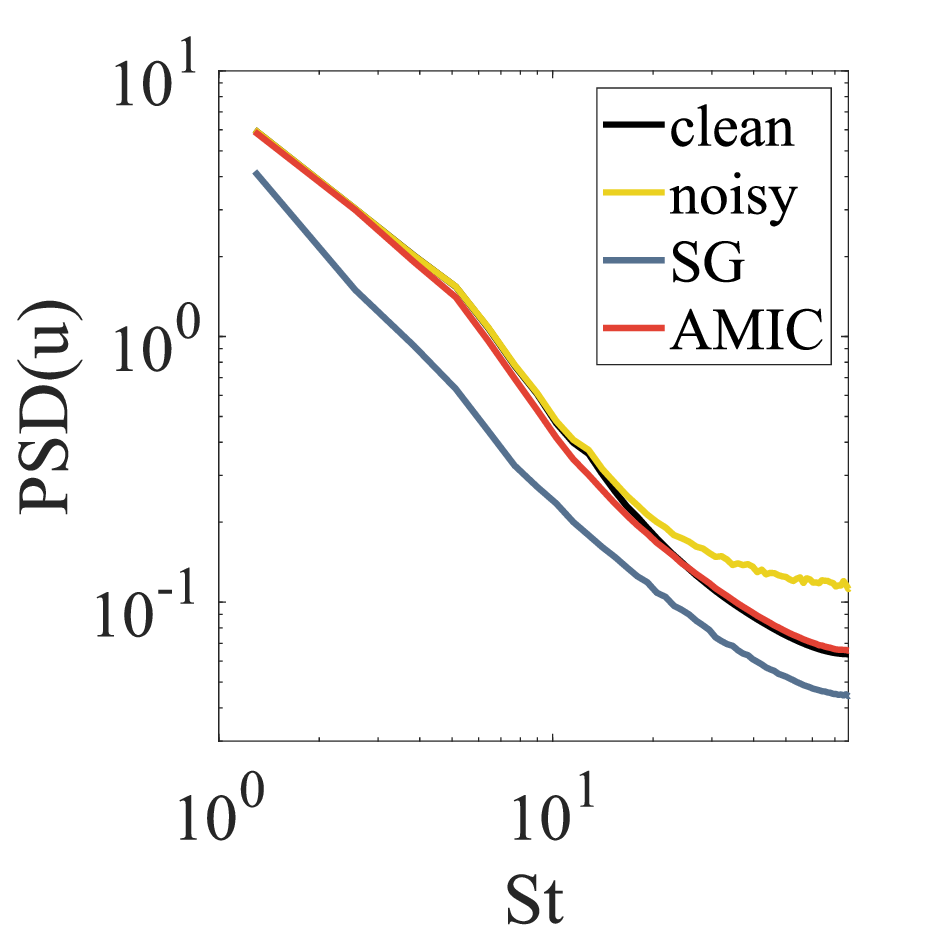}
    \end{subfigure}
    \hspace{-1mm}
    \begin{subfigure}
        \centering
        \includegraphics[width=0.48\linewidth,trim=0mm 0mm 8mm 6mm,clip]{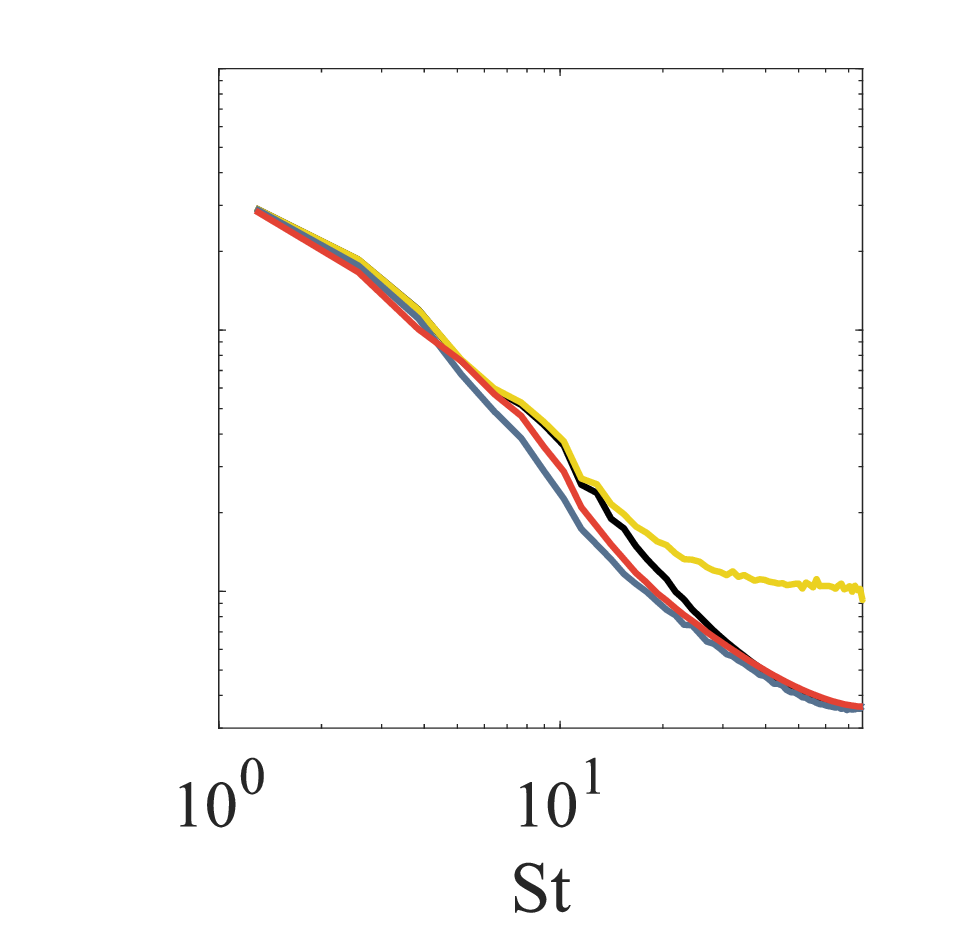}
    \end{subfigure}
    \caption{The power spectrum on the streamwise velocity at $y/h = 0.02$ (left) and $y/h = 0.2$ (right) on the noise type of GWN. The frequencies are presented in terms of Strouhal number $St=fh/U_b$.}
    \label{fig:psd}
\end{figure}

The instantaneous streamwise component of the velocity field shows similar features and it is not included here for conciseness. It must be remarked, however, that the SG-filtered fields have significantly larger errors than AMIC in the near-wall region. To quantify this, the profile of RMS error on the streamwise component of the velocity fields under GWN along the wall-normal direction is reported in Fig. \ref{fig:profile}. SG filter performs slightly better compared to AMIC when sufficiently far from the wall ($y/h > 0.2$), However, the error significantly increases when close to the wall. There are two main reasons for the higher error of SG-filtered field near the wall. One factor is the edge effect of the convolution that results in bias errors due to the strong local velocity gradient. Another factor is the larger presence of small-scale motions near the wall, thus spotlighting the stronger modulation of SG with respect to AMIC. On the other hand, since AMIC filters the velocity field along the additional dimension across different time series, it can deal with such a situation better than the SG filter, thus delivering better near-wall results. This intuition is also confirmed from the power spectrum of the streamwise velocity component at $y/h = 0.02$ and $y/h = 0.2$ in Fig. \ref{fig:psd}, where the spectra of AMIC-processed fields are closer to those from the clean field at the lower height. Instead, the spectrum of the SG-filtered field at $y/h = 0.02$ has deviations across the entire frequency range compared to the clean one, indicating the information loss from the SG-filtered field near the wall.

\begin{figure}
\flushleft
\begin{minipage}{0.96\linewidth}
    \,\,\,\,\,\,\,\,\,\,
    \includegraphics[width=0.4\linewidth,trim=4mm 130mm 4mm 4mm,clip]{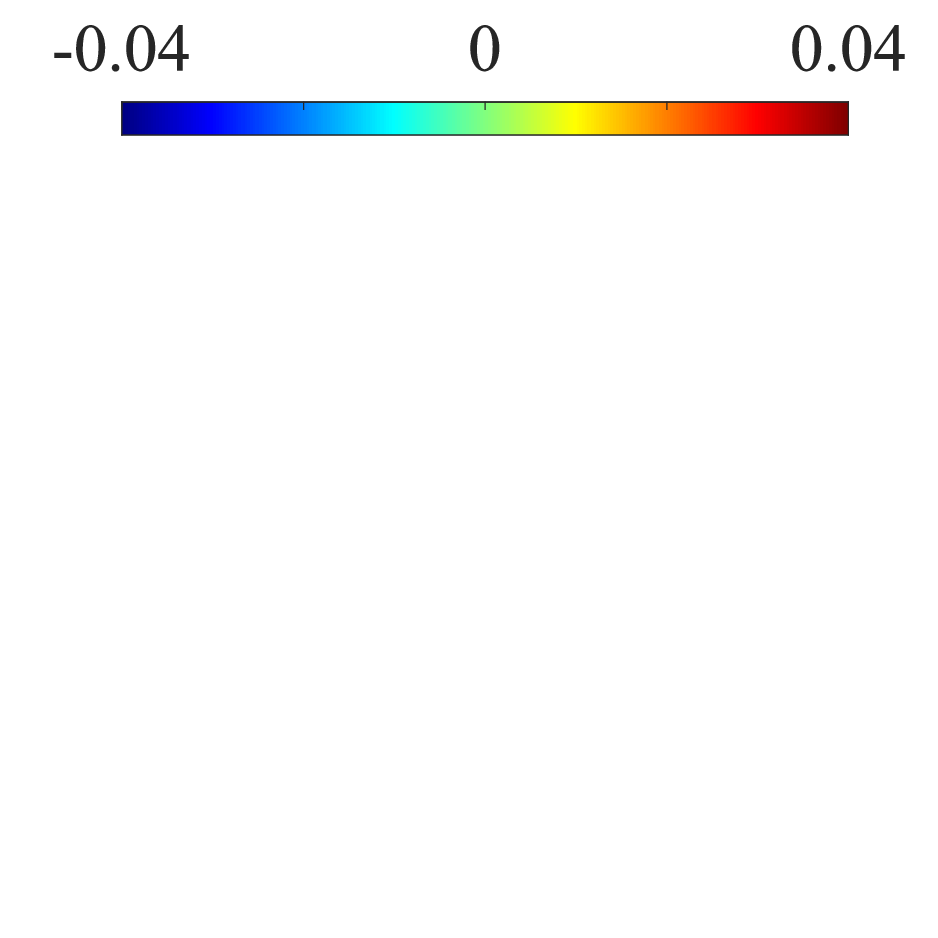}
    \,\,\,\,\,\,
    \includegraphics[width=0.4\linewidth,trim=4mm 130mm 4mm 4mm,clip]{fig/bar6.eps}
\end{minipage}
\flushleft
\begin{minipage}{0.03\linewidth}
    \rotatebox{90}{\hspace{5mm}clean field}
\end{minipage}
\begin{minipage}{0.96\linewidth}
    \begin{subfigure}
        \centering
        \includegraphics[width=0.45\linewidth,trim=0mm 20mm 8mm 20mm,clip]{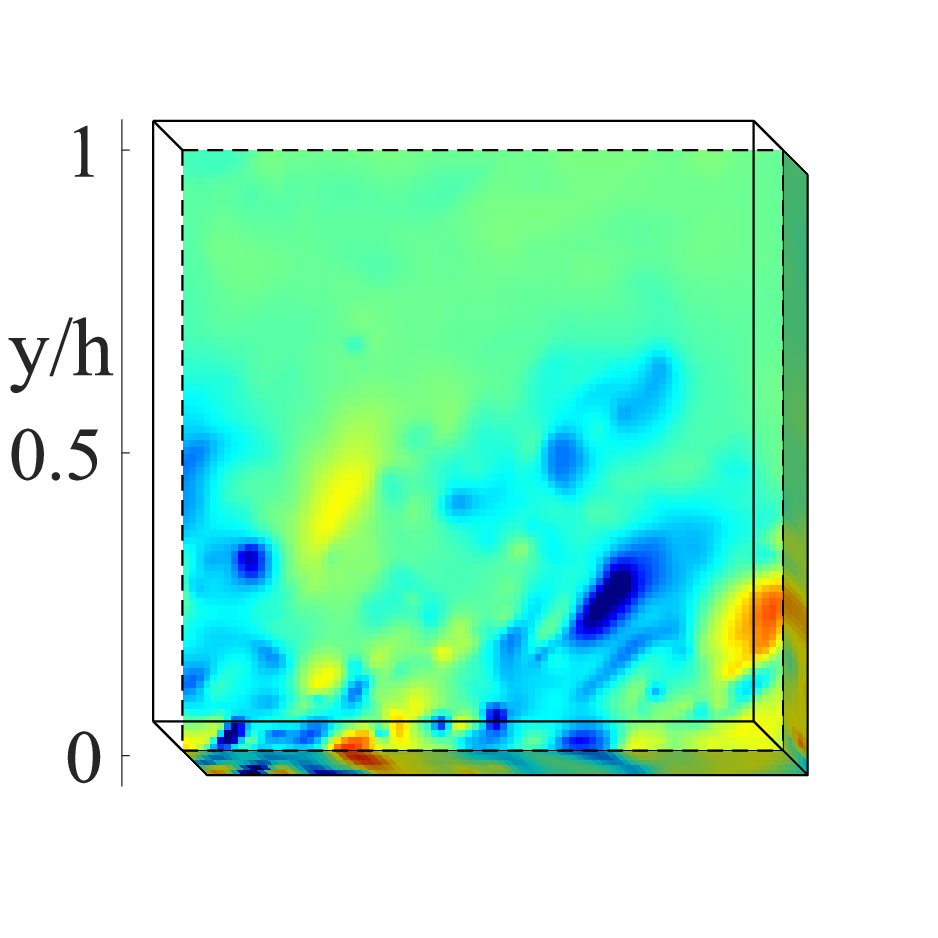}
    \end{subfigure}
\end{minipage}
\begin{minipage}{0.49\linewidth}\,\,\,\,\,\centering GWN\end{minipage}
\begin{minipage}{0.49\linewidth}\centering CGWN\,\,\,\,\,\,\,\,\end{minipage}
\flushleft
\begin{minipage}{0.03\linewidth}
    \rotatebox{90}{\hspace{5mm}noisy field}
\end{minipage}
\begin{minipage}{0.96\linewidth}
    \begin{subfigure}
        \centering
        \includegraphics[width=0.45\linewidth,trim=0mm 20mm 8mm 18mm,clip]{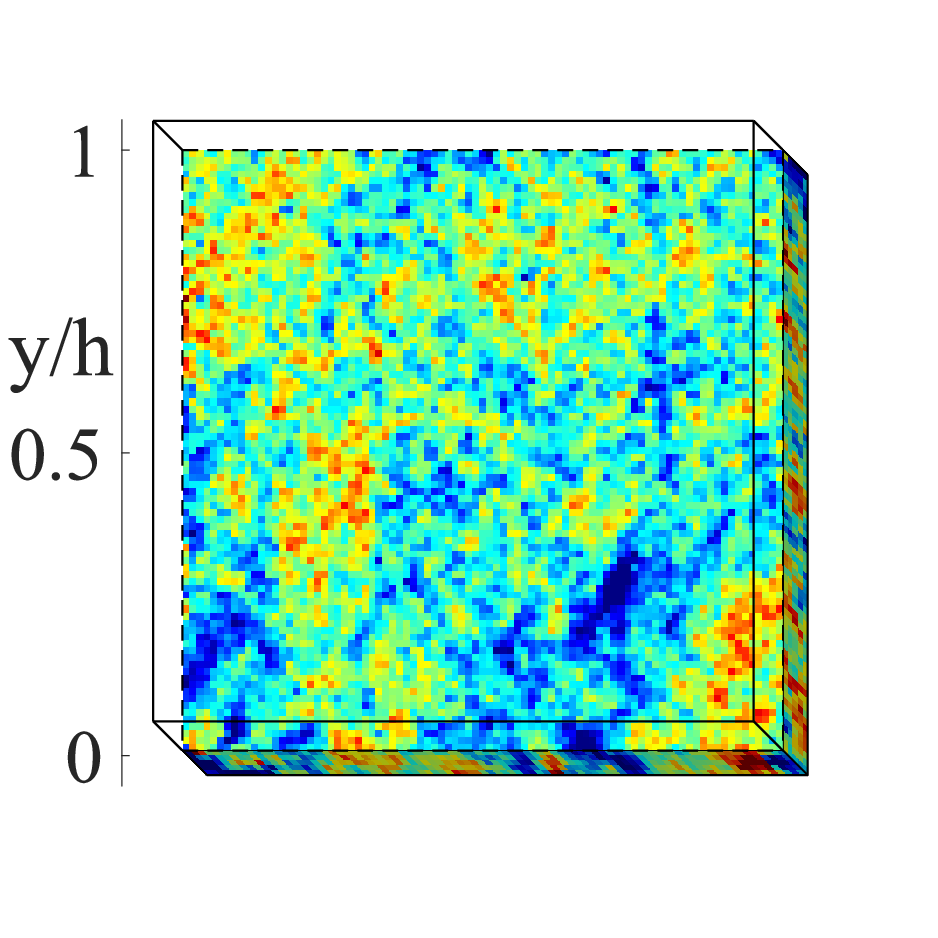}
    \end{subfigure}
    \begin{subfigure}
        \centering
        \includegraphics[width=0.45\linewidth,trim=0mm 20mm 8mm 18mm,clip]{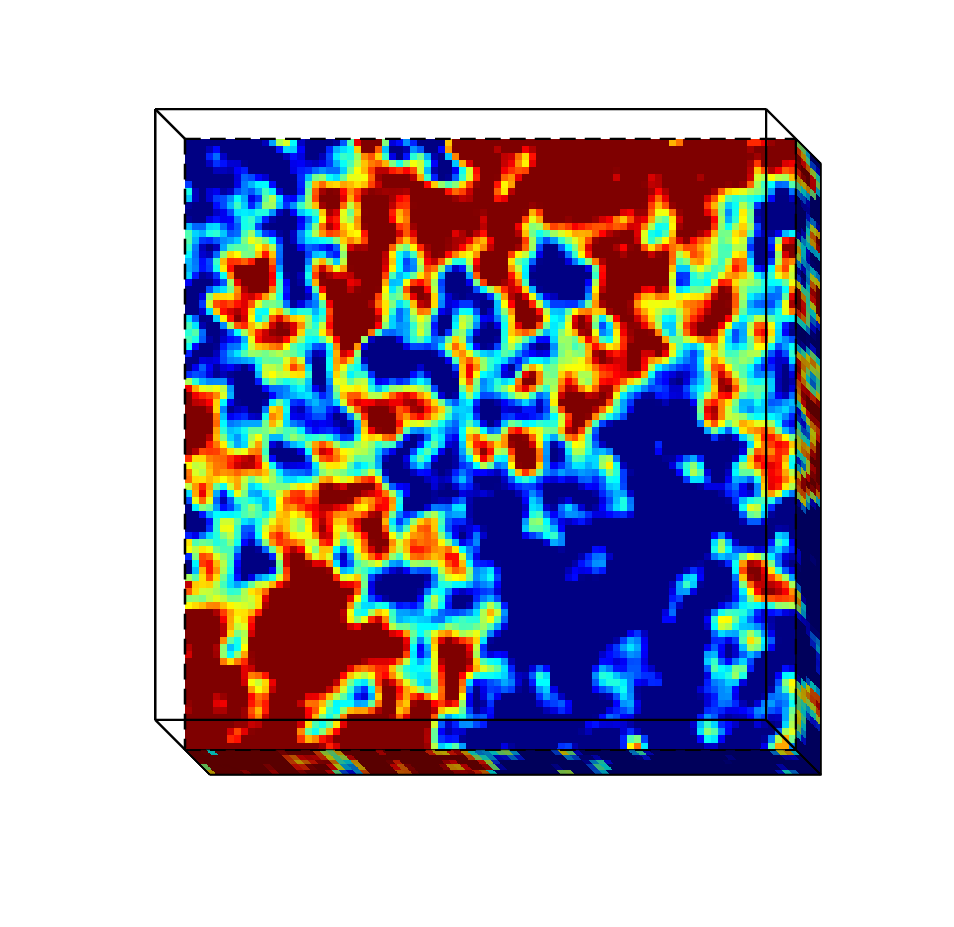}
    \end{subfigure}
\end{minipage}
\flushleft
\begin{minipage}{0.03\linewidth}
    \rotatebox{90}{\hspace{5mm}SG}
\end{minipage}
\begin{minipage}{0.96\linewidth}
    \begin{subfigure}
        \centering
        \includegraphics[width=0.45\linewidth,trim=0mm 20mm 8mm 20mm,clip]{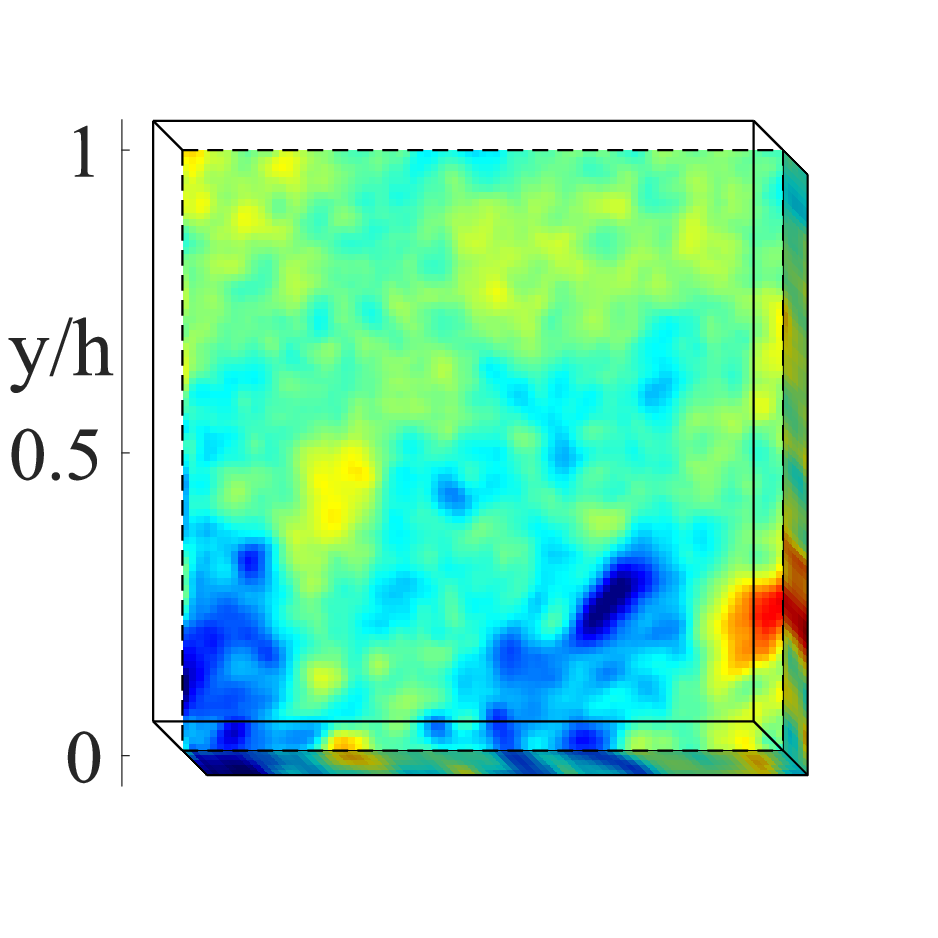}
    \end{subfigure}
    \begin{subfigure}
        \centering
        \includegraphics[width=0.45\linewidth,trim=0mm 20mm 8mm 20mm,clip]{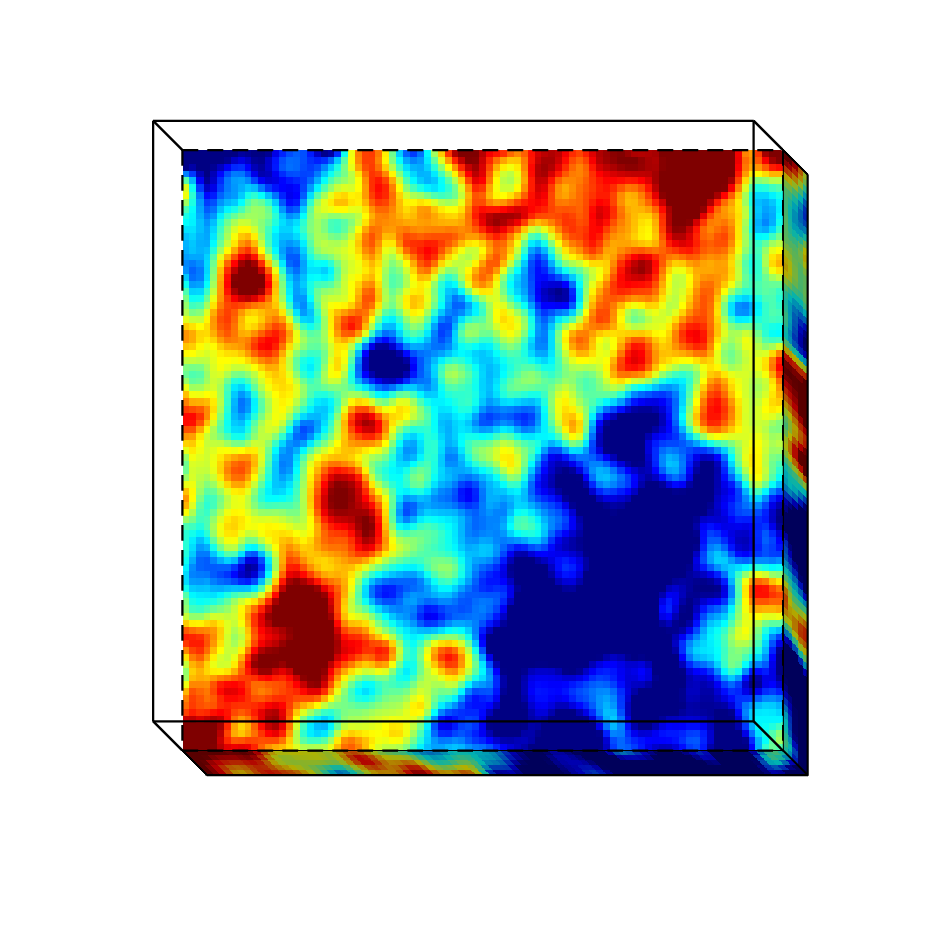}
    \end{subfigure}
\end{minipage}
\flushleft
\begin{minipage}{0.03\linewidth}
    \rotatebox{90}{\hspace{5mm}POD}
\end{minipage}
\begin{minipage}{0.96\linewidth}
    \begin{subfigure}
        \centering
        \includegraphics[width=0.45\linewidth,trim=0mm 20mm 8mm 20mm,clip]{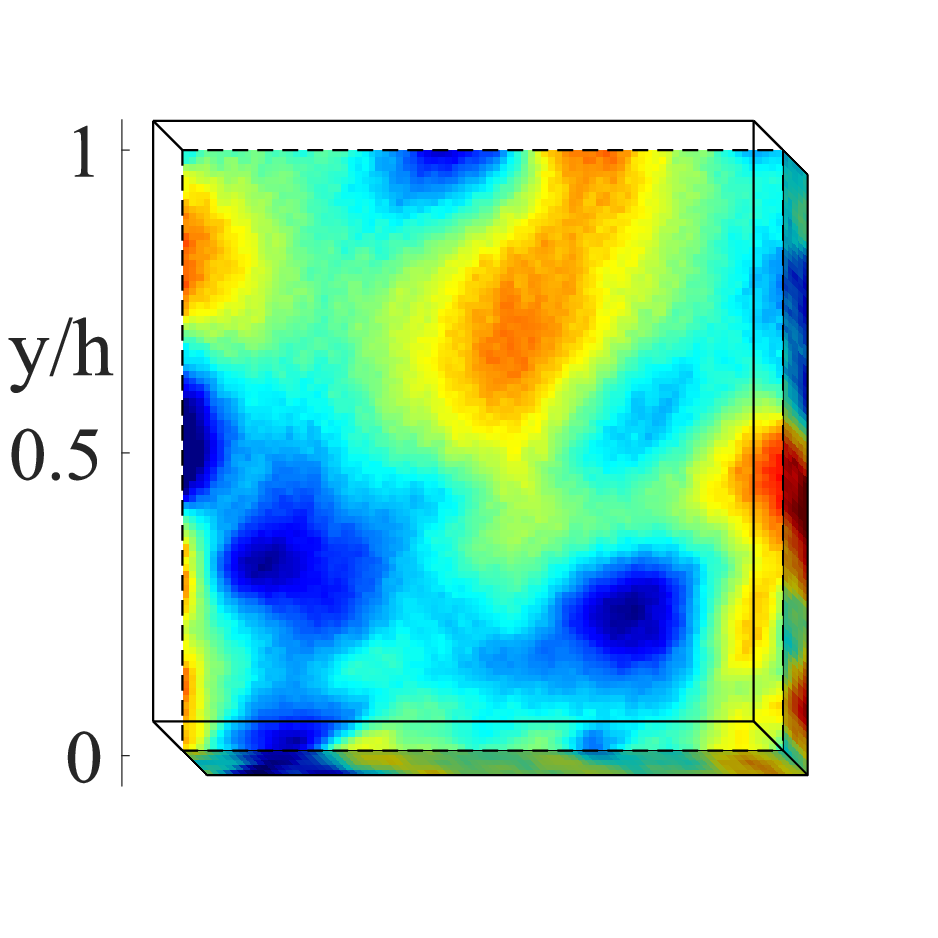}
    \end{subfigure}
    \begin{subfigure}
        \centering
        \includegraphics[width=0.45\linewidth,trim=0mm 20mm 8mm 20mm,clip]{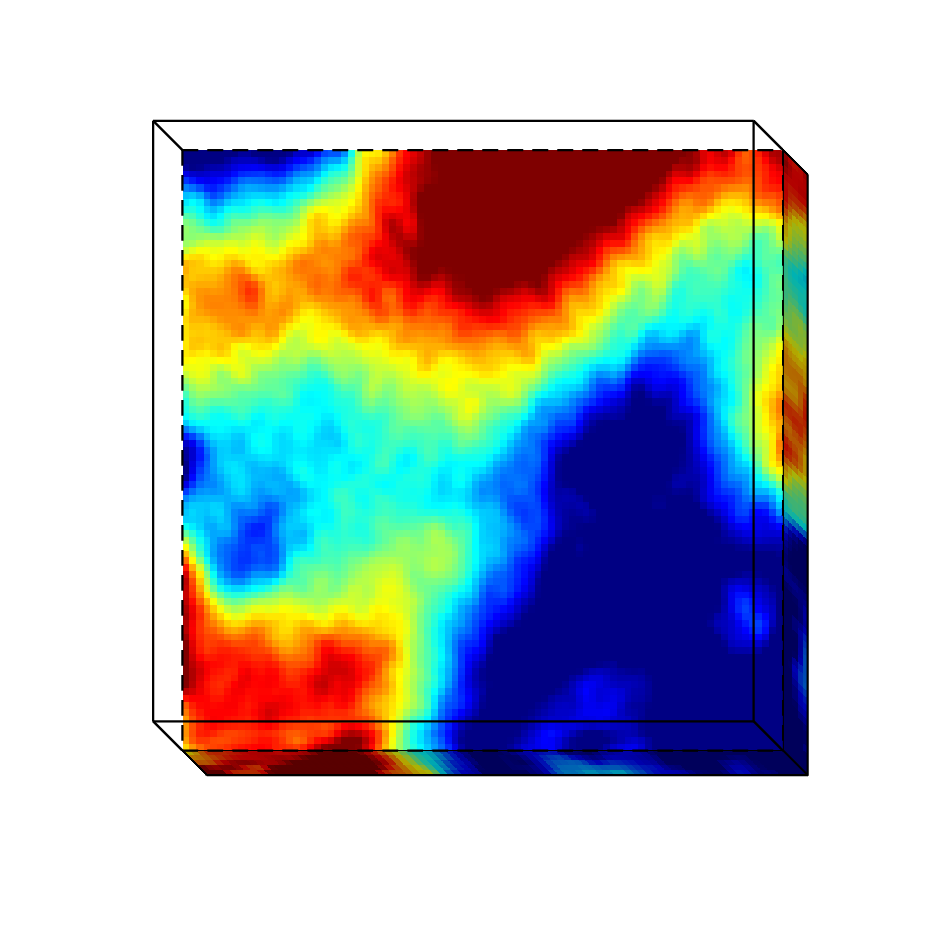}
    \end{subfigure}
\end{minipage}
\flushleft
\begin{minipage}{0.03\linewidth}
    \rotatebox{90}{\hspace{5mm}AMIC}
\end{minipage}
\begin{minipage}{0.96\linewidth}
    \begin{subfigure}
        \centering
        \includegraphics[width=0.45\linewidth,trim=0mm 0mm 8mm 20mm,clip]{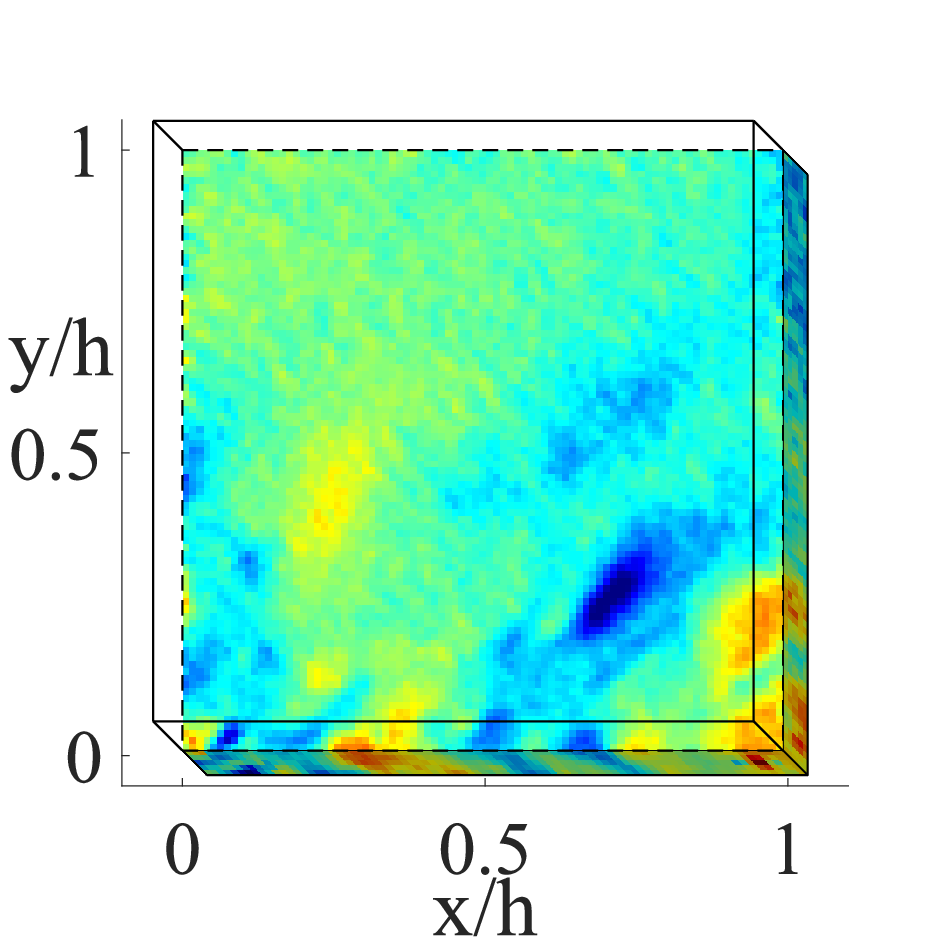}
    \end{subfigure}
    \begin{subfigure}
        \centering
        \includegraphics[width=0.45\linewidth,trim=0mm 0mm 8mm 20mm,clip]{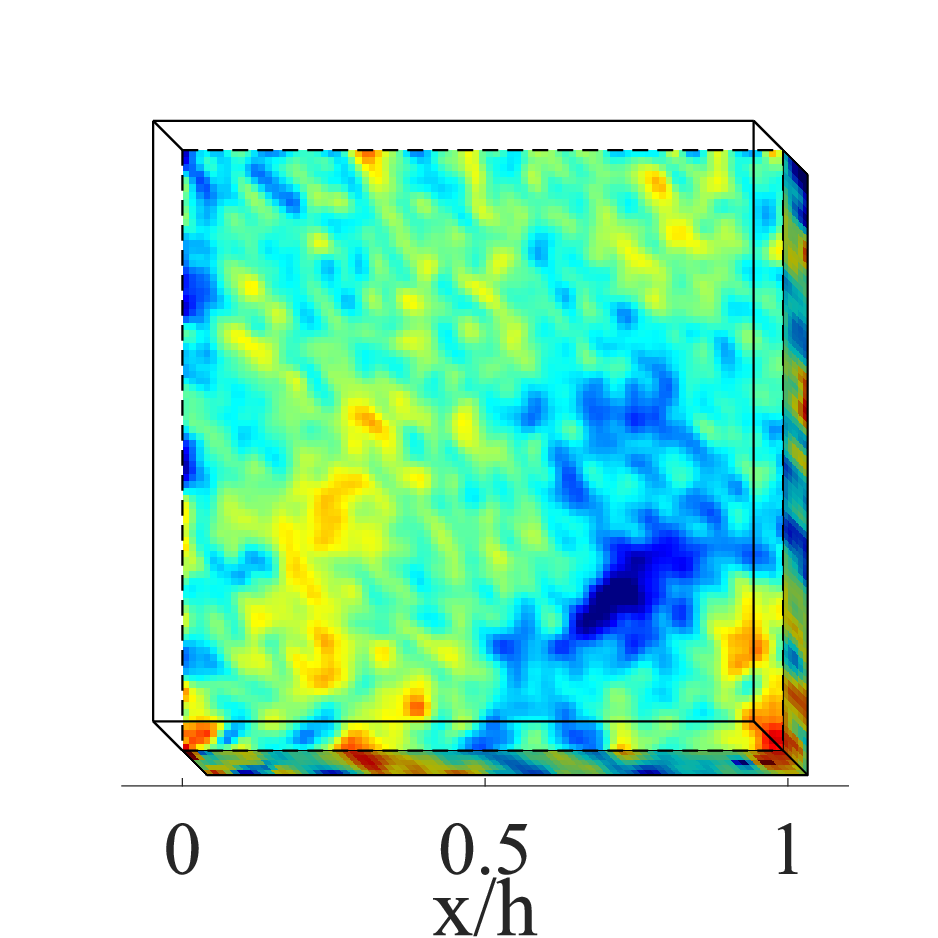}
    \end{subfigure}
\end{minipage}
\caption{Reference pressure field (1st row), and pressure fields from the noisy and processed velocity fields (top to bottom, SG, POD and AMIC), normalised by $\rho U_b^2$ with $U_b$ being the bulk velocity. The result of two types of noise are shown in the figure: GWN is shown in the left column, CGWN is shown in the right column.}
\label{fig:p}
\end{figure}

Regarding the pressure fields, the limited intensity of the fluctuations in the clean velocity field makes the integration process very sensitive to noise. Instantaneous pressure fields from the clean data, noisy (with both GWN and CGWN), SG-filtered, POD-truncated and AMIC-corrected velocity fields at the same snapshot of Fig. \ref{fig:v} are presented in Fig. \ref{fig:p}. 
The POD truncation fails to provide accurate pressure estimation because of the truncation of small-scale motions. The pressure estimation from velocity fields filtered with SG is reasonably accurate for GWN, although with slightly inferior quality than the AMIC-corrected field. Under CGWN, the estimation from the SG-filtered field cannot recover an accurate pressure distribution, while the AMIC correction provides acceptable results in terms of the position and intensity of the high- and low-pressure regions. This suggests that SG filter struggles to effectively filter spatially coherent noise, while the physical consistency imposed by AMIC is still effective.

\begin{figure}
\flushleft
\begin{minipage}{0.96\linewidth}
    \hspace{6mm}
    \includegraphics[width=0.92\linewidth,trim=4mm 100mm 4mm 0mm,clip]{fig/AF_bar1.eps}
\end{minipage}
\flushleft
\begin{minipage}{0.75\linewidth}\,\,\,\,\,\centering GWN\end{minipage}
\begin{minipage}{0.15\linewidth}\centering CGWN\,\,\,\,\,\,\,\,\end{minipage}
\flushleft
\begin{minipage}{0.03\linewidth}
    \rotatebox{90}{\hspace{0mm}SG}
\end{minipage}
\begin{minipage}{0.96\linewidth}
    \begin{subfigure}
        \centering
        \includegraphics[width=0.67\linewidth,trim=0mm 24mm 6mm 16mm,clip]{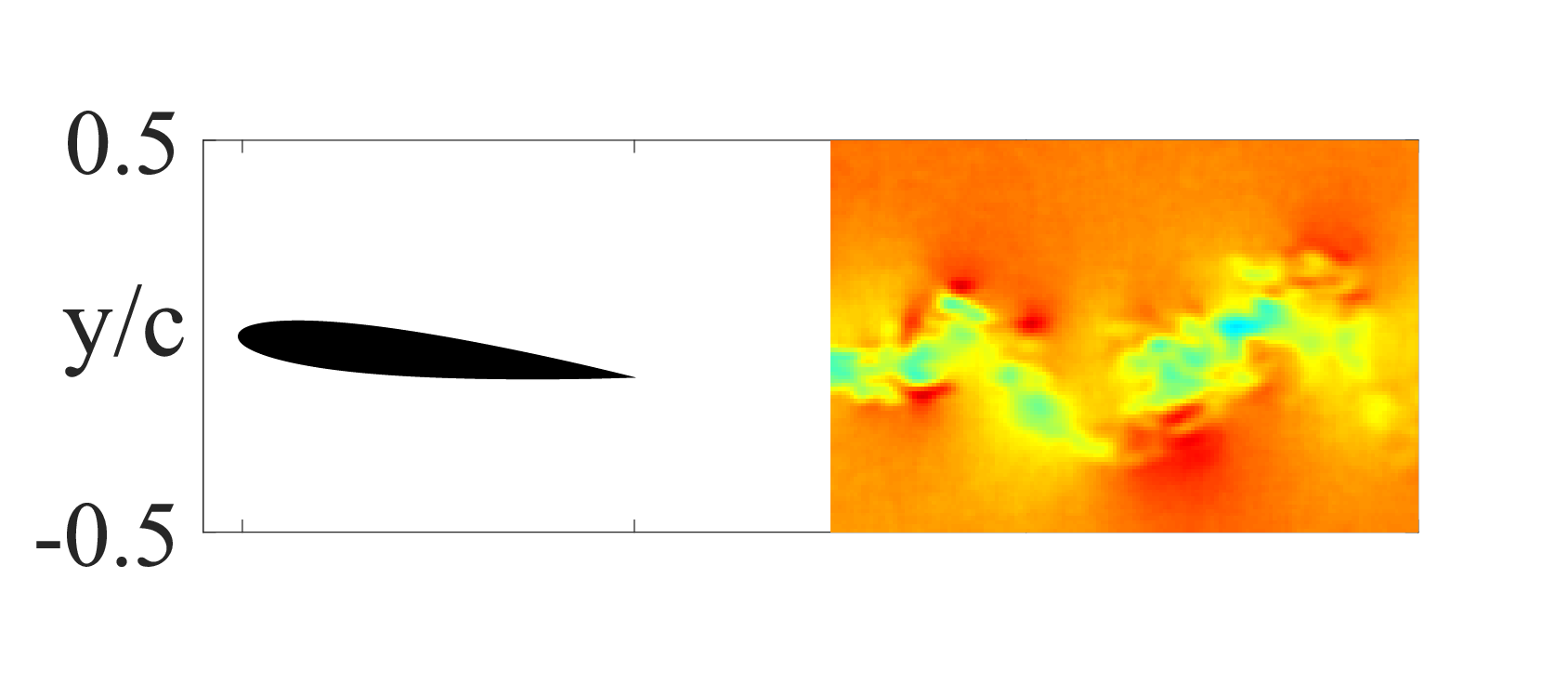}
    \end{subfigure}
    \begin{subfigure}
        \centering
        \includegraphics[width=0.29\linewidth,trim=160mm 24mm 6mm 16mm,clip]{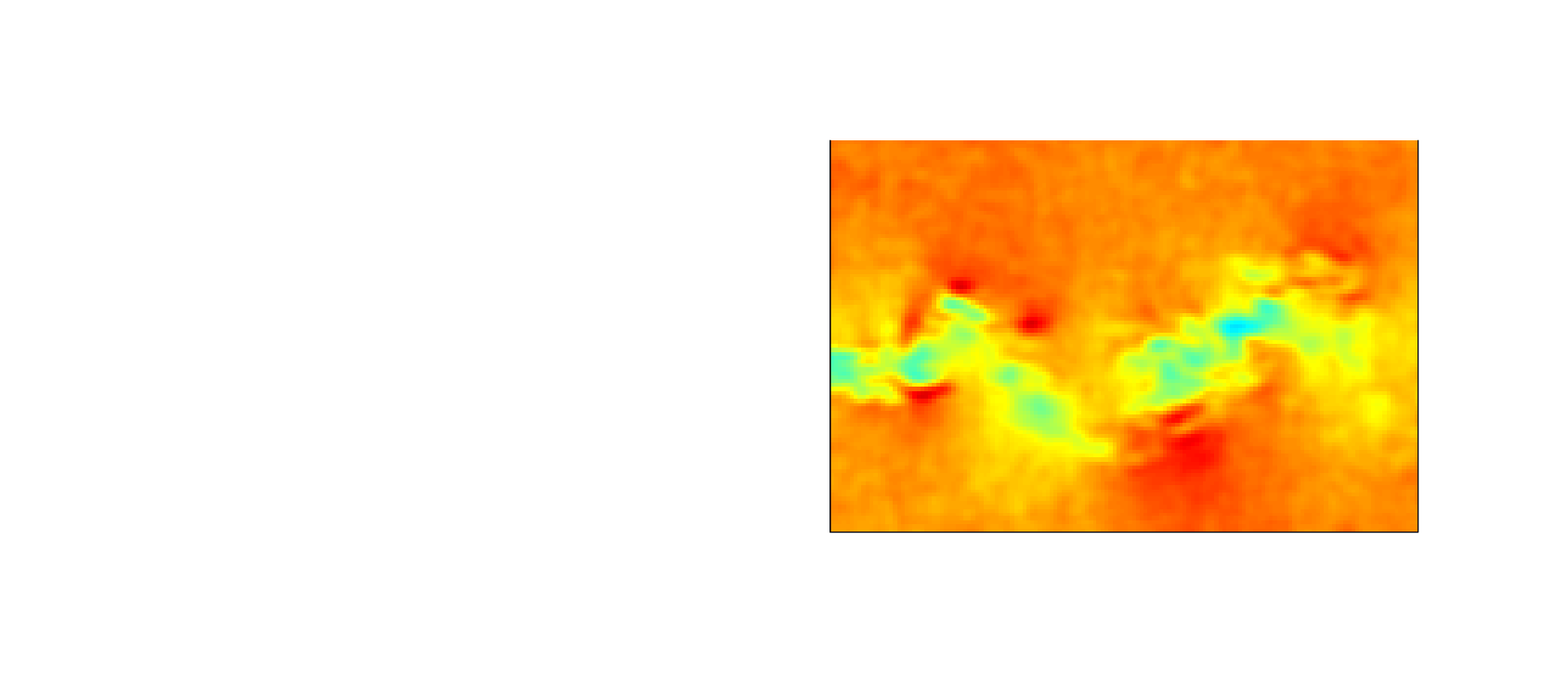}
    \end{subfigure}
\end{minipage}
\flushleft
\begin{minipage}{0.03\linewidth}
    \rotatebox{90}{\hspace{2mm}POD}
\end{minipage}
\begin{minipage}{0.96\linewidth}
    \begin{subfigure}
        \centering
        \includegraphics[width=0.67\linewidth,trim=0mm 24mm 6mm 16mm,clip]{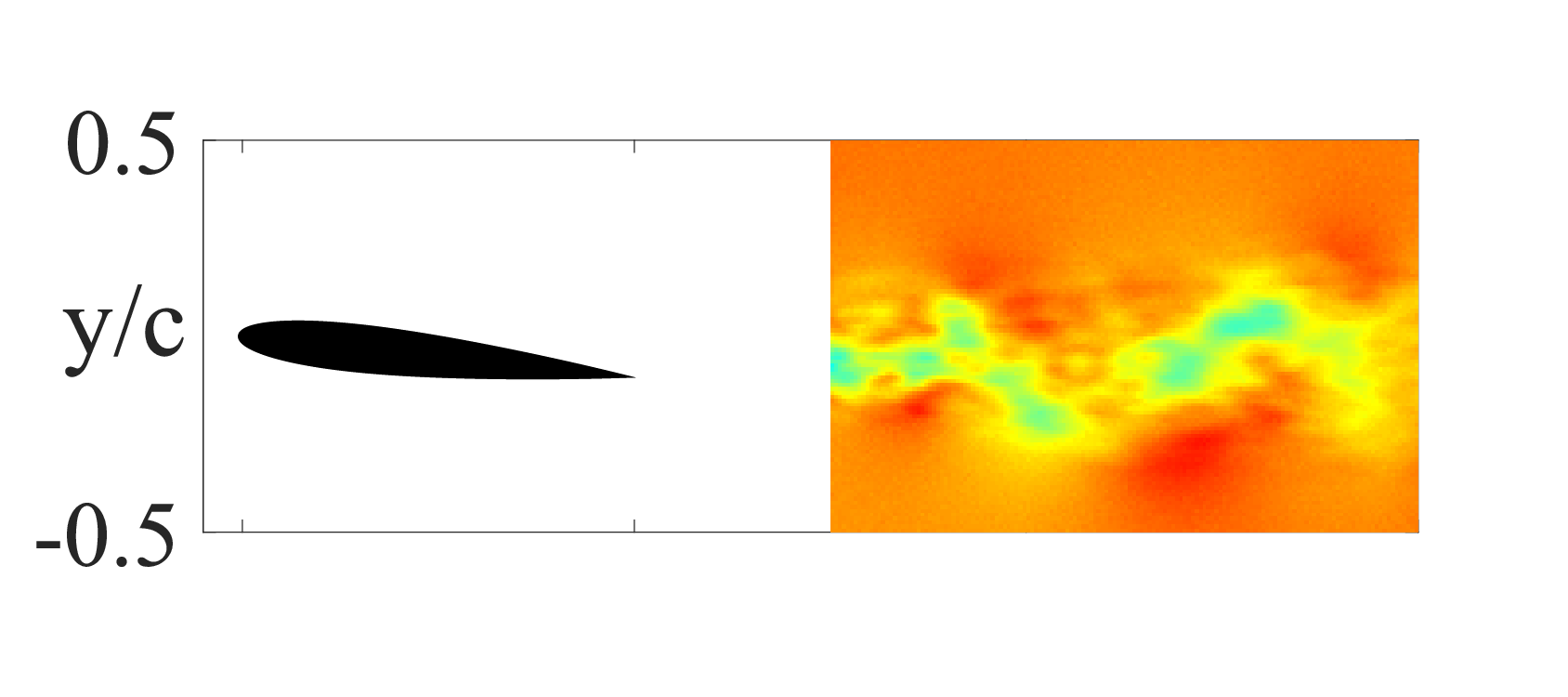}
    \end{subfigure}
    \begin{subfigure}
        \centering
        \includegraphics[width=0.29\linewidth,trim=160mm 24mm 6mm 16mm,clip]{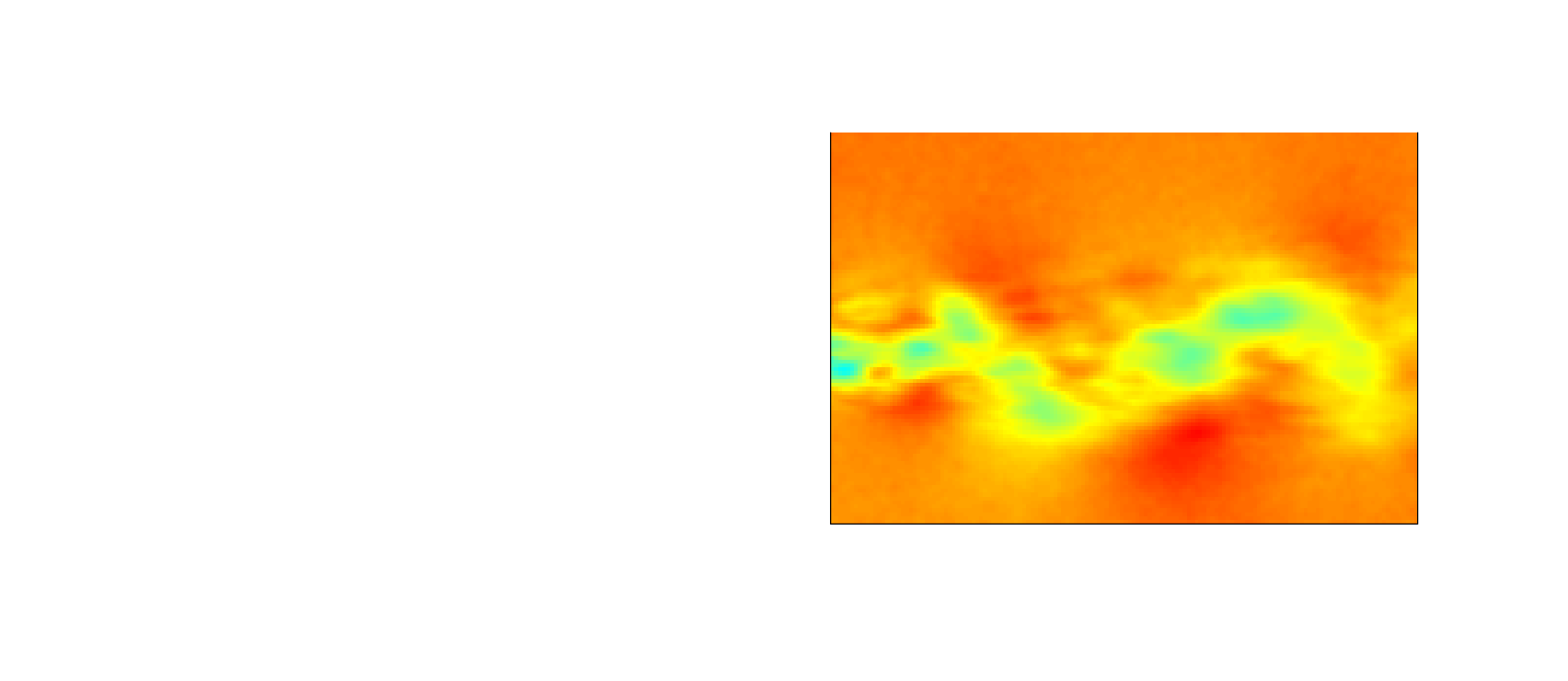}
    \end{subfigure}
\end{minipage}
\flushleft
\begin{minipage}{0.03\linewidth}
    \rotatebox{90}{\hspace{7mm}AMIC}
\end{minipage}
\begin{minipage}{0.96\linewidth}
    \begin{subfigure}
        \centering
        \includegraphics[width=0.67\linewidth,trim=0mm 0mm 6mm 16mm,clip]{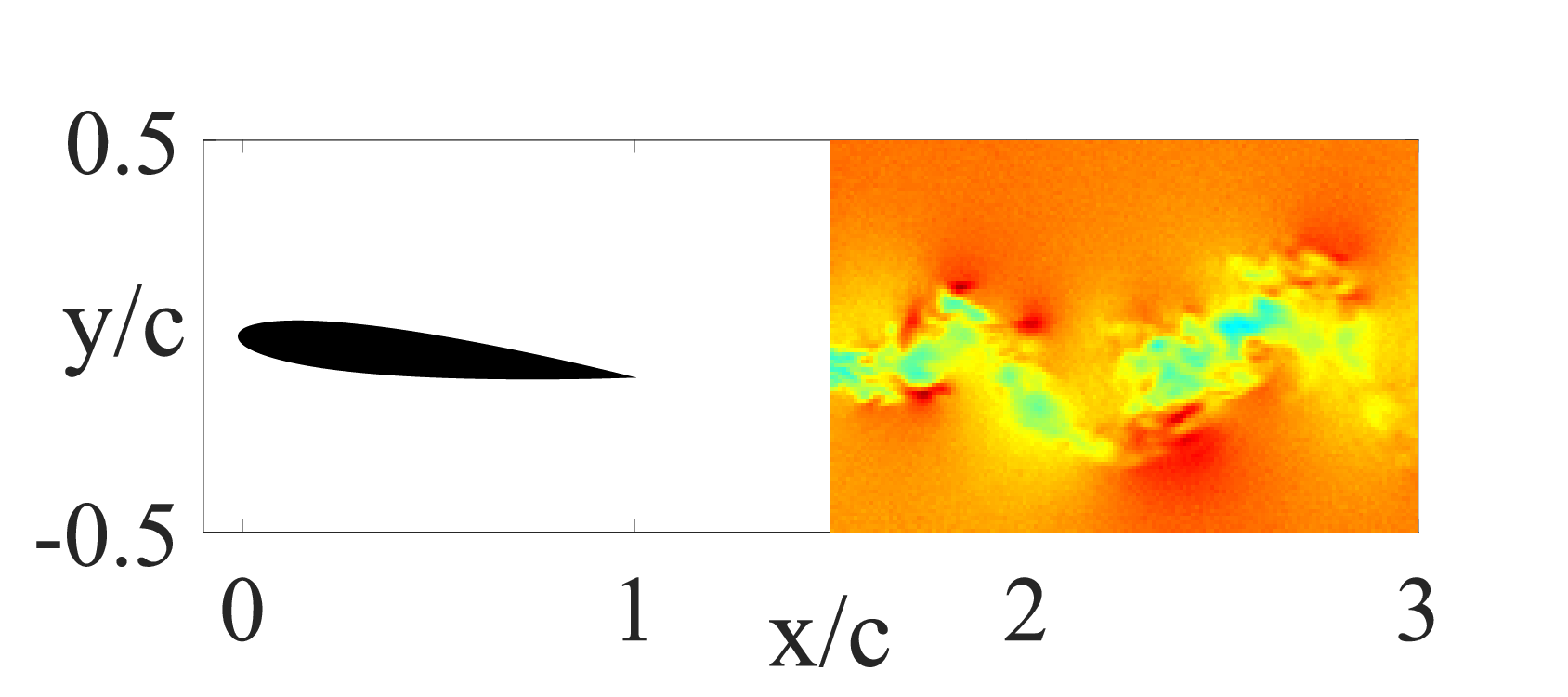}
    \end{subfigure}
    \begin{subfigure}
        \centering
        \includegraphics[width=0.29\linewidth,trim=160mm 0mm 6mm 16mm,clip]{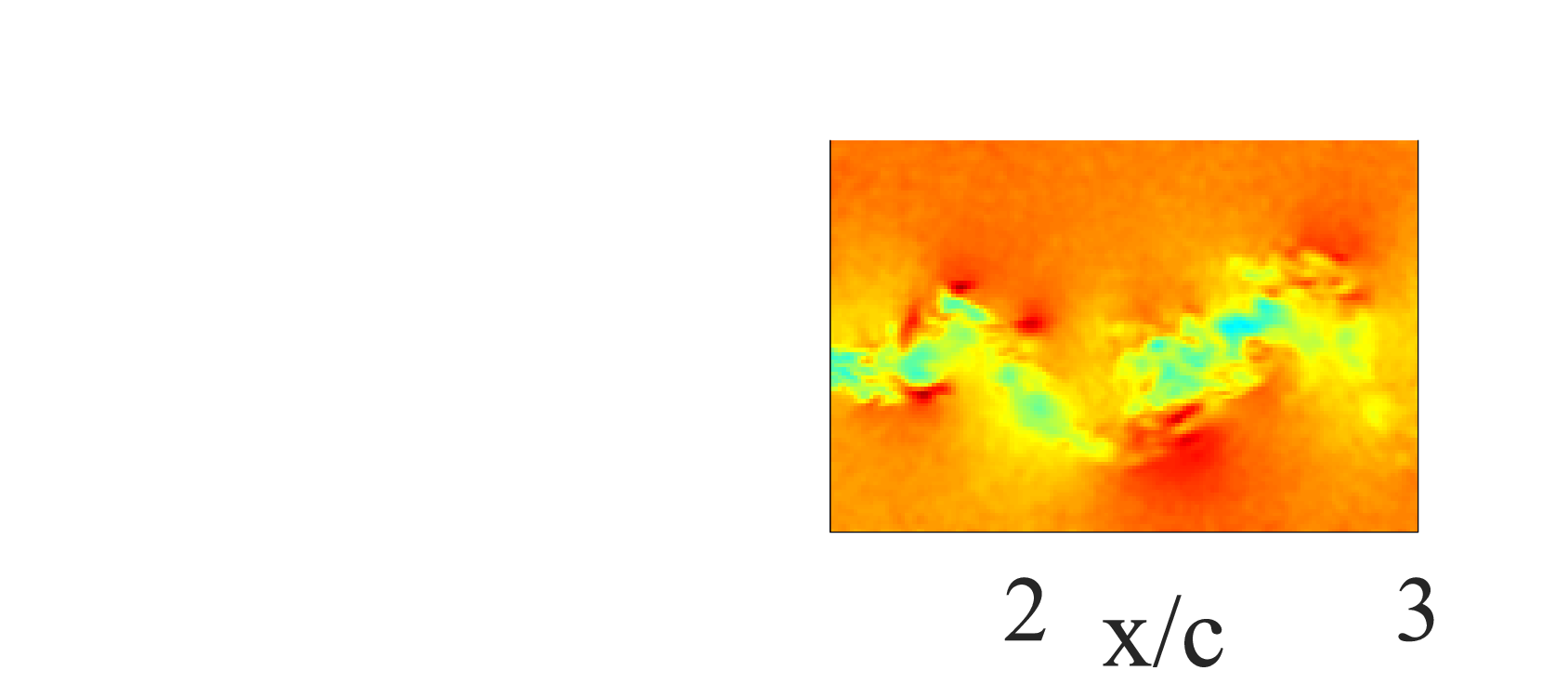}
    \end{subfigure}
\end{minipage}
\caption{Streamwise component of the velocity fields (normalized with the freestream velocity) after filtering/correcting with (from top to bottom) SG, POD and AMIC. The figure shows the processing result of two types of noise: GWN (the left column) and CGWN (the right column). The exact field for reference is included in Figure \ref{fig:AF_noise}.}
\label{fig:AF_u}
\end{figure}

\begin{figure}[h]
\flushleft
\begin{minipage}{0.96\linewidth}
    \hspace{6mm}
    \includegraphics[width=0.92\linewidth,trim=4mm 100mm 4mm 0mm,clip]{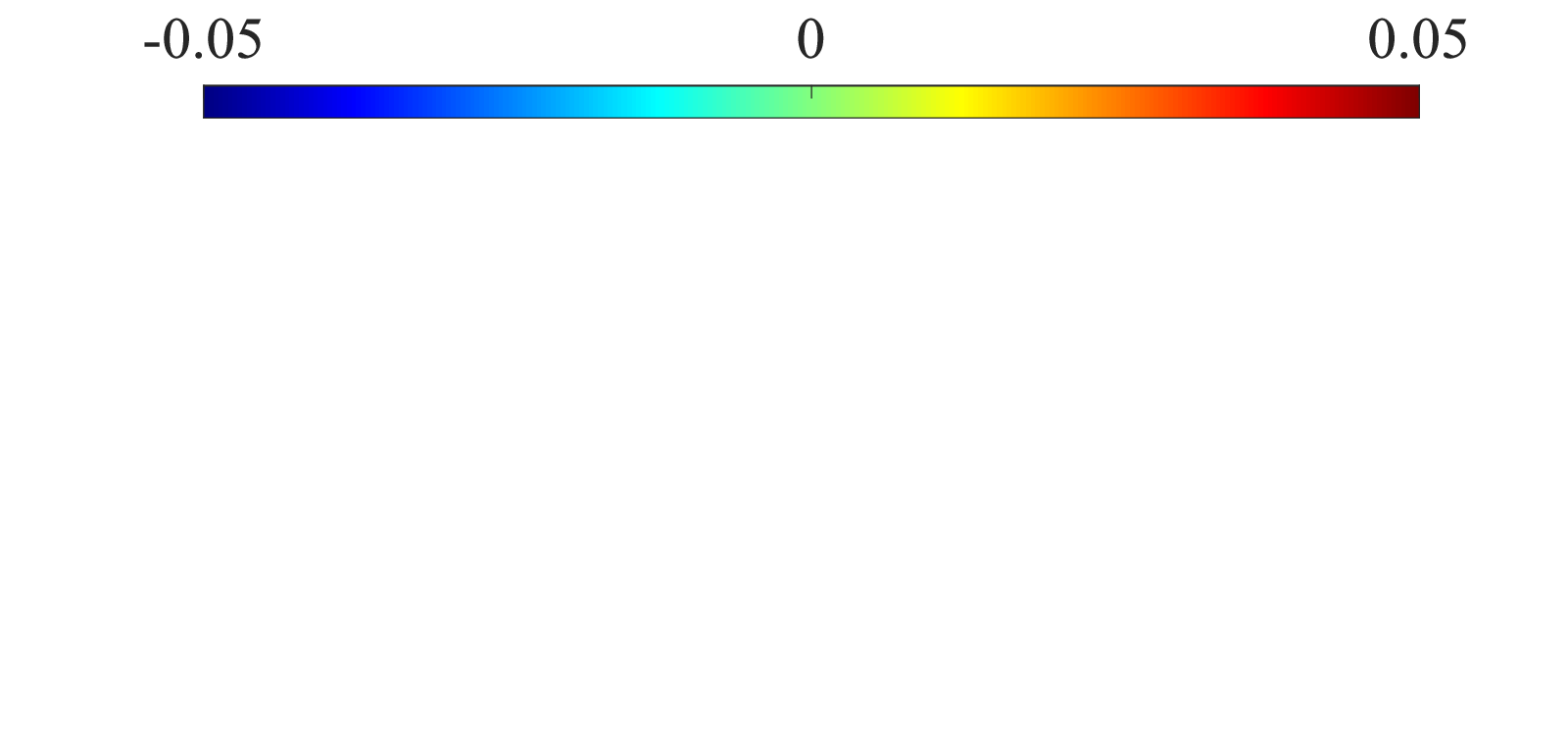}
\end{minipage}
\flushleft
\begin{minipage}{0.03\linewidth}
    \rotatebox{90}{\hspace{0mm}clean field}
\end{minipage}
\begin{minipage}{0.96\linewidth}
    \begin{subfigure}
        \centering
        \includegraphics[width=0.67\linewidth,trim=0mm 20mm 6mm 16mm,clip]{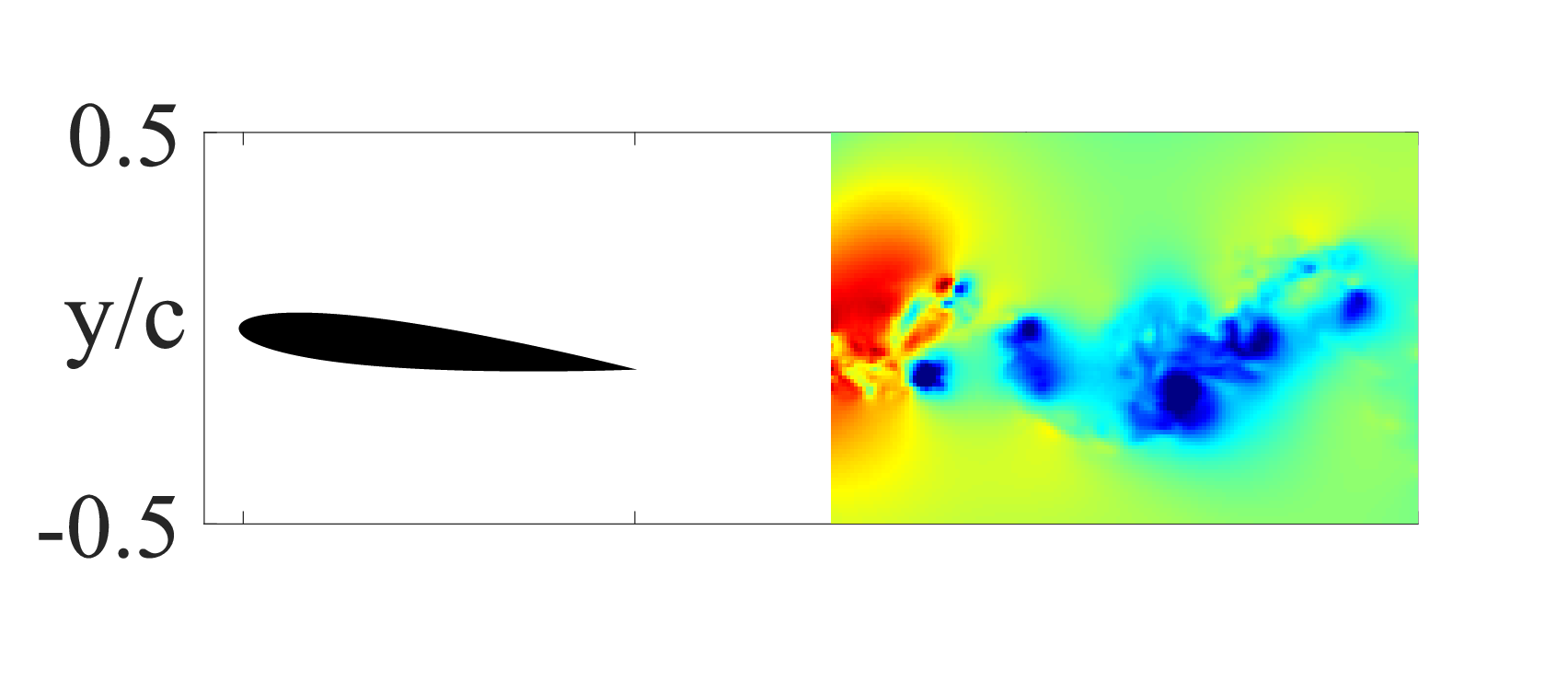}
    \end{subfigure}
\end{minipage}
\flushleft
\begin{minipage}{0.75\linewidth}\,\,\,\,\,\centering GWN\end{minipage}
\begin{minipage}{0.15\linewidth}\centering CGWN\,\,\,\,\,\,\,\,\end{minipage}
\flushleft
\begin{minipage}{0.03\linewidth}
    \rotatebox{90}{\hspace{0mm}noisy field}
\end{minipage}
\begin{minipage}{0.96\linewidth}
    \begin{subfigure}
        \centering
        \includegraphics[width=0.67\linewidth,trim=0mm 22mm 6mm 18mm,clip]{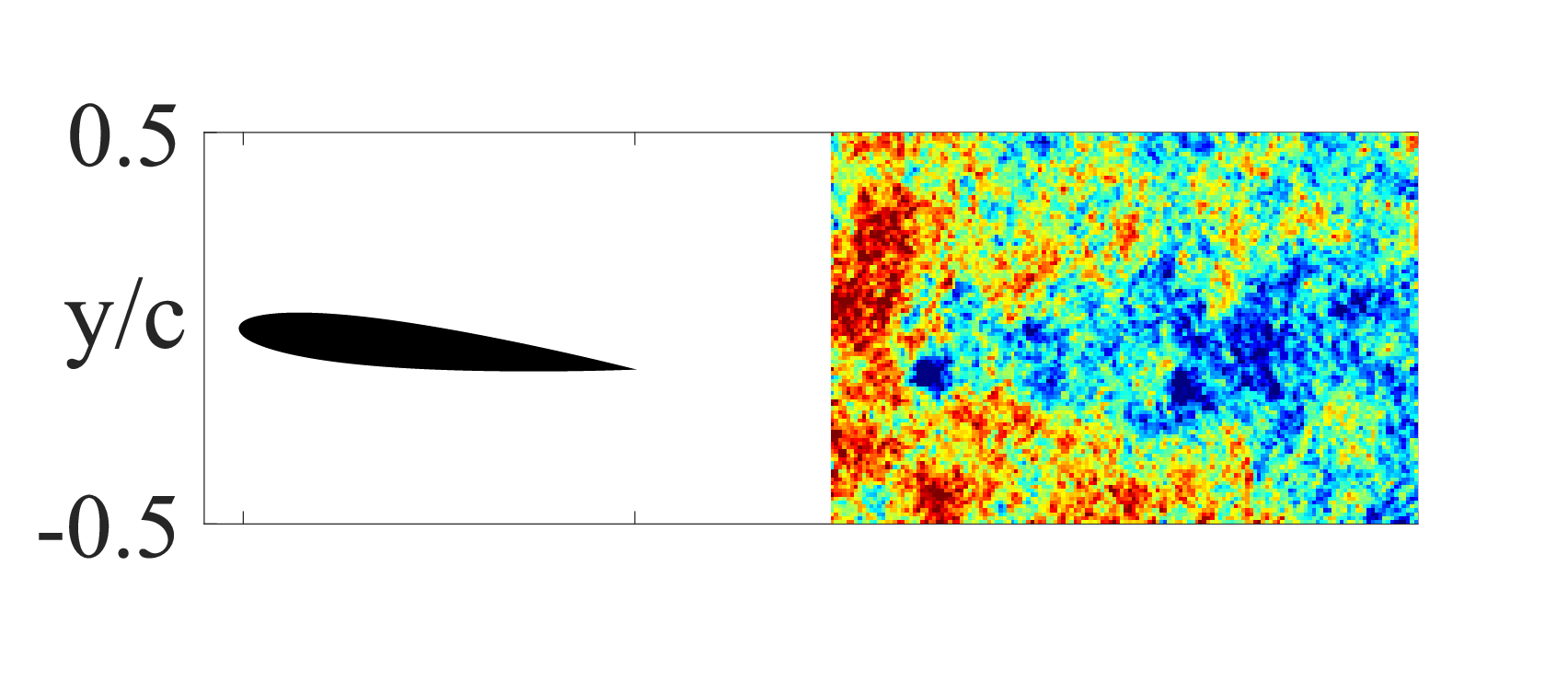}
    \end{subfigure}
    \begin{subfigure}
        \centering
        \includegraphics[width=0.29\linewidth,trim=160mm 22mm 6mm 18mm,clip]{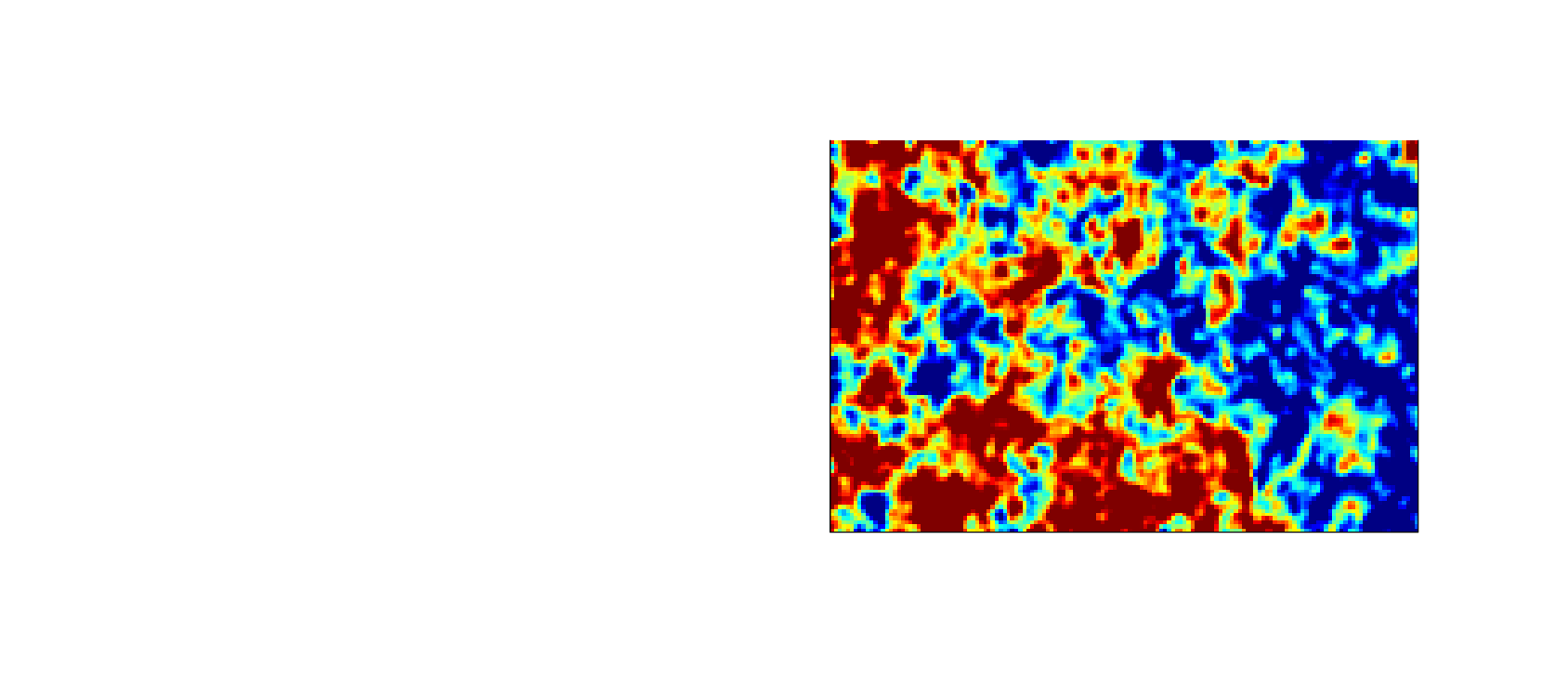}
    \end{subfigure}
\end{minipage}
\flushleft
\begin{minipage}{0.03\linewidth}
    \rotatebox{90}{\hspace{0mm}SG}
\end{minipage}
\begin{minipage}{0.96\linewidth}
    \begin{subfigure}
        \centering
        \includegraphics[width=0.67\linewidth,trim=0mm 24mm 6mm 16mm,clip]{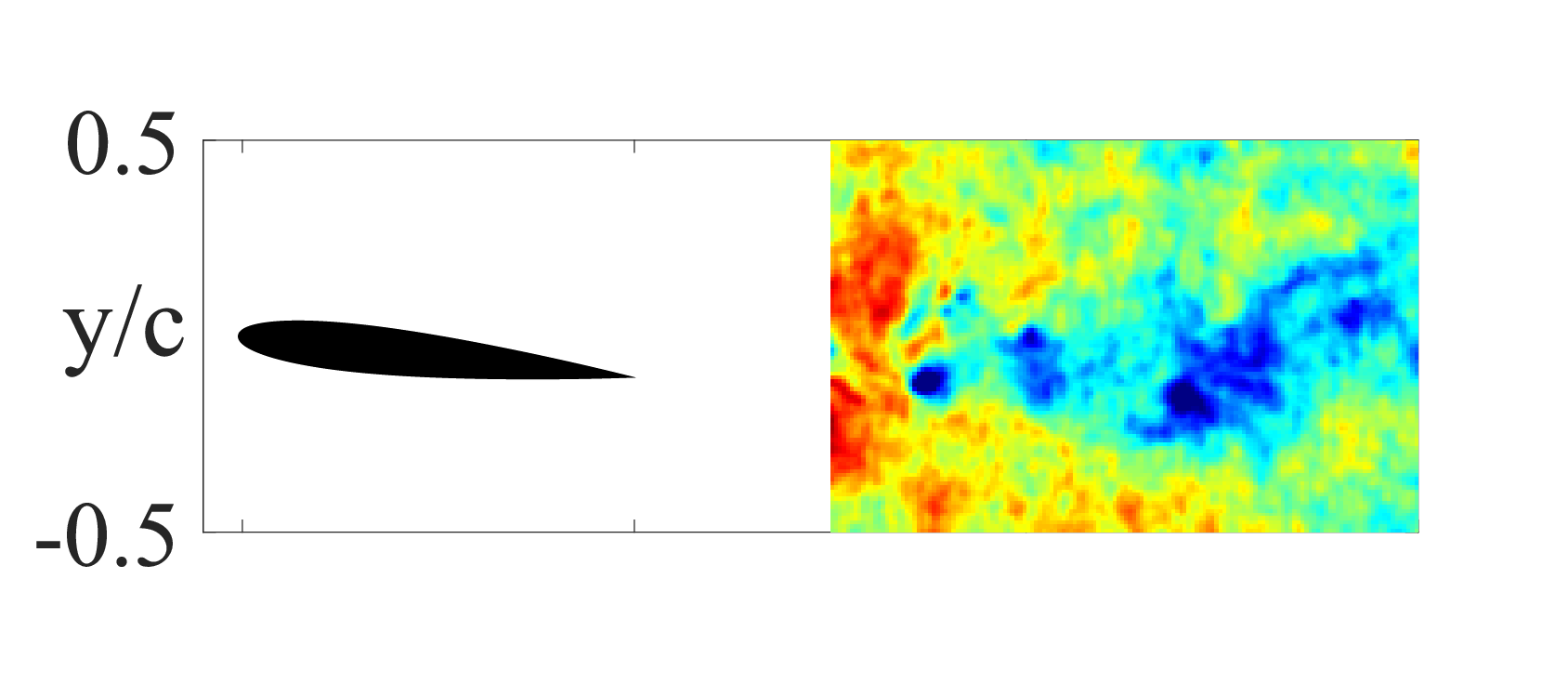}
    \end{subfigure}
    \begin{subfigure}
        \centering
        \includegraphics[width=0.29\linewidth,trim=160mm 24mm 6mm 16mm,clip]{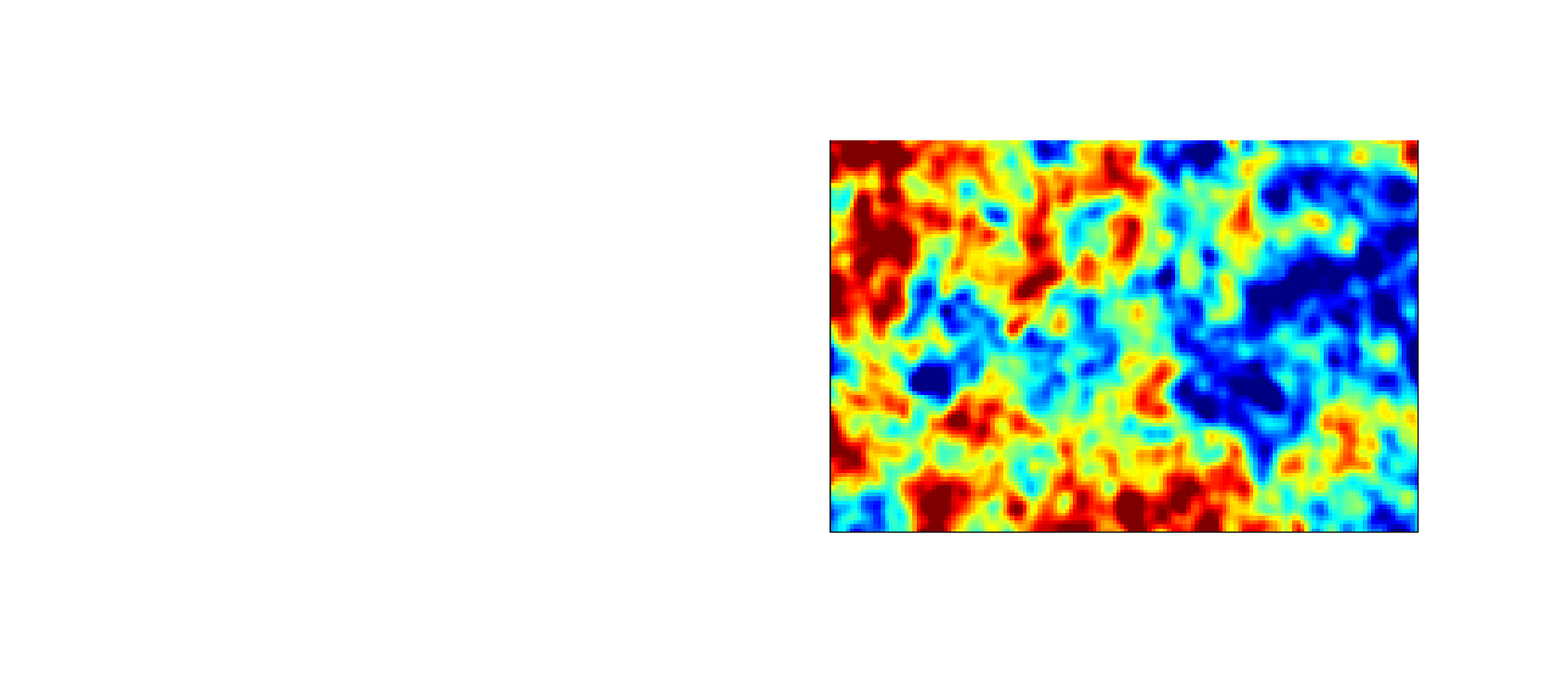}
    \end{subfigure}
\end{minipage}
\flushleft
\begin{minipage}{0.03\linewidth}
    \rotatebox{90}{\hspace{2mm}POD}
\end{minipage}
\begin{minipage}{0.96\linewidth}
    \begin{subfigure}
        \centering
        \includegraphics[width=0.67\linewidth,trim=0mm 22mm 6mm 18mm,clip]{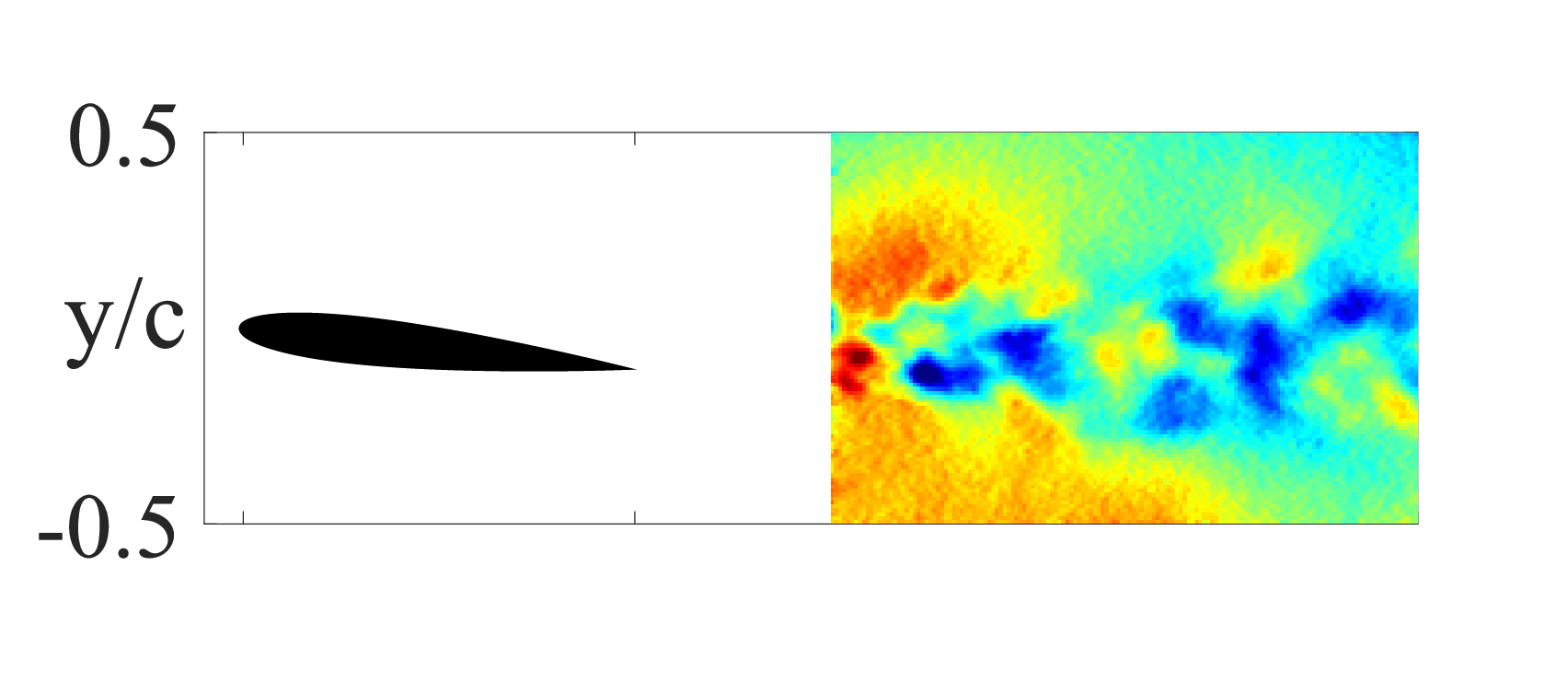}
    \end{subfigure}
    \begin{subfigure}
        \centering
        \includegraphics[width=0.29\linewidth,trim=160mm 22mm 6mm 18mm,clip]{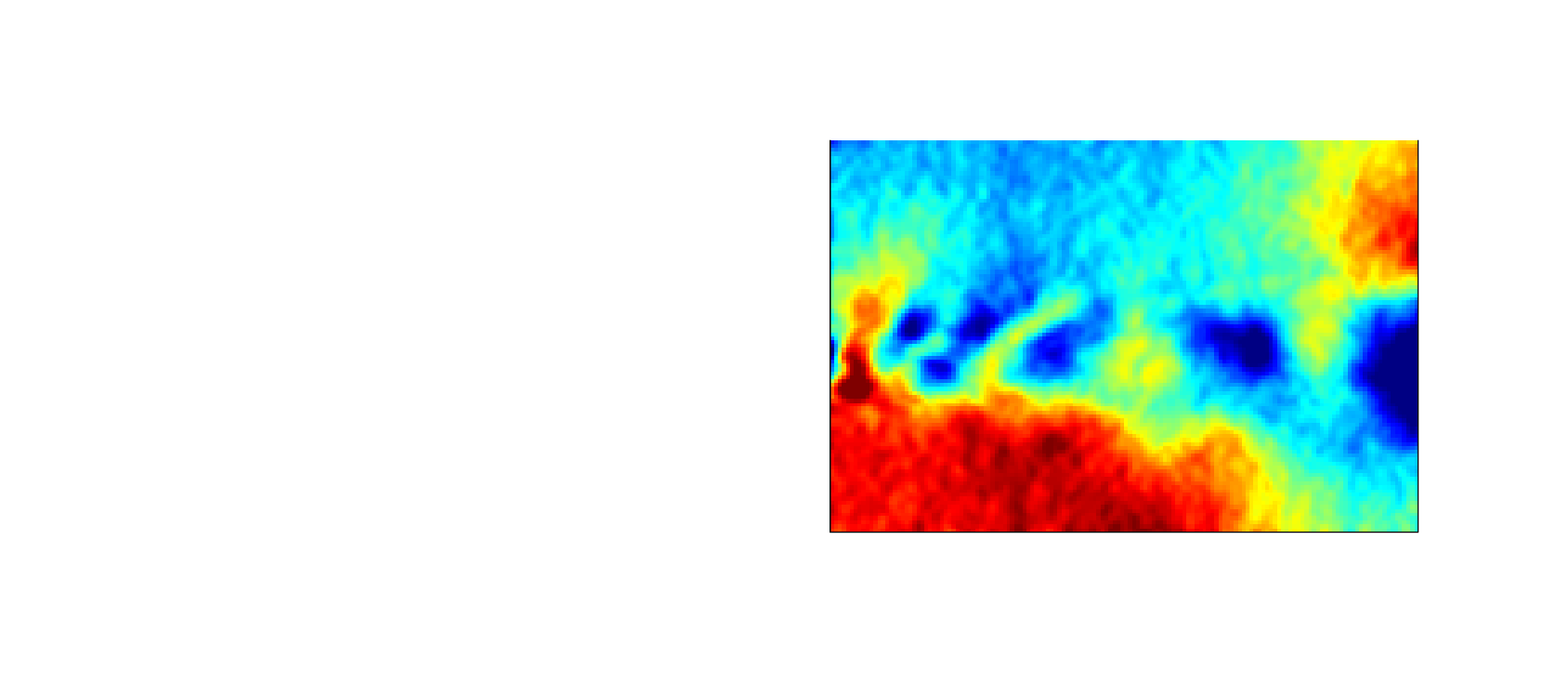}
    \end{subfigure}
\end{minipage}
\flushleft
\begin{minipage}{0.03\linewidth}
    \rotatebox{90}{\hspace{7mm}AMIC}
\end{minipage}
\begin{minipage}{0.96\linewidth}
    \begin{subfigure}
        \centering
        \includegraphics[width=0.67\linewidth,trim=0mm 0mm 6mm 16mm,clip]{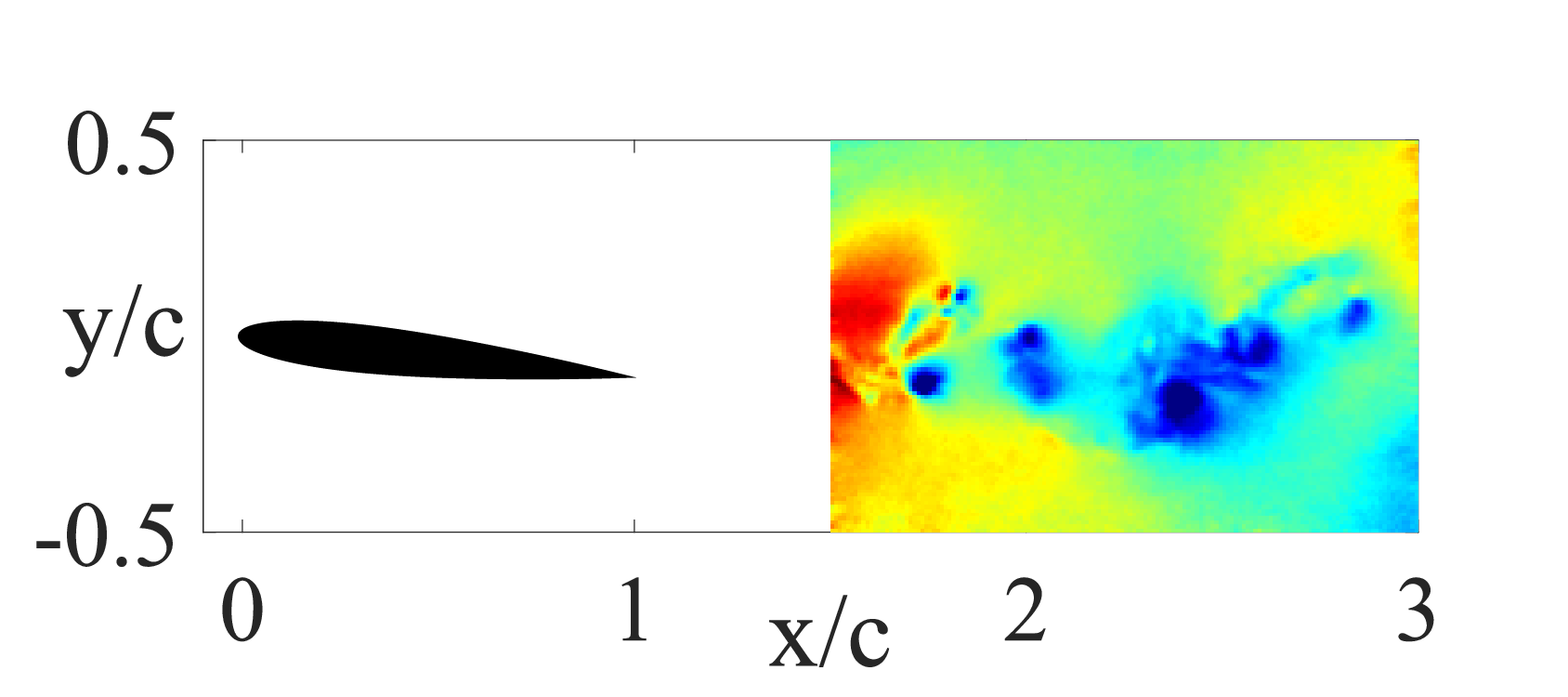}
    \end{subfigure}
    \begin{subfigure}
        \centering
        \includegraphics[width=0.29\linewidth,trim=160mm 0mm 6mm 16mm,clip]{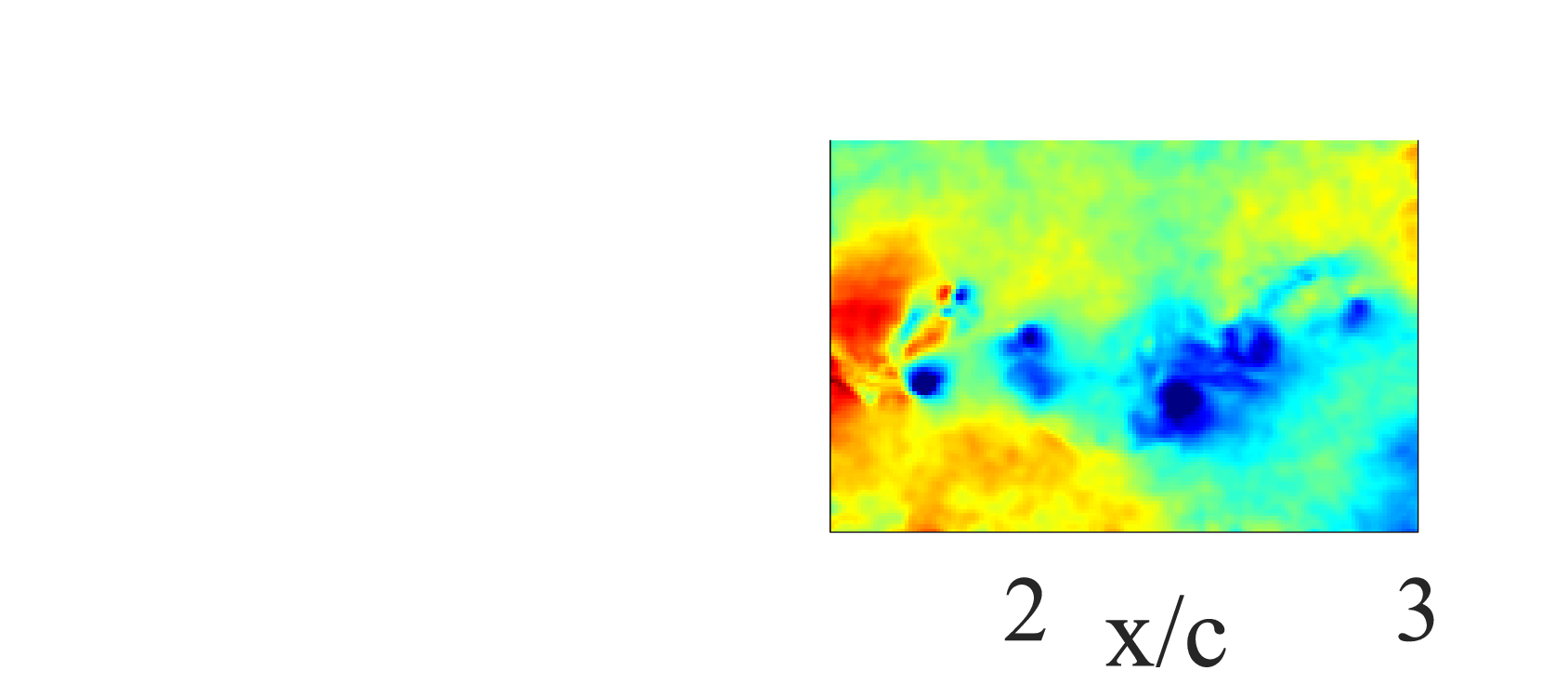}
    \end{subfigure}
\end{minipage}
\caption{Pressure fields, normalized by $\frac{1}{2}\rho U_\infty^2$, integrated from the original simulation data (1st row), from the velocity field superimposed with noise (2nd row), and after filtering/correcting it with (from 3rd row to bottom) SG filter, POD truncation and AMIC. The figure shows the processing result of two types of noise: GWN (the left column) and CGWN (the right column).}
\label{fig:AF_p}
\end{figure}

\begin{figure}[htb]
\flushleft
\begin{minipage}{0.96\linewidth}
    \hspace{6mm}
    \includegraphics[width=0.92\linewidth,trim=4mm 100mm 4mm 0mm,clip]{fig/AF_bar3.eps}
\end{minipage}
\flushleft
\begin{minipage}{0.03\linewidth}
    \rotatebox{90}{\hspace{0mm}clean field}
\end{minipage}
\begin{minipage}{0.96\linewidth}
    \begin{subfigure}
        \centering
        \includegraphics[width=0.45\linewidth,trim=0mm 24mm 6mm 2mm,clip]{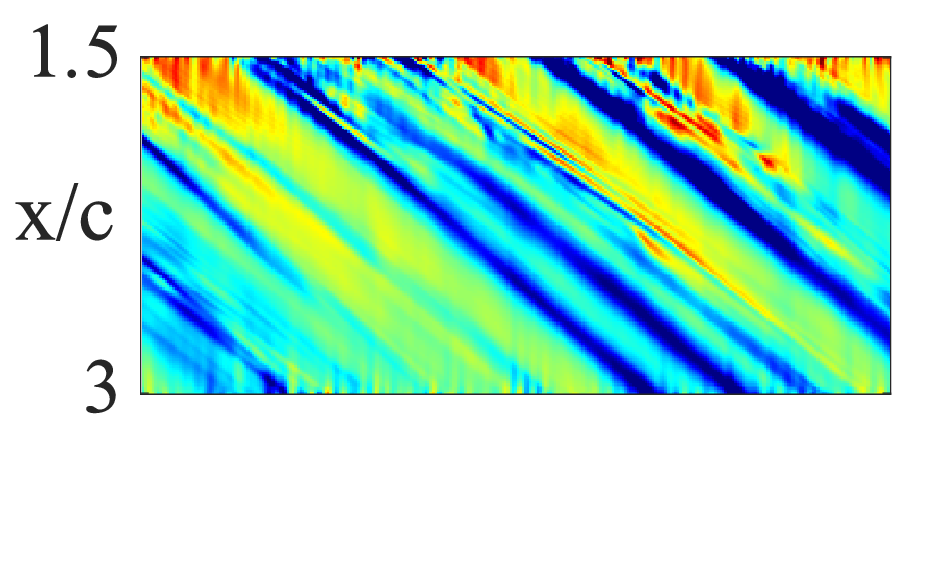}
    \end{subfigure}
\end{minipage}
\flushleft
\hspace{15mm}
\begin{minipage}{0.2\linewidth}\centering  GWN\end{minipage}
\hspace{20mm}
\begin{minipage}{0.2\linewidth}\centering CGWN\end{minipage}
\flushleft
\begin{minipage}{0.03\linewidth}
    \rotatebox{90}{\hspace{0mm}noisy field}
\end{minipage}
\begin{minipage}{0.96\linewidth}
    \begin{subfigure}
        \centering
        \includegraphics[width=0.45\linewidth,trim=0mm 24mm 6mm 2mm,clip]{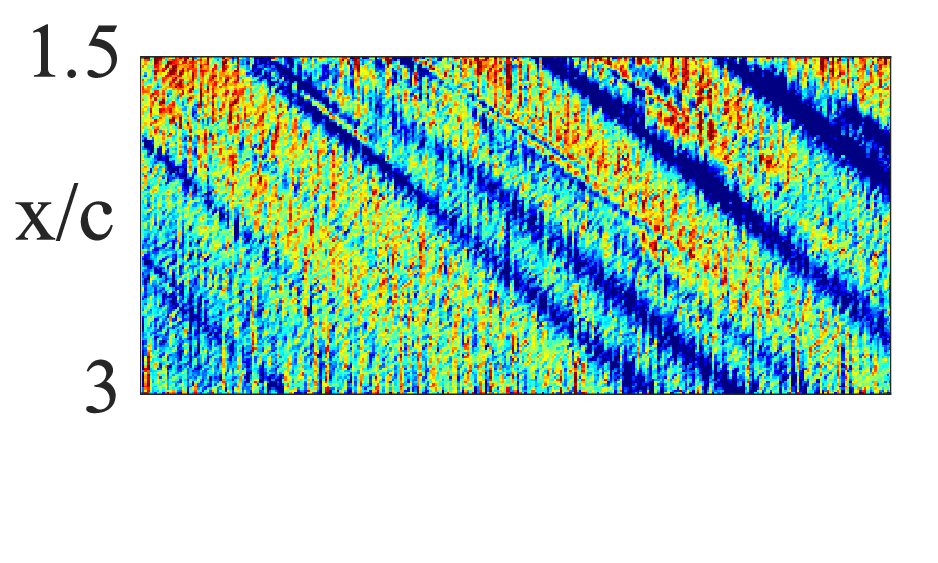}
    \end{subfigure}
    \begin{subfigure}
        \centering
        \includegraphics[width=0.45\linewidth,trim=0mm 24mm 6mm 2mm,clip]{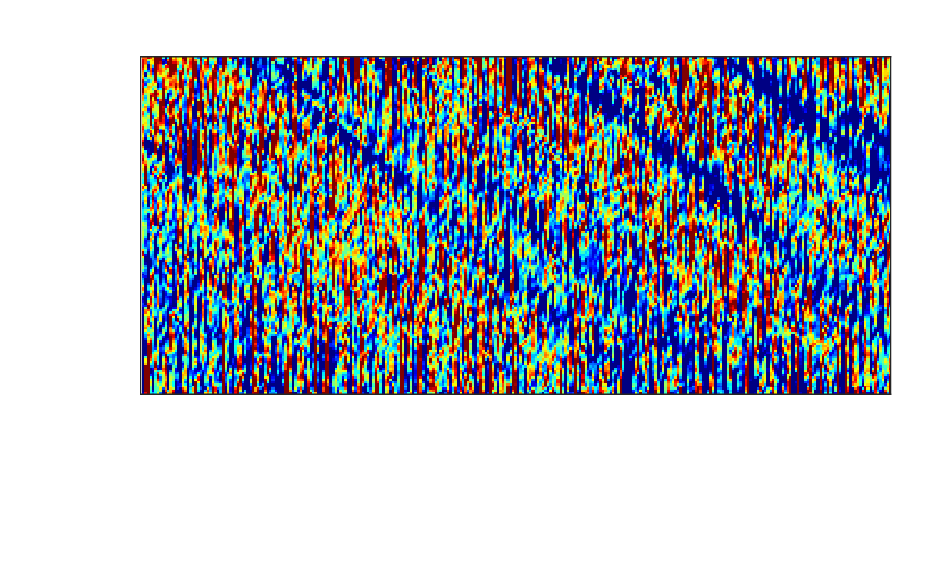}
    \end{subfigure}
\end{minipage}
\flushleft
\begin{minipage}{0.03\linewidth}
    \rotatebox{90}{\hspace{0mm}SG}
\end{minipage}
\begin{minipage}{0.96\linewidth}
    \begin{subfigure}
        \centering
        \includegraphics[width=0.45\linewidth,trim=0mm 24mm 6mm 2mm,clip]{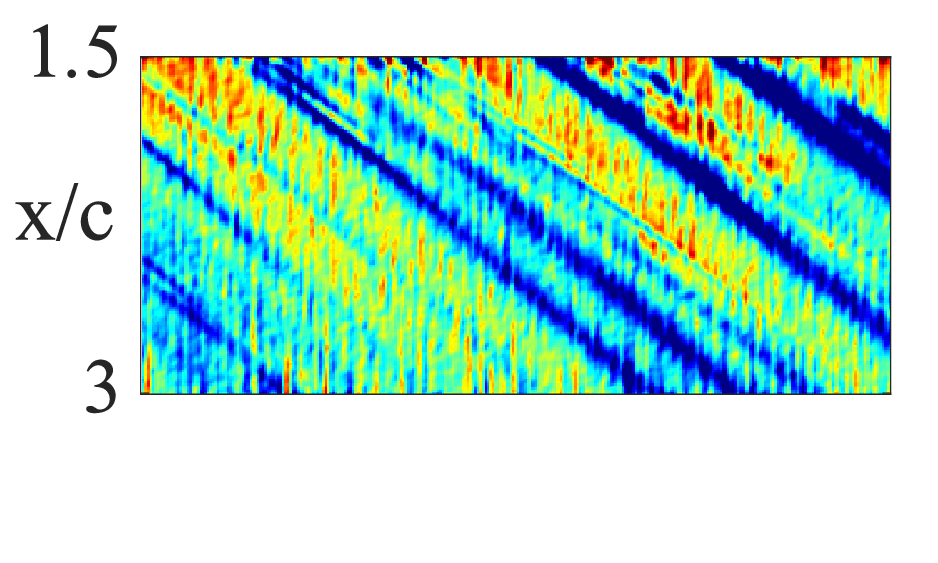}
    \end{subfigure}
    \begin{subfigure}
        \centering
        \includegraphics[width=0.45\linewidth,trim=0mm 24mm 6mm 2mm,clip]{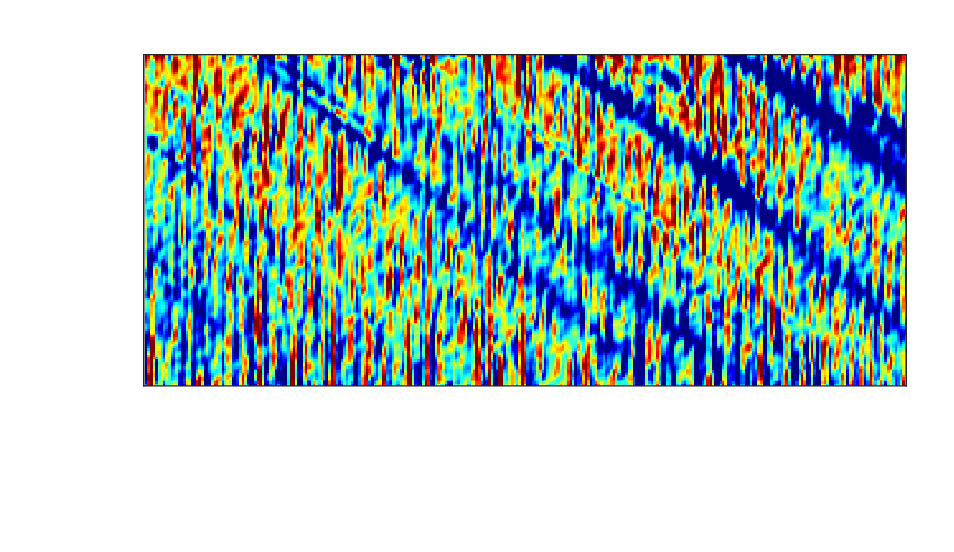}
    \end{subfigure}
\end{minipage}
\flushleft
\begin{minipage}{0.03\linewidth}
    \rotatebox{90}{\hspace{2mm}POD}
\end{minipage}
\begin{minipage}{0.96\linewidth}
    \begin{subfigure}
        \centering
        \includegraphics[width=0.45\linewidth,trim=0mm 24mm 6mm 2mm,clip]{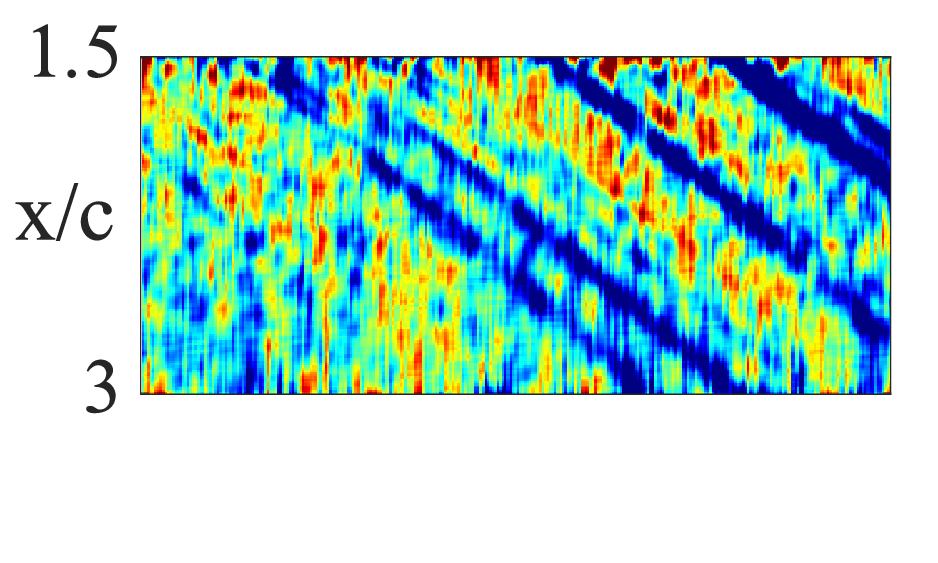}
    \end{subfigure}
    \begin{subfigure}
        \centering
        \includegraphics[width=0.45\linewidth,trim=0mm 24mm 6mm 2mm,clip]{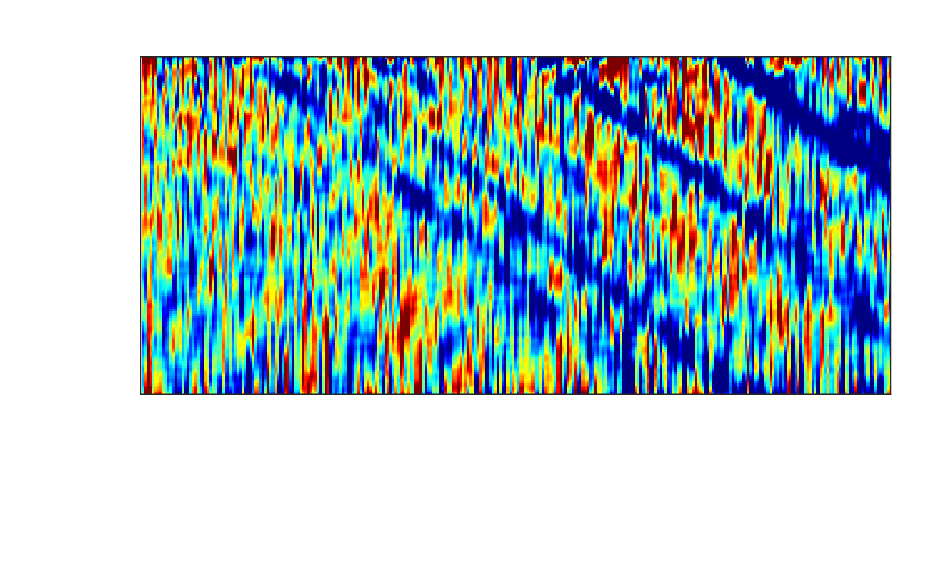}
    \end{subfigure}
\end{minipage}
\flushleft
\begin{minipage}{0.03\linewidth}
    \rotatebox{90}{\hspace{7mm}AMIC}
\end{minipage}
\begin{minipage}{0.96\linewidth}
    \begin{subfigure}
        \centering
        \includegraphics[width=0.45\linewidth,trim=0mm 4mm 6mm 2mm,clip]{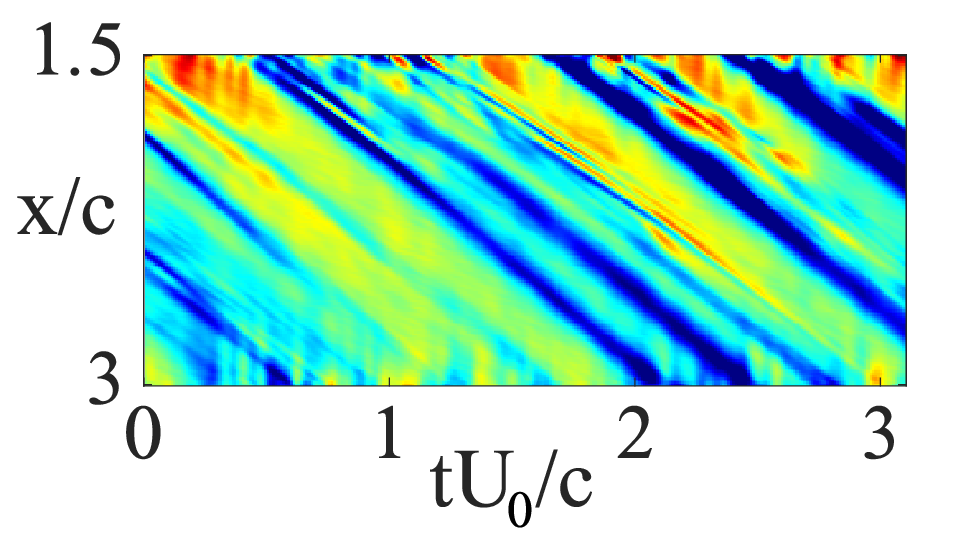}
    \end{subfigure}
    \begin{subfigure}
        \centering
        \includegraphics[width=0.45\linewidth,trim=0mm 4mm 6mm 2mm,clip]{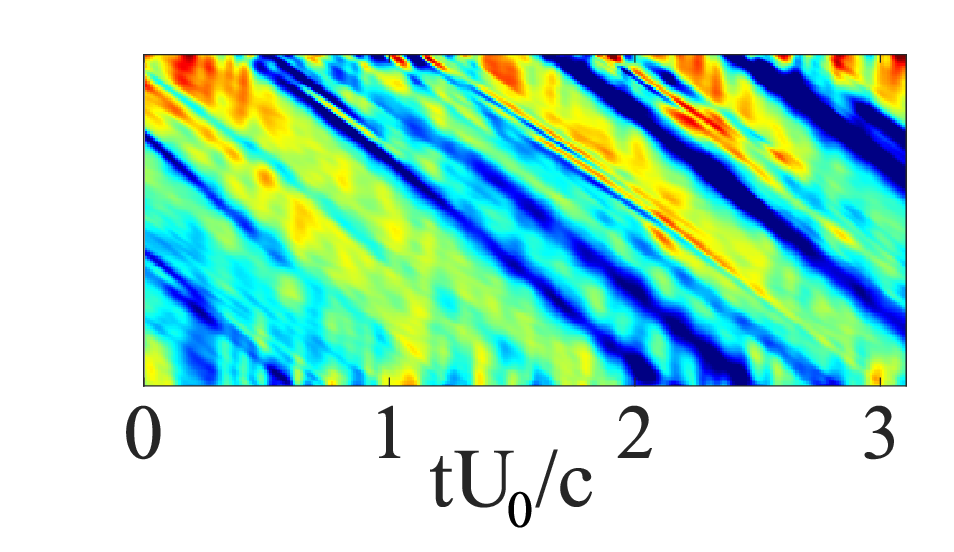}
    \end{subfigure}
\end{minipage}
\caption{The space-time diagrams of pressure field of the synthetic airfoil wake, extracted at $y/c = 0$, normalized by $\frac{1}{2}\rho U_\infty^2$, integrated from the original simulation data (1st row), from the velocity field superimposed with noise (2nd row), and after filtering/correcting it with (from 3rd row to bottom) SG filter, POD truncation and AMIC. The figure shows the processing result of two types of noise: GWN (the left column) and CGWN (the right column).}
\label{fig:AF_st}
\end{figure}

The methods are also tested on the synthetic dataset from LES of the airfoil wake, with the processed streamwise velocity field under both types of noise shown in Fig. \ref{fig:AF_u} and pressure in Fig. \ref{fig:AF_p}. Similar to the result of synthetic channel flow, the POD truncation removed some flow details alongside the noise, but preserved the wake structure, such as the meandering low-speed area and secondary structures. The SG filter effectively removed most noise when GWN was superimposed on the velocity field, although with some degree of modulation of the small scales. However, when the noise type changes to CGWN, the SG filter becomes less effective due to the spatial coherence of the noise, which can be partly fitted by the local polynomials. In contrast, AMIC did not thoroughly remove the noise under GWN but retained more details compared to the SG filter. Under CGWN, AMIC provides better noise filtering than SG, while preserving more details. The improvement in the spatiotemporal consistency can be observed from the space-time diagram in Fig. \ref{fig:AF_st}. Both SG filter and POD truncation can retrieve the pressure field from noisy velocity fields, but their capability to retain a consistent temporal behaviour is limited, especially under CGWN. AMIC, conversely, almost recovers the correct space-time relation of the clean field, showing much better temporal consistency than the other baseline methods.

Unlike the channel flow, the wake of the airfoil exhibits more intense velocity fluctuations. This allows pressure fields to be more easily predicted from noise-contaminated velocity fields. The large-scale high-pressure and low-pressure areas are reasonably captured by all the methods. Again, AMIC shows the least difference compared to the pressure calculated from the clean field. For the case of the CGWN corrupted fields, both SG filter and POD truncation exhibit large errors, while AMIC still provides a good estimation of the pressure field, albeit with a slight shift in the corners.

The overall RMS error and cosine similarity of velocity and pressure for both noisy or filtered fields, computed over $100$ frames for the channel flow and $300$ frames for the airfoil wake, are reported in Tab. \ref{tab:tab_v} and Tab. \ref{tab:tab}, respectively. The velocity RMS error in Eq. \ref{eq: error} is normalized by the bulk velocity $U_b$ for the channel case and by the freestream velocity $U_\infty$ for the airfoil case. The pressure RMS error is normalized by $\frac{1}{2}\rho U^2$, where $U$ is $U_b$ or $U_\infty$ for the channel and airfoil cases, respectively. Simultaneously, the pressure results are compared in terms of cosine similarity, defined by
\begin{equation}
    S_{cos} = \bigg|\frac{\langle p,p_{ref}\rangle}{\|p\|_2\|p_{ref}\|_2}\bigg|_t
\end{equation}
where the reference pressure $p_{ref}$ computed from the noise-free field and both parties of pressure are reshaped into column vectors, and $\langle  \cdot ,\cdot \rangle$ being the inner product. The $S_{cos}$ ranges between $[-1, 1]$, with $1$ indicating maximum alignment between original and reconstructed pressure fields. This equation is also applied to the velocity fields in Tab. \ref{tab:tab_v} and Tab. \ref{tab:tab}. The trend of RMS error and cosine similarity coincide with the observations for a single frame discussed above. POD truncation fails for both types of noise. The SG filter works well with GWN, but performs worse under CGWN. The AMIC provides the best pressure estimation among the methods by showing the least RMS error and the highest cosine similarity, even under CGWN.

\begin{table}[htb]
    \centering
    \caption{The RMS error and cosine similarity of velocity fields.}
    \label{tab:tab_v}
    \begin{tabular}{c|c|c|c|c|c}
        \hline
        noise type & metric & noisy & SG & POD & AMIC \\
        \hline
        \multicolumn{6}{c}{synthetic channel flow}\\
        \hline
        \multirow{2}{*}{GWN} & RMS & 0.00974 & 0.0119 & 0.0188 & 0.00835 \\
        \cline{2-6}
        & $S_{cos}$ & 0.99986 & 0.99980 & 0.99949 & 0.99990 \\
        \hline
        \multirow{2}{*}{CGWN} & RMS & 0.00997 & 0.0124 & 0.0182 & 0.00898 \\
        \cline{2-6}
        & $S_{cos}$ & 0.99986 & 0.99978 & 0.99953 & 0.99988 \\
        \hline
        \multicolumn{6}{c}{synthetic airfoil wake}\\
        \hline
        \multirow{2}{*}{GWN} & RMS & 0.0175 & 0.00978 & 0.0236 & 0.00690 \\
        \cline{2-6}
        & $S_{cos}$ & 0.99967 & 0.99990 & 0.99940 & 0.99995 \\
        \hline
        \multirow{2}{*}{CGWN} & RMS & 0.0179 & 0.0125 & 0.0277 & 0.00715 \\
        \cline{2-6}
        & $S_{cos}$ & 0.99966 & 0.99983 & 0.99917 & 0.99995 \\
        \hline
    \end{tabular}
\vspace{1cm}
    \centering
    \caption{The RMS error and cosine similarity of pressure fields obtained from integration of noisy and filtered velocity fields.}
    \label{tab:tab}
    \begin{tabular}{c|c|c|c|c|c}
        \hline
        noise type & metric & noisy & SG & POD & AMIC \\
        \hline
        \multicolumn{6}{c}{synthetic channel flow}\\
        \hline
        \multirow{2}{*}{GWN} & RMS & 0.0118 & 0.00554 & 0.0141 & 0.00360 \\
        \cline{2-6}
        & $S_{cos}$ & 0.533 & 0.780 & 0.433 & 0.893 \\
        \hline
        \multirow{2}{*}{CGWN} & RMS & 0.0450 & 0.0246 & 0.0288 & 0.00736 \\
        \cline{2-6}
        & $S_{cos}$ & 0.162 & 0.267 & 0.236 & 0.689 \\
        \hline
        \multicolumn{6}{c}{synthetic airfoil wake}\\
        \hline
        \multirow{2}{*}{GWN} & RMS & 0.0286 & 0.0156 & 0.0234 & 0.00927 \\
        \cline{2-6}
        & $S_{cos}$ & 0.666 & 0.845 & 0.695 & 0.937 \\
        \hline
        \multirow{2}{*}{CGWN} & RMS & 0.0784 & 0.0484 & 0.0420 & 0.0116 \\
        \cline{2-6}
        & $S_{cos}$ & 0.315 & 0.459 & 0.466 & 0.905 \\
        \hline
    \end{tabular}
\end{table}

\section{Evaluation of pressure reconstruction quality without knowing the true value of pressure field}
\label{sec:index}

In this section, we explore sanity checks that can be used to assess the estimation of time sequences of pressure fields even without knowing a ground truth reference (which is the situation to be expected in an experiment). 

\begin{figure}[]
\begin{subfigure}[]
    \centering
    \includegraphics[width=0.45\linewidth,trim=0mm 0mm 12mm 0mm,clip]{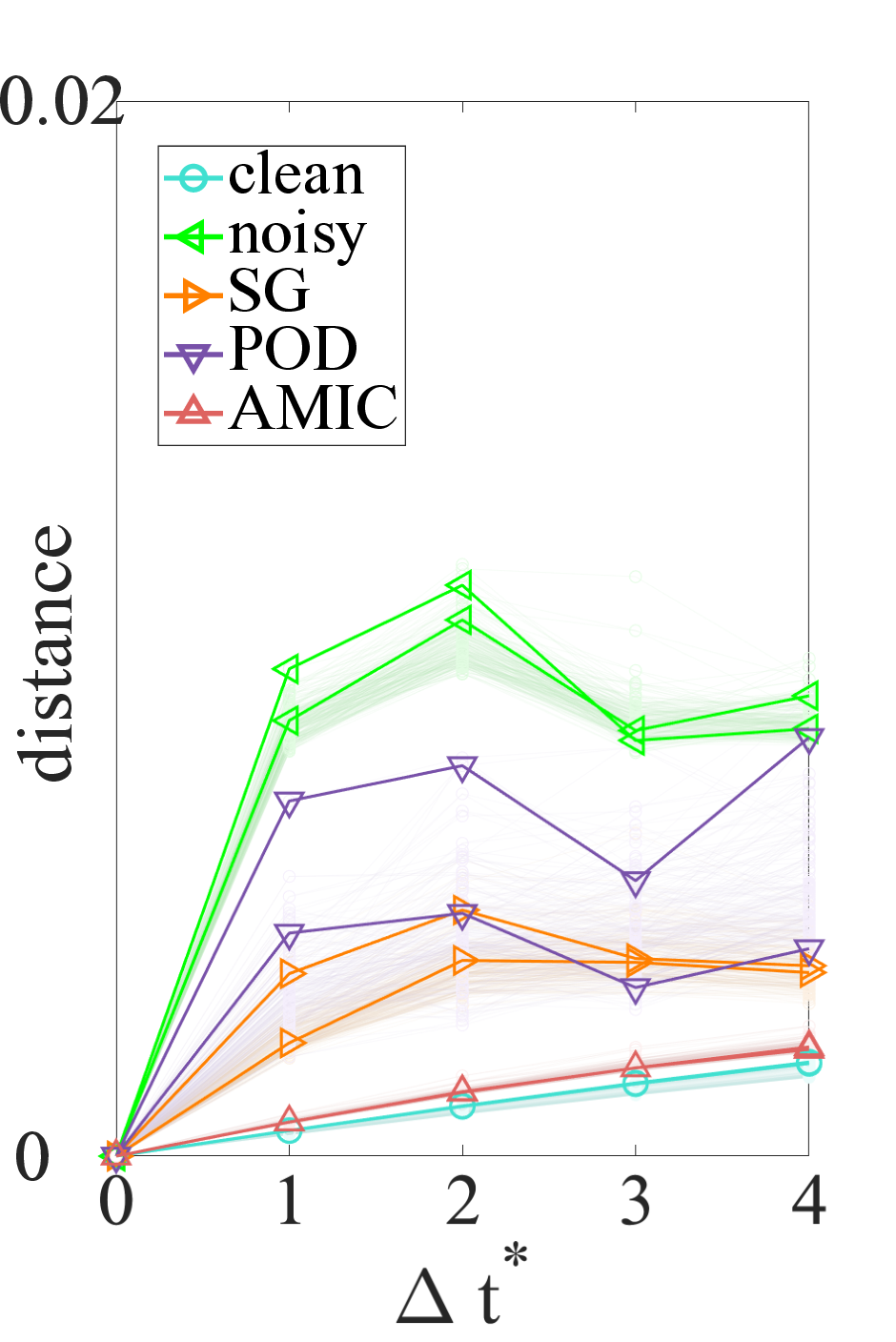}
\end{subfigure}
\begin{subfigure}[]
    \centering
    \includegraphics[width=0.45\linewidth,trim=0mm 0mm 12mm 0mm,clip]{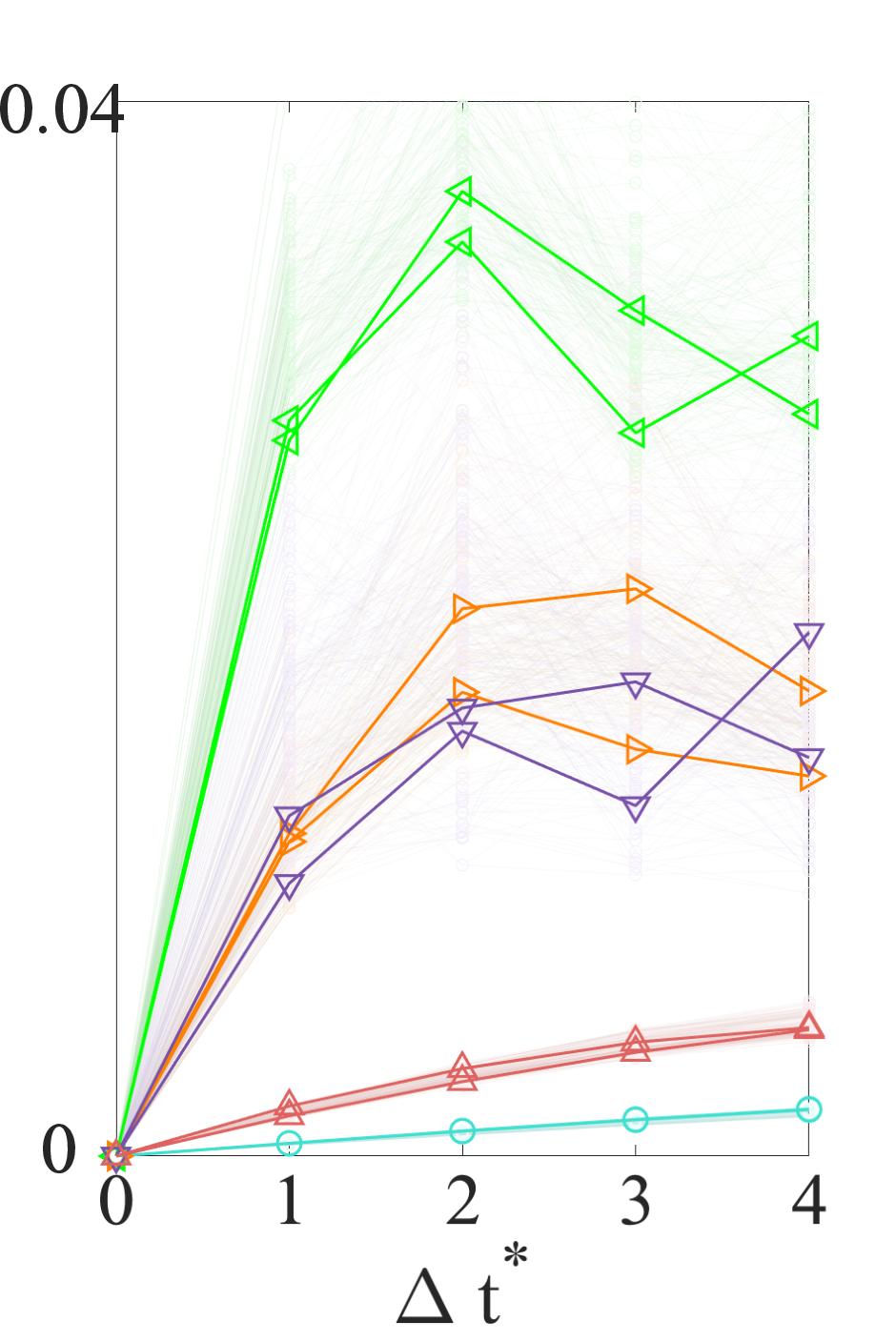}
\end{subfigure}
\caption{The distance of pressure field between adjacent frames versus the frame increments from the channel dataset, where $\Delta t^*$ is the $\Delta t$ normalized by the sampling time, with the (\textbf{a}) GWN and (\textbf{b}) CGWN, closer to original is better, with two curves for each method to be highlighted.}
\label{fig:p_goodness}
\end{figure}

Generally, uncertainties in the boundary conditions and originating from ineffective filtering of noise determine a nonphysical flickering of the temporal sequence of pressure fields. However, unless a discontinuity of the pressure field is present for any physical reasons (e.g. appearing/disappearing shock waves, expansion fans or moving objects), a smooth temporal behaviour is generally to be expected. The pressure field can be expanded to Taylor series,
\begin{equation}
\begin{aligned}
p(\mathbf{x},t+\Delta t) = &p(\mathbf{x},t) + \frac{\partial p(\mathbf{x},t)}{\partial t}\Delta t\\
+ &\frac{1}{2}\frac{\partial^2 p(\mathbf{x},t)}{\partial t^2}\Delta t^2 + \frac{1}{6}\frac{\partial^3 p(\mathbf{x},t)}{\partial t^3}\Delta t^3+o(\Delta t^3)
\end{aligned}
\end{equation}
For sufficiently short time increments $\Delta t$, the expansion can be reasonably truncated to the first order, with negligible error due to higher order terms,
\begin{equation}
    p(\mathbf{x},t+\Delta t) - p(\mathbf{x},t) \approx \frac{\partial p(\mathbf{x},t)}{\partial t}\Delta t
\end{equation}
The linear approximation requires the sample rate of time-resolved PIV to be sufficiently large compared to the timescale of the flow. Errors due to truncation are expected to be much smaller than the time jitter of pressure observed between frames. The L2-norm of the distance between consecutive pressure fields should be reasonably approximated as 
\begin{equation}
    \left\|p(\mathbf{x},t+\Delta t) - p(\mathbf{x},t)\right\|_2 \approx \left\|\frac{\partial p(\mathbf{x},t)}{\partial t}\right\|_2\Delta t
    \label{eq:dis}
\end{equation}
We plot the distance versus temporal interval for the channel dataset from $0$ to $4$ frames separation in Fig. \ref{fig:p_goodness}. Each shaded curve bundle is composed of $24$ independent realizations, with $2$ randomly selected among them being highlighted. The lines of the pressure field from the original and AMIC velocity fields under GWN are nearly coincident. On the other hand, the distance among consecutive pressure fields experiences a sudden increase and large variability in the case of noisy and POD-filtered velocity fields. This might be ascribed to flickering of the pressure fields during the state transitions from the $0^{th}$ to the $1^{st}$ frame and from the $1^{st}$ to the $2^{nd}$ frame. The SG filter seems to reduce this effect, although less effectively than AMIC. When the noise type switched to spatially coherent CGWN, the curves of noisy, SG-filtered, and POD-filtered fields exceeded the scale limits of the figure due to large errors. In contrast, the curve of AMIC remained linear, albeit with a higher slope compared to the clean field. This behaviour can be explained due to some residual noise of the filtered pressure fields, as observed in Fig. \ref{fig:p}. Nonetheless, the estimated pressure fields with AMIC still exhibit good temporal evolution.

Since the pressure field distance is proportional to the temporal interval, a quantitative measure of the linearity over time can be obtained from a linear regression. The line can be fit via linear regression to the relation $d = \alpha t$, where $d$ is the distance, $\alpha$ is the slope and $t$ is the temporal interval. Using the least square method, the slope is
\begin{equation}
    \alpha = \frac{\mathbf{T}^T\mathbf{D}}{\mathbf{T}^T\mathbf{T}}
\end{equation}
where $\mathbf{T}$ is a column vector containing temporal ticks (i.e. $[0,1,2,3,4]^T$), and $\mathbf{D}$ is a column vector containing distance. Then goodness of fitting \citep{draper1998applied} is given by
\begin{equation}
    R^2 = 1-\frac{\Sigma(d-\hat{d})^2}{\Sigma(d-\overline{d})^2}
    \label{eq:r2}
\end{equation}
where $\hat{d}$ is the prediction from linear regression, and $\overline{d}$ is the average of $d$. The goodness ranges from $0$ to $1$, with values closer to $1$ indicating a better fit. The goodness is calculated over $192$ 5-frame sequences individually and then averaged for the synthetic channel flow and over $592$ 5-frame sequences individually and then averaged for the synthetic airfoil wake. As listed in Tab. \ref{tab:R2}, the values obtained from the AMIC processed field under both types of noise are close to $1$, suggesting a strong linear behaviour. In contrast, the values obtained from the noisy, SG-filtered, and POD-filtered fields deviate significantly from $1$, indicating a poorer fit and strong flickering.

\begin{table}[]
  \centering
  \caption{The average $R^2$ value of fitting of pressure distance from original, noisy and processed velocity field}
  \label{tab:R2}
  \begin{tabular}{c|c|c|c|c|c}
    \hline
    clean & noise type & noisy & SG & POD & AMIC \\
    \hline
    \multicolumn{6}{c}{synthetic channel flow}\\
    \hline
    \multirow{2}{*}{0.9978} & GWN  & 0.1090 & 0.5856 & 0.6091 & 0.9842 \\
    \cline{2-6}
     & CGWN & 0.0951 & 0.4546 & 0.1234 & 0.9743 \\
    \hline
    \multicolumn{6}{c}{synthetic airfoil wake}\\
    \hline
    \multirow{2}{*}{0.9735} & GWN  & 0.1441 & 0.5269 & 0.6513 & 0.9778 \\
    \cline{2-6}
     & CGWN & 0.0940 & 0.4101 & 0.2250 & 0.9811 \\
    \hline
  \end{tabular}
\end{table}

\section{Experimental application}
\label{sec:experiment}

\begin{figure}
\centering
\begin{subfigure}
    \centering
    \includegraphics[width=0.4\textwidth]{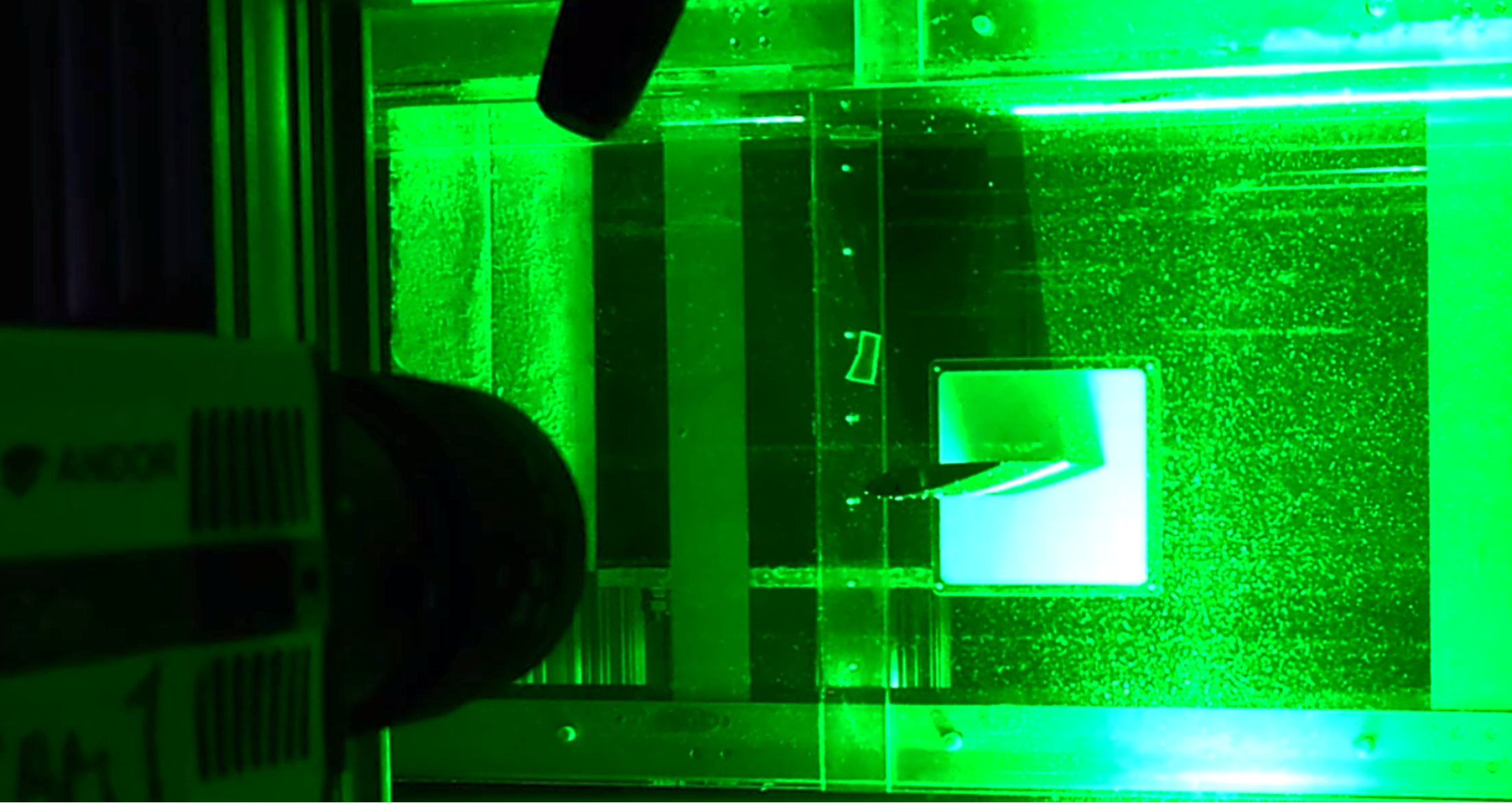}
\end{subfigure}
\caption{A photo of the experimental setup for the wake of a wing.}
\label{fig:wing_photo}
\end{figure}
AMIC performances are further evaluated using the data obtained from PIV measurements in the wake of a 2D wing model with a NACA 0018 section. A complete description of the experiment is reported in \cite{chen2022Pressure}, and here only briefly outlined. The wing is set at an angle of $10^\circ$ and has a chord of $80mm$. It is mounted in the water tunnel of Universidad Carlos III de Madrid, which has a test section of $2.5\times0.5\times0.55 m^3$. The Reynolds number, based on the chord and on the free stream velocity $U_0=0.06 m/s$, is equal to $4800$. A picture of the experimental setup is presented in Fig. \ref{fig:wing_photo}. 

The domain of interest is illuminated by a dual cavity pulsed Nd:YAG Quantel Evergreen. The recording is carried out with an Andor sCMOS camera ($2560\times2160 px^2$ sensor), equipped with a $50 mm$ focal length objective. The image resolution is set to $8.3 px/mm$. Time-resolved measurements are obtained with a sampling frequency of $30Hz$. The processing consists of a sliding correlation with a 3-frame kernel. The final interrogation window size is of $40px$ with $75\%$ overlap, leading to a vector spacing of $1.20 mm$. The vector fields consist of $110\times70$ vectors. The origin of the reference system is set at the trailing edge of the airfoil.

The statistics of error are computed over $240$ frames, although the filtering based on POD is trained on a larger dataset containing $3600$ frames to ensure convergence of the basis. The $\lambda$ for AMIC is set at $(0.1, 0.05)$, while the number of iterations is to be $8$, determined by Eq. \ref{eq:noiter}.
\begin{figure}[htb]
\flushleft
\begin{minipage}{0.96\linewidth}
    \hspace{3mm}
    \includegraphics[width=0.92\linewidth,trim=4mm 100mm 4mm 0mm,clip]{fig/AF_bar1.eps}
\end{minipage}
\flushleft
\begin{minipage}{0.03\linewidth}
    \rotatebox{90}{\hspace{7mm}PIV}
\end{minipage}
\begin{minipage}{0.92\linewidth}
    \begin{subfigure}
        \centering
        \includegraphics[width=0.96\linewidth,trim=0mm 24mm 6mm 24mm,clip]{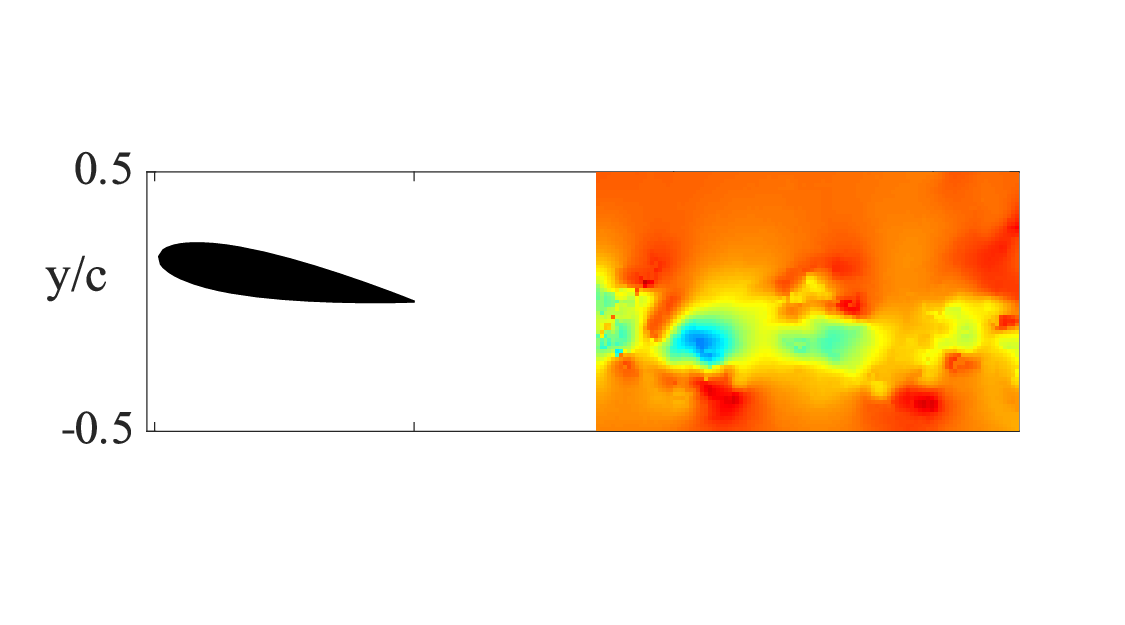}
    \end{subfigure}
\end{minipage}
\flushleft
\begin{minipage}{0.03\linewidth}
    \rotatebox{90}{\hspace{7mm}SG}
\end{minipage}
\begin{minipage}{0.92\linewidth}
    \begin{subfigure}
        \centering
        \includegraphics[width=0.96\linewidth,trim=0mm 24mm 6mm 24mm,clip]{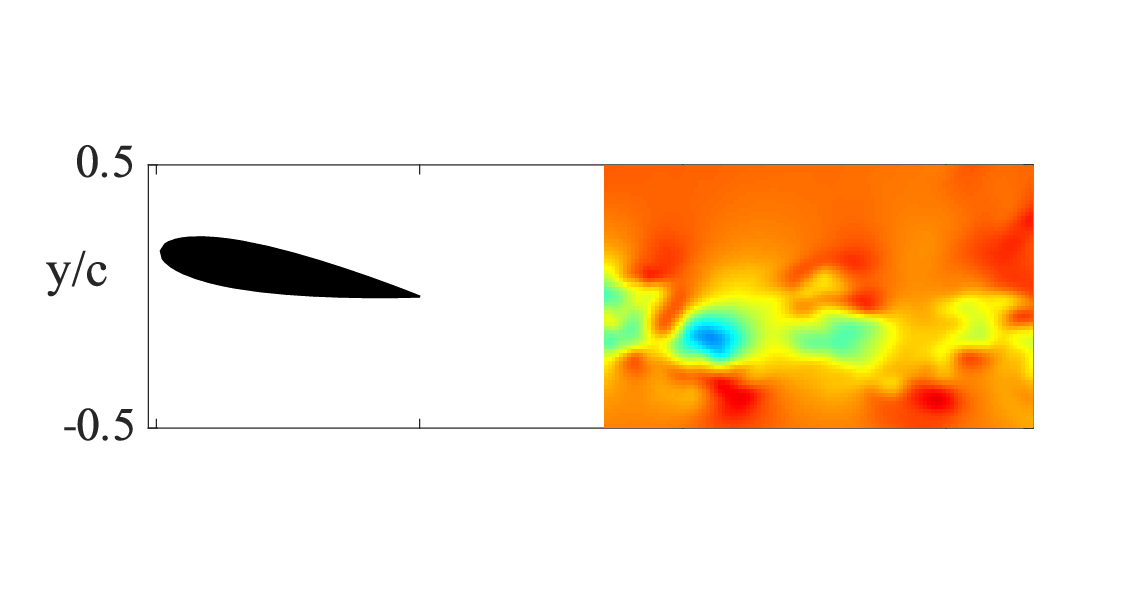}
    \end{subfigure}
\end{minipage}
\flushleft
\flushleft
\begin{minipage}{0.03\linewidth}
    \rotatebox{90}{\hspace{7mm}POD}
\end{minipage}
\begin{minipage}{0.92\linewidth}
    \begin{subfigure}
        \centering
        \includegraphics[width=0.96\linewidth,trim=0mm 24mm 6mm 24mm,clip]{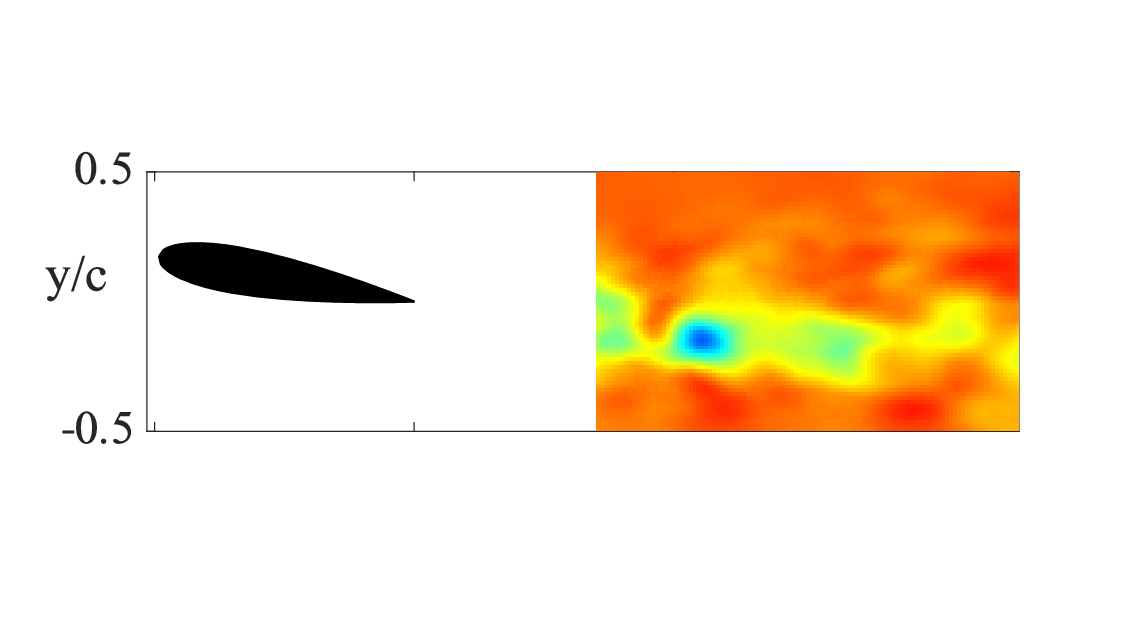}
    \end{subfigure}
\end{minipage}
\flushleft
\begin{minipage}{0.03\linewidth}
    \rotatebox{90}{\hspace{15mm}AMIC}
\end{minipage}
\begin{minipage}{0.92\linewidth}
    \begin{subfigure}
        \centering
        \includegraphics[width=0.96\linewidth,trim=0mm 0mm 6mm 24mm,clip]{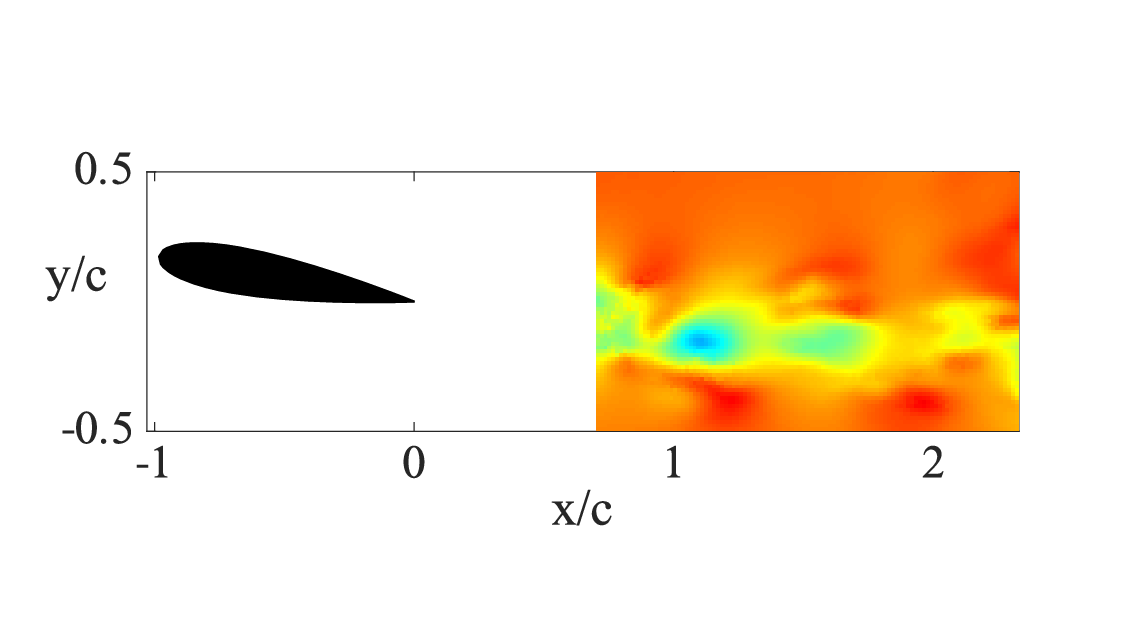}
    \end{subfigure}
\end{minipage}
\caption{Streamwise velocity field in the wake of an airfoil from PIV dataset, normalized by $U_0$. From top to bottom, unfiltered PIV result, SG, POD and AMIC.}
\label{fig:EXP_u}
\end{figure}
\begin{figure}[htb]
\flushleft
\begin{minipage}{0.96\linewidth}
    \hspace{3mm}
    \includegraphics[width=0.92\linewidth,trim=4mm 100mm 4mm 0mm,clip]{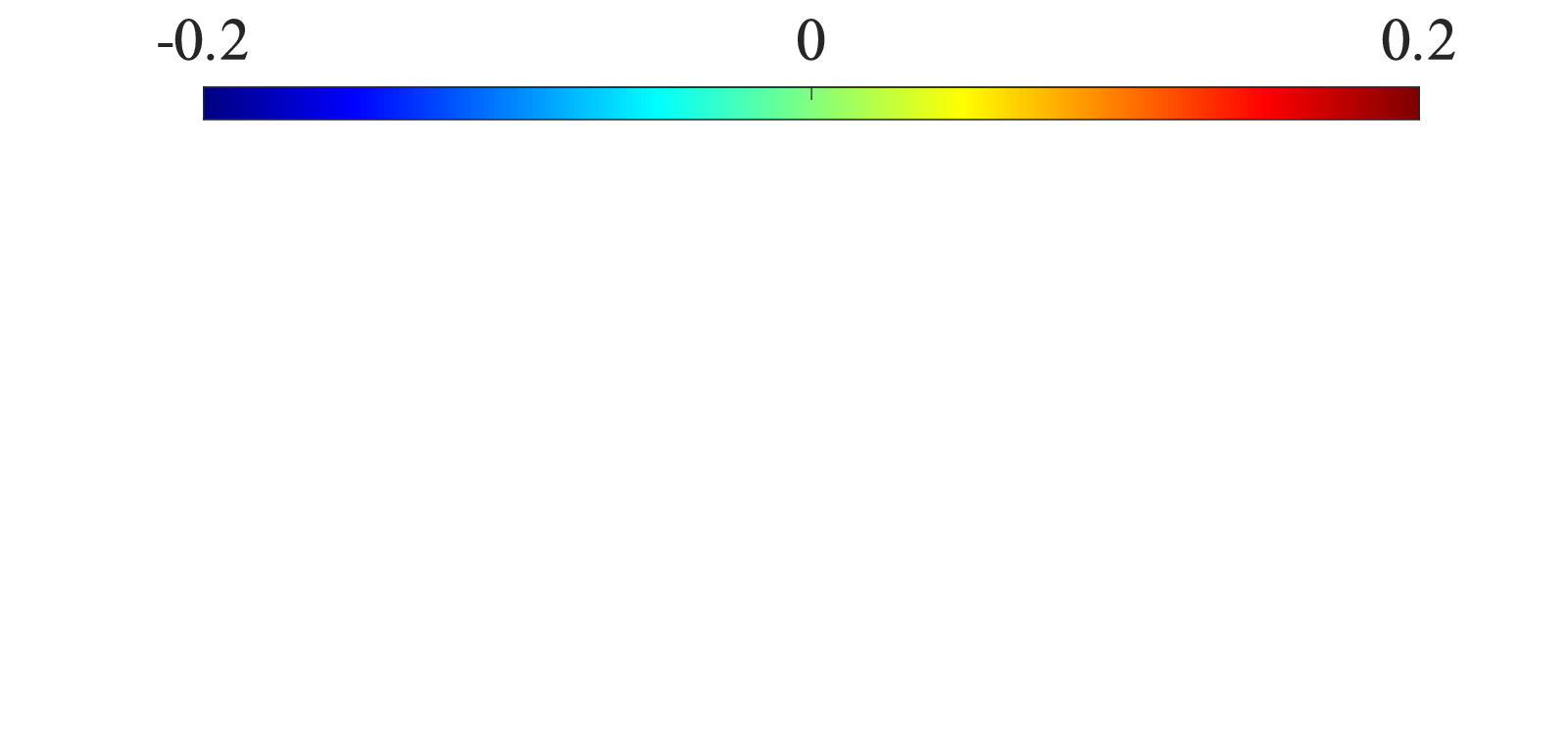}
\end{minipage}
\flushleft
\begin{minipage}{0.03\linewidth}
    \rotatebox{90}{\hspace{7mm}PIV}
\end{minipage}
\begin{minipage}{0.92\linewidth}
    \begin{subfigure}
        \centering
        \includegraphics[width=0.96\linewidth,trim=0mm 24mm 6mm 24mm,clip]{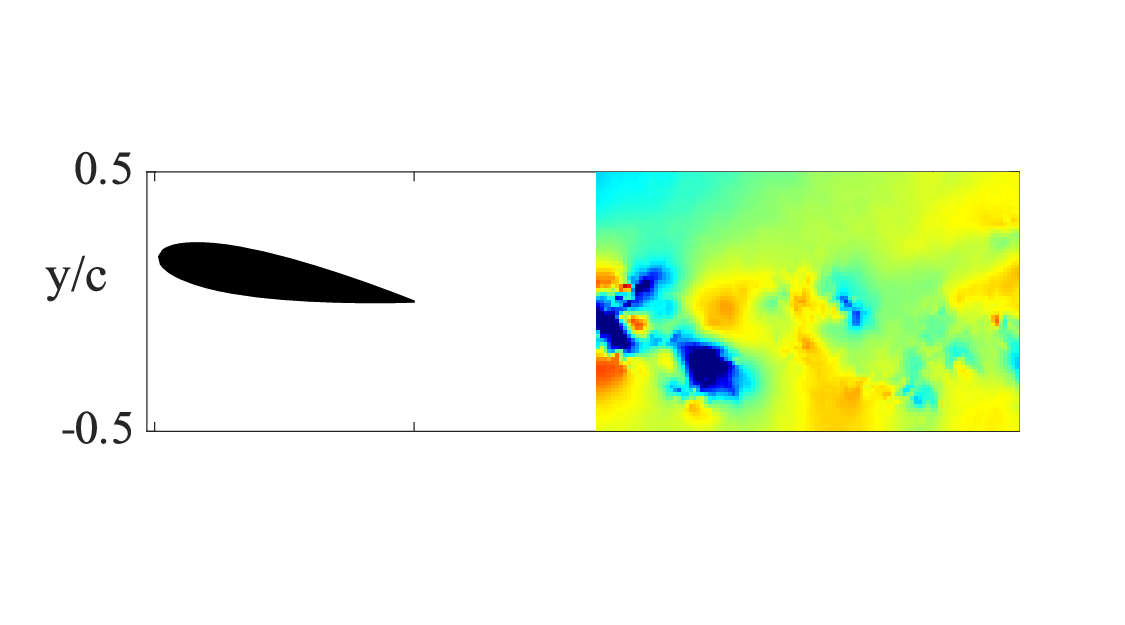}
    \end{subfigure}
\end{minipage}
\flushleft
\begin{minipage}{0.03\linewidth}
    \rotatebox{90}{\hspace{7mm}SG}
\end{minipage}
\begin{minipage}{0.92\linewidth}
    \begin{subfigure}
        \centering
        \includegraphics[width=0.96\linewidth,trim=0mm 24mm 6mm 24mm,clip]{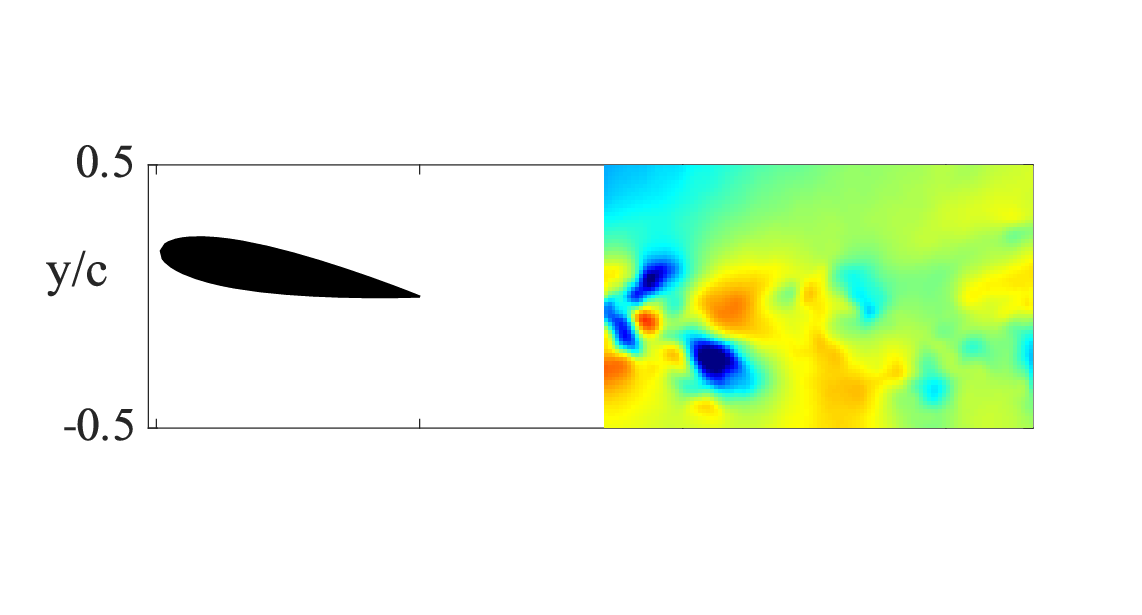}
    \end{subfigure}
\end{minipage}
\flushleft
\flushleft
\begin{minipage}{0.03\linewidth}
    \rotatebox{90}{\hspace{7mm}POD}
\end{minipage}
\begin{minipage}{0.92\linewidth}
    \begin{subfigure}
        \centering
        \includegraphics[width=0.96\linewidth,trim=0mm 24mm 6mm 24mm,clip]{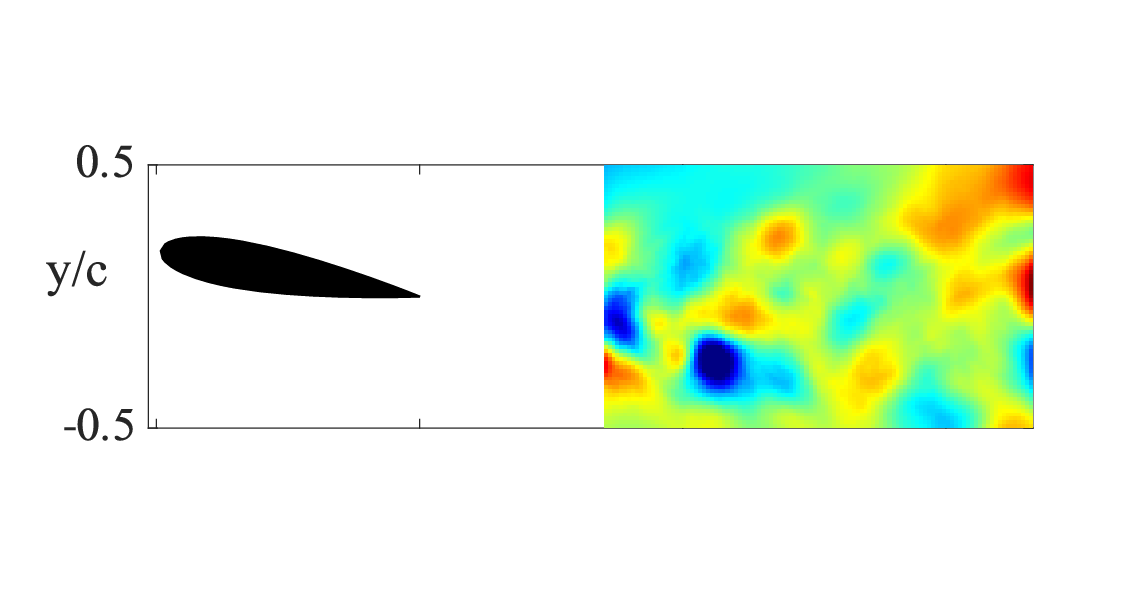}
    \end{subfigure}
\end{minipage}
\flushleft
\begin{minipage}{0.03\linewidth}
    \rotatebox{90}{\hspace{15mm}AMIC}
\end{minipage}
\begin{minipage}{0.92\linewidth}
    \begin{subfigure}
        \centering
        \includegraphics[width=0.96\linewidth,trim=0mm 0mm 6mm 24mm,clip]{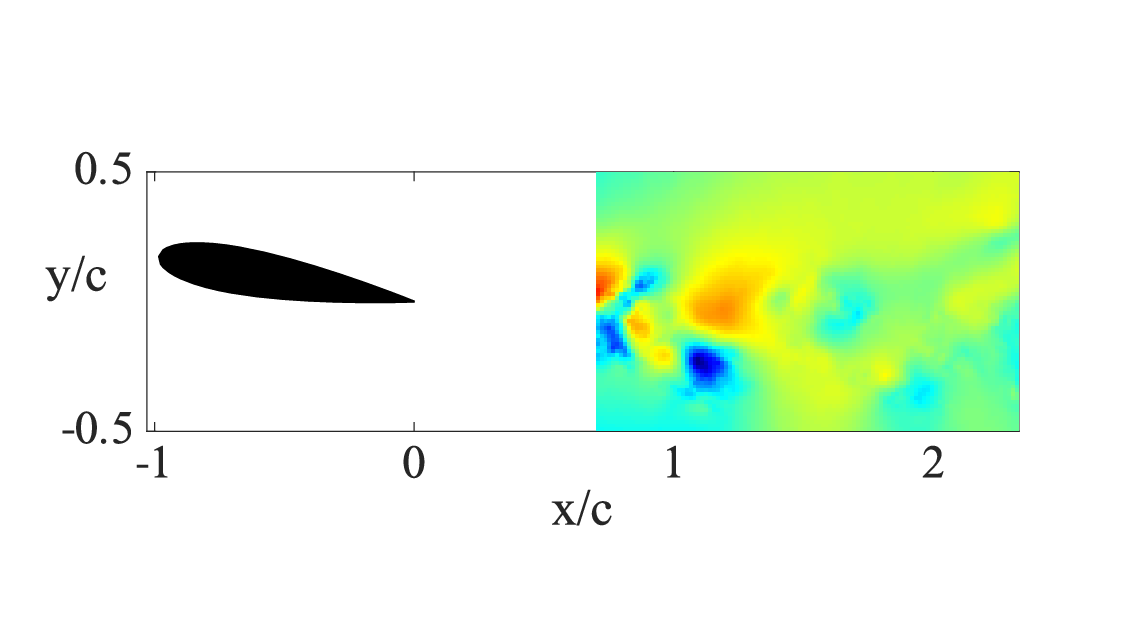}
    \end{subfigure}
\end{minipage}
\caption{Pressure field in the wake of an airfoil from PIV dataset, normalized by $\frac{1}{2}\rho U_0^2$ with $\rho$ being the density of the fluid. From top to bottom, unfiltered PIV result, SG, POD and AMIC.}
\label{fig:EXP_p}
\end{figure}
\begin{figure}[htbp]
\flushleft
\begin{minipage}{0.96\linewidth}
    \hspace{1mm}
    \includegraphics[width=0.92\linewidth,trim=4mm 100mm 4mm 0mm,clip]{fig/AF_bar1.eps}
\end{minipage}
\flushleft
\begin{minipage}{0.03\linewidth}
    \rotatebox{90}{\hspace{2mm}PIV}
\end{minipage}
\hspace{2mm}
\begin{minipage}{0.92\linewidth}
    \begin{subfigure}
        \centering
        \includegraphics[width=0.82\linewidth,trim=0mm 0mm 6mm 4mm,clip]{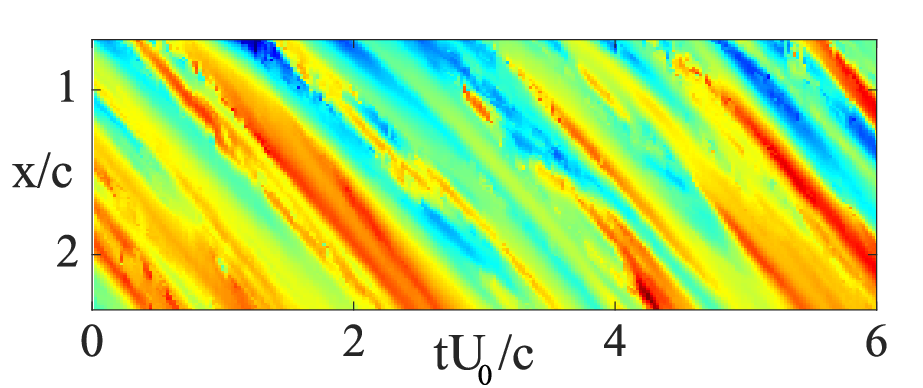}
    \end{subfigure}
\end{minipage}
\flushleft
\begin{minipage}{0.96\linewidth}
    \hspace{1mm}
    \includegraphics[width=0.92\linewidth,trim=4mm 100mm 4mm 0mm,clip]{fig/bar7.eps}
\end{minipage}
\flushleft
\begin{minipage}{0.03\linewidth}
    \rotatebox{90}{\hspace{2mm}PIV}
\end{minipage}
\hspace{2mm}
\begin{minipage}{0.92\linewidth}
    \begin{subfigure}
        \centering
        \includegraphics[width=0.82\linewidth,trim=0mm 12mm 6mm 4mm,clip]{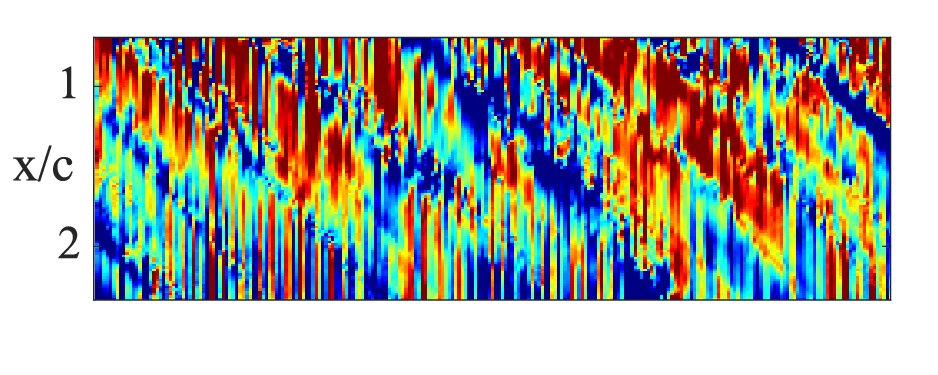}
    \end{subfigure}
\end{minipage}
\flushleft
\begin{minipage}{0.03\linewidth}
    \rotatebox{90}{\hspace{2mm}SG}
\end{minipage}
\hspace{2mm}
\begin{minipage}{0.92\linewidth}
    \begin{subfigure}
        \centering
        \includegraphics[width=0.82\linewidth,trim=0mm 12mm 6mm 4mm,clip]{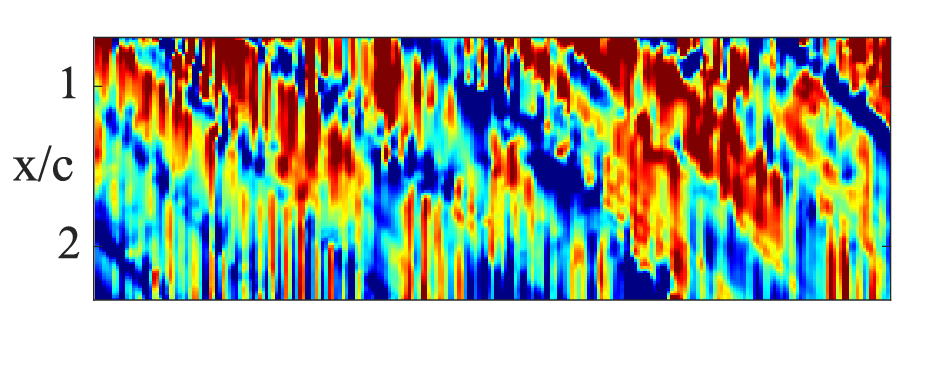}
    \end{subfigure}
\end{minipage}
\flushleft
\flushleft
\begin{minipage}{0.03\linewidth}
    \rotatebox{90}{\hspace{2mm}POD}
\end{minipage}
\hspace{2mm}
\begin{minipage}{0.92\linewidth}
    \begin{subfigure}
        \centering
        \includegraphics[width=0.82\linewidth,trim=0mm 12mm 6mm 4mm,clip]{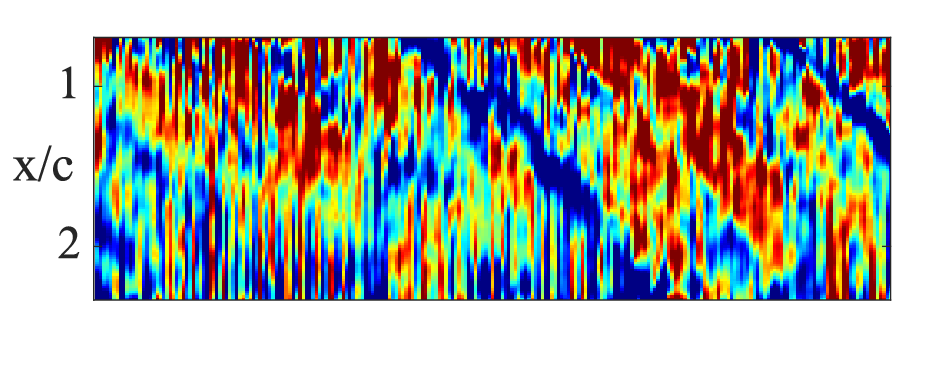}
    \end{subfigure}
\end{minipage}
\flushleft
\begin{minipage}{0.03\linewidth}
    \rotatebox{90}{\hspace{8mm}AMIC}
\end{minipage}
\hspace{2mm}
\begin{minipage}{0.92\linewidth}
    \begin{subfigure}
        \centering
        \includegraphics[width=0.82\linewidth,trim=0mm 0mm 6mm 4mm,clip]{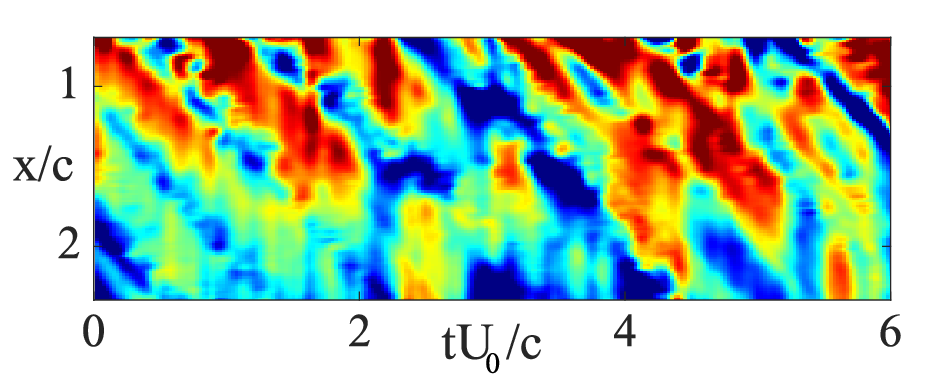}
    \end{subfigure}
\end{minipage}
\caption{The space-time diagram of streamwise velocity field (the upper block) and pressure field (the lower block) at $y/c = -0.07$ in the wake of an airfoil from PIV dataset, normalized by $U_0$ for the velocity field and $\frac{1}{2}\rho U_0^2$ for the pressure field with $\rho$ being the density of the fluid. The pressure diagram is displayed from top to bottom, unfiltered PIV result, SG, POD and AMIC.}
\label{fig:EXP_st}
\end{figure}

Fig.~\ref{fig:EXP_u} presents the velocity field from the original PIV dataset of the airfoil wake, alongside the results processed using the three aforementioned methods. The filtered velocity field seems to have no significant difference among them, apart from the POD truncated field which appears to lose some of the finer details. This trend is also visible in the pressure field calculated from the velocity data, as illustrated in Fig.~\ref{fig:EXP_p}. Here, high- and low-pressure regions appear in similar locations across both original and processed fields, with the primary differences occurring in the pressure values.

The spatiotemporal consistency of the fluctuations can be evaluated from the space-time diagrams in Fig. \ref{fig:EXP_st}. The diagrams of pressure from unfiltered PIV field, SG filtered and POD truncated field show strong flickering in time, even though this effect is not observed in the diagram of the velocity fields. This suggests that small errors in the velocity fields are strongly amplified in the pressure computation, producing the unphysical flickering. On the other hand, the AMIC process proves to be capable of suppressing it.

\begin{figure}[htbp]
\centering
\begin{subfigure}
    \centering
    \includegraphics[width=0.4\textwidth]{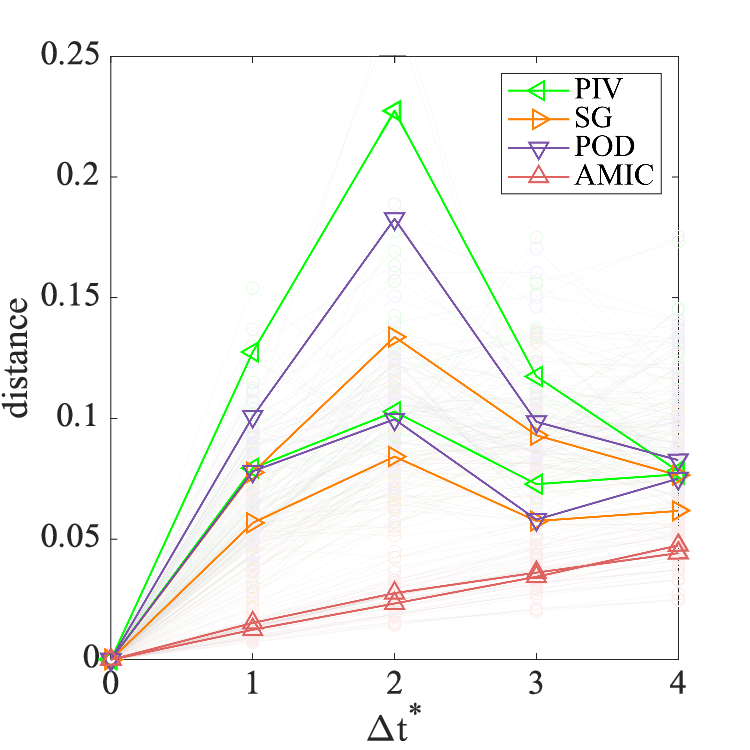}
\end{subfigure}
\caption{The distance of pressure field between adjacent frames versus the frame increments from the experimental dataset, where $\Delta t^*$ is the $\Delta t$ normalized by the sampling time, straighter is better, with two curves for each method to be highlighted.}
\label{fig:EXP_index}
\end{figure}

Given the lack of a ground truth for the pressure field in the present experimental dataset, a more quantitative evaluation of the filters can be obtained by applying the evaluation method introduced in Sec. \ref{sec:index}.
The graph of distance versus temporal interval, following Eq. \ref{eq:dis}, is plotted in Fig. \ref{fig:EXP_index} to observe how much the pressure fields follow a local linear relation. Each shaded curve bundle is composed of $50$ independent realizations, with $2$ among them being highlighted. The trend in the PIV results resembles the noise-superimposed fields from the synthetic cases, as pressure computation amplifies PIV uncertainties. The POD truncation has little improvement and the SG filter achieves more notable gains, however, a linear relationship between distance and temporal interval is not observed, indicating flickering in the pressure field over time. The AMIC processed field, on the other hand, displays nearly straight curves, aligning with the approximation of local linearity. The performance is further quantified using averages $R^2$ in Eq. \ref{eq:r2}. As listed in Tab. \ref{tab:EXP_R2}, the result from AMIC processed field is very close to $1$, providing further proof of the alignment of AMIC with the time-consistency sanity check we proposed.

\begin{table}[htb]
  \centering
  \caption{The average $R^2$ value of fitting of pressure distance from original and processed velocity field from the PIV dataset.}
  \label{tab:EXP_R2}
  \begin{tabular}{c|c|c|c}
    \hline
    PIV & SG & POD & AMIC \\
    \hline
    0.3878 & 0.6319 & 0.4525 & 0.9859 \\
    \hline
  \end{tabular}
\end{table}

\section{Conclusion}

We proposed an iterative correction based on an advection model to correct noisy time-resolved PIV fields. The method builds upon Taylor's hypothesis, i.e. small-scale motions are advected by large-scale motions. The correction is applied iteratively over multiple frames to tune the velocity field and reduce the error. In every iteration, the time-series of the velocity field is propagated by several frames forward and backwards according to the physical model, then the velocity field is updated by the weighted average of the original and propagated ones. Although not included for brevity, the computational cost is rather small (of the order of 1 minute per sample for the synthetic 3D channel flow) and scales approximately linearly with the number of vectors.

The AMIC correction provides two benefits. Firstly, this method smooths the velocity field across the dimension of different propagated time-series, thus the correction preserves most of the flow details in comparison to the conventional filters. Secondly, the AMIC correction leverages temporal information to refine spatial distribution and employs spatial information to enhance temporal evolution. This dual correction strategy minimizes errors in both temporal and spatial derivatives within the Navier-Stokes equation, yielding higher accuracy in pressure estimation from noisy velocity fields.

In its current implementation, AMIC is valid for flows in which advection is dominant. However, it can be easily modified to include other physical models. The simplistic advection-based model can replaced by more refined advection models, VIC or Navier-Stokes equations. This would allow the method to be tuned depending on the studied flow and to reduce the error in the velocity and pressure field estimation in specific cases.

\section*{Acknowledgements}
This project has received funding from the European Research Council (ERC) under the European Union’s Horizon 2020 research and innovation programme (grant agreement No 949085). Views and opinions expressed are however those of the authors only and do not necessarily reflect those of the European Union or the European Research Council. Neither the European Union nor the granting authority can be held responsible for them.  

\appendix

\section{The convective and fluctuating velocity in Taylor's hypothesis}

The advection-based model (see Eq.\ref{eqn:TH}) implemented in this work heavily relies on the estimate of the convective velocity and of the fluctuating velocity. Properly setting these quantities can help to improve the accuracy of velocity field propagation using Taylor's hypothesis. 
Different methods to estimate the convective velocity field can be found in the literature. Here we list the two main ways to get the convective velocity: \begin{itemize}
    \item using the temporal averaged velocity field, which is the most commonly used; 
    \item using a low-pass-filtered instantaneous velocity.
\end{itemize}
The second method requires to set a filter size, but on the upside it is more applicable to the situation of large-scale vortex shedding or intermittent flow when the temporal average is not necessarily representative of the instantaneous advection. 

In this paper, we use the second definition for the convection velocity field, where a 3D Gaussian filter with $\sigma = 7$ is used. We also apply a near-wall velocity profile correction to the streamwise velocity (considering $y$ is the wall-normal direction) to cancel the edge effect of filters, which is
\begin{equation}
    \mathbf{u}_c(x, y, z) = \mathbf{u}_f(x, y, z) - \overline{\mathbf{u}_f(x, y, z)}\mid_{x, z} + \mathbf{u}_{prof}(y)
\end{equation}
where $\mathbf{u}_{prof}(y)$ is the velocity profile averaged from the instantaneous velocity field, $\mathbf{u}_f$ is the filtered velocity, $\overline{\mathbf{u}_f}$ is the average of $\mathbf{u}_f$ at every layer of $y$ and $\mathbf{u}_c$ is the convective velocity as output.

\bibliographystyle{elsarticle-harv} 
\bibliography{reference}

\begin{thebibliography}{30}
\expandafter\ifx\csname natexlab\endcsname\relax\def\natexlab#1{#1}\fi
\providecommand{\url}[1]{\texttt{#1}}
\providecommand{\href}[2]{#2}
\providecommand{\path}[1]{#1}
\providecommand{\DOIprefix}{doi:}
\providecommand{\ArXivprefix}{arXiv:}
\providecommand{\URLprefix}{URL: }
\providecommand{\Pubmedprefix}{pmid:}
\providecommand{\doi}[1]{\href{http://dx.doi.org/#1}{\path{#1}}}
\providecommand{\Pubmed}[1]{\href{pmid:#1}{\path{#1}}}
\providecommand{\bibinfo}[2]{#2}
\ifx\xfnm\relax \def\xfnm[#1]{\unskip,\space#1}\fi
\bibitem[{Azijli and Dwight(2015)}]{azijli2015solenoidal}
\bibinfo{author}{Azijli, I.}, \bibinfo{author}{Dwight, R.P.}, \bibinfo{year}{2015}.
\newblock \bibinfo{title}{Solenoidal filtering of volumetric velocity measurements using gaussian process regression}.
\newblock \bibinfo{journal}{Experiments in fluids} \bibinfo{volume}{56}, \bibinfo{pages}{198}.
\bibitem[{Azijli et~al.(2016)Azijli, Sciacchitano, Ragni, Palha and Dwight}]{azijli2016posteriori}
\bibinfo{author}{Azijli, I.}, \bibinfo{author}{Sciacchitano, A.}, \bibinfo{author}{Ragni, D.}, \bibinfo{author}{Palha, A.}, \bibinfo{author}{Dwight, R.P.}, \bibinfo{year}{2016}.
\newblock \bibinfo{title}{A posteriori uncertainty quantification of piv-based pressure data}.
\newblock \bibinfo{journal}{Experiments in Fluids} \bibinfo{volume}{57}, \bibinfo{pages}{1--15}.
\bibitem[{Brindise and Vlachos(2017)}]{brindise2017proper}
\bibinfo{author}{Brindise, M.C.}, \bibinfo{author}{Vlachos, P.P.}, \bibinfo{year}{2017}.
\newblock \bibinfo{title}{Proper orthogonal decomposition truncation method for data denoising and order reduction}.
\newblock \bibinfo{journal}{Experiments in Fluids} \bibinfo{volume}{58}, \bibinfo{pages}{1--18}.
\bibitem[{Chen et~al.(2022)Chen, Raiola and Discetti}]{chen2022Pressure}
\bibinfo{author}{Chen, J.}, \bibinfo{author}{Raiola, M.}, \bibinfo{author}{Discetti, S.}, \bibinfo{year}{2022}.
\newblock \bibinfo{title}{Pressure from data-driven estimation of velocity fields using snapshot piv and fast probes}.
\newblock \bibinfo{journal}{Experimental Thermal and Fluid Science: International Journal of Experimental Heat Transfer, Thermodynamics, and Fluid Mechanics} , \bibinfo{pages}{136}.
\bibitem[{Christiansen(1973)}]{christiansen1973numerical}
\bibinfo{author}{Christiansen, I.}, \bibinfo{year}{1973}.
\newblock \bibinfo{title}{Numerical simulation of hydrodynamics by the method of point vortices}.
\newblock \bibinfo{journal}{Journal of Computational Physics} \bibinfo{volume}{13}, \bibinfo{pages}{363--379}.
\bibitem[{Draper and Smith(1998)}]{draper1998applied}
\bibinfo{author}{Draper, N.R.}, \bibinfo{author}{Smith, H.}, \bibinfo{year}{1998}.
\newblock \bibinfo{title}{Applied regression analysis}. volume \bibinfo{volume}{326}.
\newblock \bibinfo{publisher}{John Wiley \& Sons}.
\bibitem[{Epps and Krivitzky(2019)}]{epps2019singular}
\bibinfo{author}{Epps, B.P.}, \bibinfo{author}{Krivitzky, E.M.}, \bibinfo{year}{2019}.
\newblock \bibinfo{title}{Singular value decomposition of noisy data: noise filtering}.
\newblock \bibinfo{journal}{Experiments in Fluids} \bibinfo{volume}{60}, \bibinfo{pages}{1--23}.
\bibitem[{Foucaut and Stanislas(2002)}]{foucaut2002some}
\bibinfo{author}{Foucaut, J.M.}, \bibinfo{author}{Stanislas, M.}, \bibinfo{year}{2002}.
\newblock \bibinfo{title}{Some considerations on the accuracy and frequency response of some derivative filters applied to particle image velocimetry vector fields}.
\newblock \bibinfo{journal}{Measurement Science and Technology} \bibinfo{volume}{13}, \bibinfo{pages}{1058}.
\bibitem[{Gesemann(2015)}]{gesemann2015particle}
\bibinfo{author}{Gesemann, S.}, \bibinfo{year}{2015}.
\newblock \bibinfo{title}{From particle tracks to velocity and acceleration fields using b-splines and penalties}.
\newblock \bibinfo{journal}{arXiv preprint arXiv:1510.09034} .
\bibitem[{He et~al.(2024)He, Zeng, Wang, Wen and Liu}]{he2024four}
\bibinfo{author}{He, C.}, \bibinfo{author}{Zeng, X.}, \bibinfo{author}{Wang, P.}, \bibinfo{author}{Wen, X.}, \bibinfo{author}{Liu, Y.}, \bibinfo{year}{2024}.
\newblock \bibinfo{title}{Four-dimensional variational data assimilation of a turbulent jet for super-temporal-resolution reconstruction}.
\newblock \bibinfo{journal}{Journal of Fluid Mechanics} \bibinfo{volume}{978}, \bibinfo{pages}{A14}.
\bibitem[{de~Kat and Ganapathisubramani(2012)}]{de2012pressure}
\bibinfo{author}{de~Kat, R.}, \bibinfo{author}{Ganapathisubramani, B.}, \bibinfo{year}{2012}.
\newblock \bibinfo{title}{Pressure from particle image velocimetry for convective flows: a taylor’s hypothesis approach}.
\newblock \bibinfo{journal}{Measurement Science and Technology} \bibinfo{volume}{24}, \bibinfo{pages}{024002}.
\bibitem[{Van~der Kindere et~al.(2019)Van~der Kindere, Laskari, Ganapathisubramani and De~Kat}]{van2019pressure}
\bibinfo{author}{Van~der Kindere, J.}, \bibinfo{author}{Laskari, A.}, \bibinfo{author}{Ganapathisubramani, B.}, \bibinfo{author}{De~Kat, R.}, \bibinfo{year}{2019}.
\newblock \bibinfo{title}{Pressure from 2d snapshot piv}.
\newblock \bibinfo{journal}{Experiments in fluids} \bibinfo{volume}{60}, \bibinfo{pages}{1--18}.
\bibitem[{Lemke and Sesterhenn(2016)}]{lemke2016adjoint}
\bibinfo{author}{Lemke, M.}, \bibinfo{author}{Sesterhenn, J.}, \bibinfo{year}{2016}.
\newblock \bibinfo{title}{Adjoint-based pressure determination from piv data in compressible flows—validation and assessment based on synthetic data}.
\newblock \bibinfo{journal}{European Journal of Mechanics-B/Fluids} \bibinfo{volume}{58}, \bibinfo{pages}{29--38}.
\bibitem[{Li et~al.(2008)Li, Perlman, Wan, Yang, Meneveau, Burns, Chen, Szalay and Eyink}]{li2008JHTDB}
\bibinfo{author}{Li, Y.}, \bibinfo{author}{Perlman, E.}, \bibinfo{author}{Wan, M.}, \bibinfo{author}{Yang, Y.}, \bibinfo{author}{Meneveau, C.}, \bibinfo{author}{Burns, R.}, \bibinfo{author}{Chen, S.}, \bibinfo{author}{Szalay, A.}, \bibinfo{author}{Eyink, G.}, \bibinfo{year}{2008}.
\newblock \bibinfo{title}{A public turbulence database cluster and applications to study lagrangian evolution of velocity increments in turbulence}.
\newblock \bibinfo{journal}{Journal of Turbulence} \bibinfo{volume}{9}.
\bibitem[{Liu and Moreto(2020)}]{liu2020error}
\bibinfo{author}{Liu, X.}, \bibinfo{author}{Moreto, J.R.}, \bibinfo{year}{2020}.
\newblock \bibinfo{title}{Error propagation from the piv-based pressure gradient to the integrated pressure by the omnidirectional integration method}.
\newblock \bibinfo{journal}{Measurement Science and Technology} \bibinfo{volume}{31}, \bibinfo{pages}{055301}.
\bibitem[{McClure and Yarusevych(2017)}]{mcclure2017instantaneous}
\bibinfo{author}{McClure, J.}, \bibinfo{author}{Yarusevych, S.}, \bibinfo{year}{2017}.
\newblock \bibinfo{title}{Instantaneous piv/ptv-based pressure gradient estimation: a framework for error analysis and correction}.
\newblock \bibinfo{journal}{Experiments in Fluids} \bibinfo{volume}{58}, \bibinfo{pages}{1--18}.
\bibitem[{Pan et~al.(2016)Pan, Whitehead, Thomson and Truscott}]{pan2016error}
\bibinfo{author}{Pan, Z.}, \bibinfo{author}{Whitehead, J.}, \bibinfo{author}{Thomson, S.}, \bibinfo{author}{Truscott, T.}, \bibinfo{year}{2016}.
\newblock \bibinfo{title}{Error propagation dynamics of piv-based pressure field calculations: how well does the pressure poisson solver perform inherently?}
\newblock \bibinfo{journal}{Measurement Science and Technology} \bibinfo{volume}{27}, \bibinfo{pages}{084012}.
\bibitem[{Raiola et~al.(2015)Raiola, Discetti and Ianiro}]{Raiola2015POD}
\bibinfo{author}{Raiola, M.}, \bibinfo{author}{Discetti, S.}, \bibinfo{author}{Ianiro, A.}, \bibinfo{year}{2015}.
\newblock \bibinfo{title}{On piv random error minimization with optimal pod-based low-order reconstruction}.
\newblock \bibinfo{journal}{Experiments in fluids} \bibinfo{volume}{56}.
\bibitem[{Richardson(1911)}]{richardson1911ix}
\bibinfo{author}{Richardson, L.F.}, \bibinfo{year}{1911}.
\newblock \bibinfo{title}{Ix. the approximate arithmetical solution by finite differences of physical problems involving differential equations, with an application to the stresses in a masonry dam}.
\newblock \bibinfo{journal}{Philosophical Transactions of the Royal Society of London. Series A, Containing Papers of a Mathematical or Physical Character} \bibinfo{volume}{210}, \bibinfo{pages}{307--357}.
\bibitem[{Savitzky and Golay(1964)}]{savitzky1964smoothing}
\bibinfo{author}{Savitzky, A.}, \bibinfo{author}{Golay, M.J.}, \bibinfo{year}{1964}.
\newblock \bibinfo{title}{Smoothing and differentiation of data by simplified least squares procedures.}
\newblock \bibinfo{journal}{Analytical chemistry} \bibinfo{volume}{36}, \bibinfo{pages}{1627--1639}.
\bibitem[{Scarano et~al.(2022)Scarano, Schneiders, Saiz and Sciacchitano}]{scarano2022dense}
\bibinfo{author}{Scarano, F.}, \bibinfo{author}{Schneiders, J.F.}, \bibinfo{author}{Saiz, G.G.}, \bibinfo{author}{Sciacchitano, A.}, \bibinfo{year}{2022}.
\newblock \bibinfo{title}{Dense velocity reconstruction with vic-based time-segment assimilation}.
\newblock \bibinfo{journal}{Experiments in Fluids} \bibinfo{volume}{63}, \bibinfo{pages}{96}.
\bibitem[{Schiavazzi et~al.(2014)Schiavazzi, Coletti, Iaccarino and Eaton}]{schiavazzi2014matching}
\bibinfo{author}{Schiavazzi, D.}, \bibinfo{author}{Coletti, F.}, \bibinfo{author}{Iaccarino, G.}, \bibinfo{author}{Eaton, J.K.}, \bibinfo{year}{2014}.
\newblock \bibinfo{title}{A matching pursuit approach to solenoidal filtering of three-dimensional velocity measurements}.
\newblock \bibinfo{journal}{Journal of Computational Physics} \bibinfo{volume}{263}, \bibinfo{pages}{206--221}.
\bibitem[{Schneiders et~al.(2014)Schneiders, Dwight and Scarano}]{schneiders2014Time}
\bibinfo{author}{Schneiders, J.F.G.}, \bibinfo{author}{Dwight, R.P.}, \bibinfo{author}{Scarano, F.}, \bibinfo{year}{2014}.
\newblock \bibinfo{title}{Time-supersampling of 3d-piv measurements with vortex-in-cell simulation}.
\newblock \bibinfo{journal}{Experiments in Fluids} \bibinfo{volume}{55}, \bibinfo{pages}{1692}.
\bibitem[{Sciacchitano(2019)}]{sciacchitano2019uncertainty}
\bibinfo{author}{Sciacchitano, A.}, \bibinfo{year}{2019}.
\newblock \bibinfo{title}{Uncertainty quantification in particle image velocimetry}.
\newblock \bibinfo{journal}{Measurement Science and Technology} \bibinfo{volume}{30}, \bibinfo{pages}{092001}.
\bibitem[{de~Silva et~al.(2013)de~Silva, Philip and Marusic}]{silva2013minimization}
\bibinfo{author}{de~Silva, C.M.}, \bibinfo{author}{Philip, J.}, \bibinfo{author}{Marusic, I.}, \bibinfo{year}{2013}.
\newblock \bibinfo{title}{Minimization of divergence error in volumetric velocity measurements and implications for turbulence statistics}.
\newblock \bibinfo{journal}{Experiments in fluids} \bibinfo{volume}{54}, \bibinfo{pages}{1--17}.
\bibitem[{Sperotto et~al.(2022)Sperotto, Pieraccini and Mendez}]{sperotto2022meshless}
\bibinfo{author}{Sperotto, P.}, \bibinfo{author}{Pieraccini, S.}, \bibinfo{author}{Mendez, M.A.}, \bibinfo{year}{2022}.
\newblock \bibinfo{title}{A meshless method to compute pressure fields from image velocimetry}.
\newblock \bibinfo{journal}{Measurement Science and Technology} \bibinfo{volume}{33}, \bibinfo{pages}{094005}.
\bibitem[{Towne et~al.(2023)Towne, Dawson, Br{\`e}s, Lozano-Dur{\'a}n, Saxton-Fox, Parthasarathy, Jones, Biler, Yeh, Patel et~al.}]{towne2023database}
\bibinfo{author}{Towne, A.}, \bibinfo{author}{Dawson, S.T.}, \bibinfo{author}{Br{\`e}s, G.A.}, \bibinfo{author}{Lozano-Dur{\'a}n, A.}, \bibinfo{author}{Saxton-Fox, T.}, \bibinfo{author}{Parthasarathy, A.}, \bibinfo{author}{Jones, A.R.}, \bibinfo{author}{Biler, H.}, \bibinfo{author}{Yeh, C.A.}, \bibinfo{author}{Patel, H.D.}, et~al., \bibinfo{year}{2023}.
\newblock \bibinfo{title}{A database for reduced-complexity modeling of fluid flows}.
\newblock \bibinfo{journal}{AIAA journal} \bibinfo{volume}{61}, \bibinfo{pages}{2867--2892}.
\bibitem[{Van~Oudheusden(2013)}]{van2013piv}
\bibinfo{author}{Van~Oudheusden, B.}, \bibinfo{year}{2013}.
\newblock \bibinfo{title}{Piv-based pressure measurement}.
\newblock \bibinfo{journal}{Measurement Science and Technology} \bibinfo{volume}{24}, \bibinfo{pages}{032001}.
\bibitem[{Wang et~al.(2016)Wang, Gao, Wang, Wei and Wang}]{wang2016Divergence}
\bibinfo{author}{Wang, C.Y.}, \bibinfo{author}{Gao, Q.}, \bibinfo{author}{Wang, H.P.}, \bibinfo{author}{Wei, R.J.}, \bibinfo{author}{Wang, J.J.}, \bibinfo{year}{2016}.
\newblock \bibinfo{title}{Divergence-free smoothing for volumetric piv data}.
\newblock \bibinfo{journal}{Experiments in Fluids} \bibinfo{volume}{57}, \bibinfo{pages}{1--23}.
\bibitem[{Zhang et~al.(2022)Zhang, Bhattacharya and Vlachos}]{zhang2022uncertainty}
\bibinfo{author}{Zhang, J.}, \bibinfo{author}{Bhattacharya, S.}, \bibinfo{author}{Vlachos, P.P.}, \bibinfo{year}{2022}.
\newblock \bibinfo{title}{Uncertainty of piv/ptv based eulerian pressure estimation using velocity uncertainty}.
\newblock \bibinfo{journal}{Measurement Science and Technology} \bibinfo{volume}{33}, \bibinfo{pages}{065303}.

\end{thebibliography}






\end{document}